\documentclass[12pt]{article}
\usepackage[paper=a4paper,margin=1in]{geometry}
\usepackage{t1enc}      

\usepackage{amsmath}
\usepackage{amsfonts}
\usepackage{amssymb}
\usepackage{amsthm}
\usepackage{graphicx}
\usepackage{mathrsfs}  

\newcommand{\edth} {\mbox{\symbol{'360}}}

\newcommand{\bi}{\bf i}
\newcommand{\bj}{\bf j}

\newtheorem*{theorem*}{Theorem}

\providecommand{\keywords}[1]
{\small	\textbf{\textit{Keywords:}} #1 }

\numberwithin{equation}{section}

\begin{document}
\bibliographystyle{unsrt}

\title{Minkowski space from quantum mechanics}
\author{L\'aszl\'o B. Szabados \\
Wigner Research Centre for Physics, \\
H-1525 Budapest 114, P. O. Box 49, EU, \\
E-mail: lbszab@rmki.kfki.hu}

\maketitle

\begin{abstract}
Penrose's Spin Geometry Theorem is extended further, from $SU(2)$ and 
$E(3)$ (Euclidean) to $E(1,3)$ (Poincar\'e) invariant elementary quantum 
mechanical systems. The Lorentzian spatial distance between any two 
non-parallel timelike straight lines of Minkowski space, considered to be 
the centre-of-mass world lines of $E(1,3)$-invariant elementary classical 
mechanical systems with positive rest mass, is expressed in terms of 
\emph{$E(1,3)$-invariant basic observables}, viz. the 4-momentum and the 
angular momentum of the systems. An analogous expression for 
\emph{$E(1,3)$-invariant elementary quantum mechanical systems} in terms 
of the \emph{basic quantum observables} in an abstract, algebraic 
formulation of quantum mechanics is given, and it is shown that, in the 
classical limit, it reproduces the Lorentzian spatial distance between 
the timelike straight lines of Minkowski space with asymptotically 
vanishing uncertainty. Thus, the \emph{metric structure} of Minkowski 
space can be recovered from quantum mechanics in the classical limit 
using only the observables of abstract quantum mechanical systems. 
\end{abstract}

\keywords{Spin-Geometry Theorem, centre-of-mass states, classical limit of 
relativistic quantum systems, empirical Lorentzian distance}


\section{Introduction}
\label{sec-1}

\subsection{The classical spacetime}
\label{sub-1.1}

An \emph{operational definition} of the classical spacetime of special and 
general relativity is based on the use of freely moving/falling point-like 
(test) particles and light rays \cite{EPS}, and to measure the distance 
between these particles idealized standard (purely classical geometric 
\cite{KH,MW} or atomic quantum \cite{Pe68}) clocks, mirrors and light rays 
are used. Thus, in particular, the classical spacetime structure is defined 
by \emph{classical point mechanical} and \emph{geometrical optical} notions. 
(Classical \emph{field theoretical} concepts, e.g. the energy-momentum 
tensor, are used only in the introduction of Einstein's field equations. 
For a strict axiomatic treatise of the ideas of \cite{EPS}, see 
\cite{Linn}.) 

Historically, however, the well known structure of the Galilei and Minkowski 
spacetime was read off from the structure of the Galilei--Newton mechanics 
and Maxwell's electrodynamics, respectively. In particular, the Galilei and 
Poincar\'e symmetries were recognized first as the symmetries of the 
equations of motion in the theories describing these particular material 
physical systems. The notion of spacetime with these symmetries was 
introduced only later as a \emph{useful notion} to specify the 
spatio-temporal location of phenomena happening with the concrete physical 
systems in a convenient and transparent way. 

Although the classical Galilei or Minkowski spacetime is defined to be the 
set of the idealized (viz. structureless, i.e. pointlike and instantaneous) 
\emph{classical} events (abstracted from the realistic events happening 
with the existing \emph{macroscopic physical object}, e.g. from collisions 
of billiard balls), this notion turned out to be very useful even in 
quantum physics \cite{StWi,Haag}. In particular, many of the quantum field 
theoretical calculations became much more transparent when special 
relativistic spacetime concepts, e.g. Lorentz covariance, are used.

\subsection{The problem and a possible resolution}
\label{sub-1.2}

However, the great successes of quantum theory should not hide the absurdity 
that, e.g. in quantum mechanics, the independent variables in the argument 
of the wave function of an electron are defined e.g. by collisions of 
classical, macroscopic billiard balls; or, in quantum electrodynamics, the 
causal structure that the propagators are based on are defined by light 
rays in the geometrical optical approximation of classical electrodynamics. 
The resolution of this contradiction is that the \emph{a priori} `spacetime', 
used as a background in the quantum physical \emph{calculations}, and the 
\emph{operationally defined} `physical spacetime' in which we arrange the 
events (e.g. in a scattering process) are \emph{not the same}. The former 
is a geometrical representation of certain aspects of the \emph{internal 
structure} of the quantum systems, viz. that it is the dual of the momentum 
space defined via the Fourier transform\footnote{
Since the momentum space is a Lorentzian \emph{vector space} with the 
\emph{physically distinguished} origin, the group of its isometries is 
only the homogeneous Lorentz (rather than the Poincar\'e) group, and its 
\emph{algebraic dual} is also a vector space. However, its dual space 
becomes an \emph{affine vector space} (i.e. without any distinguished point 
as its origin) if the dual is defined via the Fourier transform of the wave 
functions $\phi$ on the momentum space in which not only a constant, but 
even a special $p^a$-dependent phase ambiguity is allowed, i.e. when 
$\phi$ and $\tilde\phi$, defined pointwise by $\tilde\phi:p^a\mapsto\exp(
{\rm i}p_e\xi^e/\hbar)\phi(p^a)$, are considered to be equivalent for any 
$\xi^e\in\mathbb{R}^4$ (see e.g. \cite{Geroch}, and Appendix \ref{sub-A.2} 
of the present paper). This makes the dual space, i.e. the resulting 
Minkowski \emph{coordinate} space, a \emph{homogeneous space} with the 
larger Poincar\'e group as its group of isometries. Thus, in the present 
framework, while the Lorentzian symmetries of the Minkowski coordinate 
space come from the isometries of the momentum space, the translational 
symmetries have a different root; and this latter could be considered to be 
of \emph{quantum mechanical origin}.}, 
but the latter is the set of the \emph{actual events}\footnote{
In \cite{Haag90}, the \emph{irreversibility} is also attributed to the 
events. Moreover, the fact that Einstein's field equations can be derived 
from thermodynamical principles \cite{Jacobson} may strengthen further 
the view that all the structures of the `physical spacetime' are of 
emergent nature.}
that happen in the Universe. 

The theoretical significance of the clarification of the origin of the 
structures of the physical spacetime is twofold: first, this might yield 
a deeper understanding \emph{how the classical physical world that we see 
emerges from quantum theory}; and, second, if gravity is the non-triviality 
of the geometry of the \emph{operationally defined} physical spacetime (as 
we understand it according to general relativity), then these investigations 
may, ultimately, reveal \emph{how gravity, and, in particular, its universal 
nature, is rooted in the quantum world}. Thus, in a conceptually coherent 
approach, \emph{the spacetime should be re-defined by purely quantum 
physical concepts}, without using directly any classical physical structure, 
and, in particular, any \emph{a priori} notion of spacetime. 

As a first step to derive the classical spacetime geometry from purely 
quantum concepts, already in 1966, Penrose suggested the spin network 
model of spacetime \cite{Pe79}. Using combinatorial techniques, he showed 
that, in the classical limit, the geometry of directions in Euclidean 
3-space (i.e. the \emph{conformal structure} of $\mathbb{R}^3$) can be 
recovered from this model \cite{Pe79,Penrose}. Later, this result was 
derived using the more familiar formalism of quantum mechanics, and it 
became known as the Spin Geometry Theorem \cite{Mo}. 

In our previous paper \cite{Sz21c}, we re-derived the above result in an 
even simpler and more direct way in the \emph{algebraic formulation} of 
quantum mechanics, in which the system was, in fact, the formal union of 
a large number of \emph{$SU(2)$-invariant elementary quantum mechanical 
systems}. Also, in an analogous way but extending the symmetry group from 
$SU(2)$ to the quantum mechanical Euclidean group $E(3)$ (and using 
geometric methods and complex techniques developed in general relativity 
to work out the classical limit), we recovered the \emph{metric} (rather 
than only the conformal) structure of the Euclidean 3-space \cite{Sz22a}. 
The next step in this program would be the derivation of the metric of 
the flat Minkowski space. The aim of the present paper is to show that 
\emph{the metric structure of the Minkowski space can indeed be recovered 
in this way by extending the symmetry group further into the quantum 
mechanical Poincar\'e group $E(1,3)$}. (By the \emph{quantum mechanical} 
Poincar\'e group $E(1,3)$ we mean the semidirect product of the group of 
translations and $SL(2,\mathbb{C})$, i.e. the universal covering of the 
connected component of the \emph{classical} Poincar\'e group, which is 
the isometry group of the Minkowski space.)

\subsection{The strategy}
\label{sub-1.3}

The main idea is that, instead of the classical test particles in the 
operational definition of the physical spacetime, we should use 
\emph{$E(1,3)$-invariant elementary quantum mechanical systems}, and the 
various geometrical structures of the spacetime should be introduced by the 
\emph{observables} of (rather than measurements on) these quantum systems. 
Then, in the spacetime structure that emerges in this way, the intrinsic 
quantum nature of the elementary systems is manifested in a nontrivial way, 
and the various geometrical structures of the classical spacetime should 
emerge from these observables in the classical limit. 

Since the general strategy of the present approach has only been sketched 
earlier \cite{Sz21c,Sz22a}, here we summarize its key points in a more 
coherent and a bit more detailed way: 

First, following \cite{Pe79,Penrose}, our approach is also positivistic as 
it does not depend on any \emph{a priori} notion of spacetime. However, it 
is a bit `more Machian' than that of \cite{Pe79,Penrose} in the sense that 
the quantum systems are \emph{not} intended to be used to define any notion 
like `direction' or `position'. We define only \emph{relative orientations} 
by some sort of `empirical angles' and \emph{relative positions} by 
`empirical distances' between \emph{elementary subsystems of a large 
composite system}, even though the `directions' and `positions' themselves 
are \emph{not} defined at all. The `empirical geometry' of the physical 
spacetime should ultimately be synthesized from these `empirical geometrical 
quantities'. (For the idea how to introduce the geometry from empirically 
given distances between \emph{pairs of points}, see e.g. \cite{MW}.) 

Second, since at the fundamental level \emph{no} instruments (e.g. mirrors 
or clocks) exist, the notion of physical spacetime cannot be based on 
\emph{concrete experimental/measuring processes}. Rather, it should be 
based on the use of \emph{observables} of the quantum systems. The various 
geometrical quantities are built exclusively from the basic quantum 
observables in the analogous way how the corresponding classical geometrical 
quantities are built from the basic classical observables. (Hence, this 
aspect of the present approach is perhaps a bit `Platonist' rather than 
positivistic, and not `instrumentalist' at all.) 

Third, according to the abstract, algebraic formulation of quantum theory, 
a quantum system is thought to be specified completely if the algebra of 
its observables and its representation are given. In particular, 
\emph{elementary quantum mechanical systems with symmetry group $G$} are 
specified by the ($C^*$-closure of the) universal enveloping algebra 
${\cal A}$ of the Lie algebra of $G$ in which the observables are the 
self-adjoint elements, and the states belong to the carrier space of a 
\emph{unitary, irreducible representation} of $G$ on some complex, 
separable Hilbert space ${\cal H}$. This representation yields a 
representation of ${\cal A}$, too. The self-adjoint generators of the 
action of $G$ will be called the \emph{basic quantum observables}. (For the 
Poincar\'e group, this notion of elementary quantum mechanical systems was 
introduced by Newton and Wigner \cite{NeWign}.) It is this symmetry, 
attributed to the elementary quantum physical systems, that can be expected 
to emerge in some way as the symmetry of the resulting physical spacetime; 
just like the Galilei and Poincar\'e symmetries of the Galilei and Minkowski 
spacetimes that came, respectively, from the Galilei--Newton mechanics and 
Maxwell's electrodynamics (see the second paragraph above). 

Moreover, since general relativity is the counterpart of the quantum 
theory as a \emph{general framework} rather than the theory of any 
particular interaction (as it is manifested in the universality of free 
fall via the Galilei--E\"otv\"os experiments), the resulting spacetime 
structure should also share this universality. Therefore, the quantum 
observables that the construction is based on must depend only on the 
\emph{kinematical framework} of the quantum systems, but they may 
\emph{not} depend e.g. on any particular Hamiltonian. It is these 
elementary systems that replace the classical point particles in the 
operational definition of the classical physical spacetime. 

Finally, to recover the classical spacetime structure from the above model 
an appropriate notion of \emph{classical limit} is needed. Although 
there are attempts to represent the classical limit of a quantum system by 
special but otherwise completely regular states of the quantum system (e.g. 
by considering the classical theory to be the $\hbar\to0$ limit of the 
quantum theory \cite{WerLand,Feintzeig}, or by some version of the coherent 
states, see e.g. \cite{Sz21} and the references therein to a sample from 
the extended literature), here, following Wigner \cite{Wi}, Ch. 27, we adopt 
a much weaker notion of classical limit: we do \emph{not} require the states 
of the classical system to be among the states of the quantum system. This 
limit is defined through a (not necessarily convergent) sequence of quantum 
states, and only the \emph{expectation value} of the \emph{basic} quantum
observables that is  required to tend to their large classical value, 
formally to infinite, in such a way that the corresponding standard 
deviations tend to zero \emph{relative} to the expectation values. Here, 
the classical \emph{theory} is not expected to be a limit of the quantum 
\emph{theory} in any sense, and hence the states of the classical theory 
need \emph{not} to be states of the quantum system either. 

In the above notion of \emph{elementary} $G$-invariant quantum mechanical 
systems not only the algebra ${\cal A}$ of observables, but the unitary, 
irreducible representation, labeled by the value of the Casimir invariants, 
is also fixed. The definition of the distance operator and the empirical 
distance will be based on this notion in subsections \ref{sub-4.1} and 
\ref{sub-4.2}, respectively. However, in the definition of the classical 
limit we should consider a sequence of states whose elements belong to 
\emph{different} irreducible representations. Thus, we should also consider 
the more general $G$-invariant quantum mechanical systems with the 
\emph{same} algebra ${\cal A}$ of observables, but the representation is 
the direct sum/integral of all its unitary irreducible representations. 
Strictly speaking, it is this more general notion of $E(1,3)$-invariant 
quantum mechanical systems that we should use in our Theorem in subsection 
\ref{sub-4.3}.

\subsection{The key ingredients and the main result}
\label{sub-1.4}

One of the key notions of the present paper, viz. the \emph{empirical 
distance}, is based on the observation that the intrinsic spin part of 
the angular momentum of a \emph{composite} system is the sum of the spin 
of the constituent subsystems and their \emph{relative orbital angular 
momenta}; and, in the traditional formulation of quantum mechanics, the 
latter is an expression of the distance between the centre-of-mass 
(world)lines of the subsystems and their momenta. This distance could in 
fact be recovered as the classical limit of the expectation value of 
expressions of the abstract basic quantum observables of $E(3)$-invariant 
elementary quantum mechanical systems \cite{Sz22a}; and, as we will see, 
this can indeed be done in the present case, too. (The idea that the 
distance could be inferred \emph{somehow} from the angular momentum has 
already been raised by Penrose in \cite{Pe79}.) 

The other key ingredient is the appropriate mathematical form of the idea 
of the \emph{classical limit of Poincar\'e-invariant quantum systems}. This 
is introduced in two steps, generalizing Wigner's notion of the classical 
limit of $SU(2)$-invariant quantum systems first to $E(3)$-invariant ones 
(as we did it in \cite{Sz22a}), and then, in the present paper, to 
$E(1,3)$-invariant systems with positive rest mass. These notions are based 
on another notions, viz. on the canonical angular momentum states in the 
$SU(2)$-invariant case, on special centre-of-mass states in the 
$E(3)$-invariant case, and on special co-moving, centre-of-mass states in 
the present case. The latter are the states in which the expectation value 
of the centre-of-mass vector operator is vanishing, the expectation value of 
the energy-momentum vector operator has only the time component, and the 
Pauli--Lubanski spin has only a single spatial component. 

In the present paper, we found an explicit expression for the empirical 
distance in terms of the 4-momenta and angular momenta, 
and also the mathematical form of the states defining the classical limit 
of Poincar\'e-invariant quantum systems for which our main result, 
formulated mathematically in the Theorem in subsection \ref{sub-4.3}, 
is proven: we show that for any finite number of timelike straight lines 
in Minkowski space we can always find $E(1,3)$-invariant quantum 
mechanical systems such that the empirical distance between them reduces in 
the classical limit just to the classical Lorentzian distance between them. 
Therefore, \emph{the metric structure of the Minkowski space can be 
recovered from the abstract $E(1,3)$-invariant quantum mechanical systems}. 
Also, this result of the present investigations confirms that it is the 
world lines of the freely moving particles (rather than the points) that 
should be considered to be the elementary objects in spacetime, just in 
accordance with the basic idea of \cite{EPS}. The actual notion of the 
classical limit of $E(1,3)$-invariant quantum systems as well as their 
co-moving, centre-of-mass states may have significance in other problems in 
relativistic quantum theory, independently of the present context. 

Nevertheless, although the strategy of the present investigations is the 
same and simple as that in the $E(3)$-invariant case, technically the 
calculations in the present $E(1,3)$-invariant case is considerably longer 
and more complicated; but they are mostly only routine ones.

\subsection{The structure of the paper}
\label{sub-1.5}

In the next section, we summarize the key properties of the 
Poincar\'e-invariant classical mechanical systems, express the distance 
between the world lines of any two of them (with positive rest mass) by 
basic classical observables, and show that the knowledge of this distance 
between any two timelike straight lines is equivalent to the knowledge of 
the metric structure of the Minkowski space. In Section \ref{sec-3}, the 
$E(1,3)$-invariant elementary quantum mechanical systems are defined, and 
their co-moving, centre-of-mass states as well as the mathematical notion 
of their classical limit are introduced. Section \ref{sec-4} is devoted to 
composite systems consisting of two (and, in principle, any finite number 
of) $E(1,3)$-invariant elementary quantum mechanical systems. 
The empirical distance between the elementary subsystems is introduced 
and its classical limit is calculated here. The key result of the paper 
is summarized in a theorem also in this section. The main part of the 
paper concludes with some final remarks in Section \ref{sec-5}. 

Since the representation theory of the Poincar\'e group, at least in the 
form that we use in the present paper, is not very well known in the 
relativity and differential geometry communities; and also the geometric 
and complex techniques and methods developed in general relativity are 
almost completely unknown outside the relativity community, for the sake 
of completeness we summarize the key elements of these ideas and methods 
in the appendices. 

The signature of the Lorentzian metric is chosen to be $(+,-,-,-)$. Small 
Latin indices, say $a,b,c,...,h=0,...,3$ and $i,j,k,...=1,2,3$, are 
concrete tensor indices, referring to a fixed orthonormal basis in the 
momentum space; and the capital Latin indices, e.g. $A,B,C,...=0,1$, are 
concrete spinor name indices with respect to a fixed spin frame 
associated with this orthonormal vector basis. We do not use abstract 
indices. Round/square brackets around indices denote 
symmetrization/anti-symmetrization. We use the units in which $c=1$, but 
Planck's constant $\hbar$ is kept. Our standard references to the 
spinorial and complex techniques are \cite{PR,HT}, but a more concise 
summary of them is given in the appendices of \cite{Sz21}. 


\section{Poincar\'e invariant elementary classical mechanical systems}
\label{sec-2}

\subsection{The definition of the elementary systems}
\label{sub-2.1}

A classical mechanical system will be called a Poincar\'e invariant
\emph{elementary} system (or, for brevity, a single particle) if its states 
can be characterized \emph{completely} by the energy-momentum and angular 
momentum, $p^a$ and $J^{ab}$, respectively; and under a Lorentz 
transformation they transform as a Lorentzian 4-vector and anti-symmetric 
tensor, respectively, while under a translation with $\xi^a\in\mathbb{R}^4$ 
they change as $(p^a,J^{ab})\mapsto(\tilde p^a,\tilde J^{ab}):=(p^a,J^{ab}+2
\xi^{[a}p^{b]})$. $p^a$ and $J^{ab}$ are elements of the dual space of the 
commutative ideal and the $so(1,3)$ Lie sub-algebra of the Poincar\'e 
algebra $e(1,3)$, respectively. The commutative ideal is endowed with the 
Lorentzian metric $\eta_{ab}:={\rm diag}(1,-1,-1,-1)$. This metric 
identifies the space of translation generators with the momentum space, 
which becomes a Minkowski vector space $\mathbb{R}^{1,3}=(\mathbb{R}^4,
\eta_{ab})$; and it determines a metric on the whole tensor algebra over it. 
We lower and raise freely the small Latin indices $a,b,c,...$ by $\eta_{ab}$ 
and its inverse, $\eta^{ab}$, respectively. Also, this yields the Lie 
brackets on the space of the basic observables, 
\begin{eqnarray}
&{}&\bigl\{p^a,p^b\bigr\}=0, \hskip 20pt 
  \bigl\{p^a,J^{bc}\bigr\}=\eta^{ab}p^c-\eta^{ac}p^b,  \label{eq:2.1.1a}\\
&{}&\bigl\{J^{ab},J^{cd}\bigr\}=\eta^{ac}J^{db}-\eta^{ad}J^{cb}+\eta^{bd}J^{ca}-
  \eta^{bc}J^{da}, \label{eq:2.1.1c}
\end{eqnarray}
which are just those of the Poincar\'e Lie algebra $e(1,3)$. In addition to 
the conditions on the basic observables above, we assume that $p^a$ is 
\emph{non-zero, non-spacelike} and \emph{future pointing} (i.e. its timelike 
component, $p^0$, is strictly positive) with respect to $\eta_{ab}$ and a 
fixed time orientation. 

The rest mass, the Pauli--Lubanski spin and the centre-of-mass vectors are 
defined, respectively, according to 
\begin{equation}
\mu^2:=\eta_{ab}p^ap^b, \hskip 20pt
S_a:=\frac{1}{2}\varepsilon_{abcd}J^{bc}p^d, \hskip 20pt
M_a:=J_{ab}p^b. \label{eq:2.1.2}
\end{equation}
Here $\varepsilon_{abcd}$ is the natural volume 4-form on $\mathbb{R}^{1,3}$ 
determined by $\eta_{ab}$. $\mu^2$ and $S_aS^a$ are invariant with respect 
to Poincar\'e transformations; which invariance follows from 
(\ref{eq:2.1.1a})-(\ref{eq:2.1.1c}), too: $\mu^2$ and $S_aS^a$ are 
\emph{Casimir invariants}. From the definitions it follows that $S_ap^a=0$ 
and $M_ap^a=0$, and that the identity 
\begin{equation}
\mu^2J_{ab}=-\varepsilon_{abcd}S^cp^d+M_ap_b-M_bp_a \label{eq:2.1.3}
\end{equation}
holds. Since $p^a$ is non-spacelike, $\mu^2\geq0$ holds; and since $S_ap^a
=0$, $S^a$ is spacelike or null: $S_aS^a\leq0$. Since $S_a$ and $p^a$ are 
invariant with respect to translations while $M_a\mapsto\tilde M_a=M_a+
(\mu^2\eta_{ab}-p_ap_b)\xi^b$, for $\mu>0$ this identity is interpreted as the 
decomposition of the total angular momentum into the sum of its spin and 
orbital parts. (Strictly speaking, if $\mu>0$, then the dimensionally 
correct Pauli--Lubanski spin and centre-of-mass vectors are $1/\mu$-times 
and $1/\mu^2$-times the expressions above, respectively.) 

If $\mu=0$, then (\ref{eq:2.1.3}) does \emph{not} provide a decomposition 
of $J^{ab}$ into its spin and orbital parts. In this case, contracting 
(\ref{eq:2.1.3}) with $S^a$, we find that $S_aM^a=0$ holds. If $M^a$ is null, 
then by $M_ap^a=0$ it follows that $M_a=kp_a$ for some $k\in\mathbb{R}$, which 
by (\ref{eq:2.1.3}) implies $S_a=\chi p_a$ for some $\chi\in\mathbb{R}$. In 
a similar way, if $S^a$ is null, then by $S_ap^a=0$ it follows that $S_a=
\chi p_a$, and hence, also by (\ref{eq:2.1.3}), $M_a=kp_a$ follows. 
Therefore, either both $M_a$ and $S_a$ are spacelike and orthogonal to each 
other or both are proportional to the null 4-momentum $p_a$. In the latter 
case, the factor of proportionality $\chi$ turns out to be a Casimir 
invariant and it might be called the `classical helicity' of the system. 
$k$ is invariant with respect to Lorentz transformations, but, under a 
translation, it changes according to $k\mapsto k-p_a\xi^a$. 

In \cite{PeMacC72}, Penrose and MacCallum clarified under what conditions 
can we find a translation $\xi^a$ resulting $\tilde M_a=0$. The summary of 
their results are as follows. If $\mu>0$, then $S_a$ and $M_a$ are spacelike 
or zero, and $0=\tilde M_a=M_a+(\mu^2\eta_{ab}-p_ap_b)\xi^b$ can always be 
solved for $\xi^a$. The solutions form a 1-parameter family: $\xi^a=-M^a/\mu
^2+up^a/\mu$, $u\in\mathbb{R}$. Hence, as it is interpreted in Minkowski 
space in \cite{PeMacC72}, an elementary Poincar\'e-invariant system with 
$\mu>0$ can always be `localized' to a timelike straight line, viz. to 
$\gamma^a(u):=M^a/\mu^2+up^a/\mu$. The timelike straight line $\gamma^a$, 
associated with the Poincar\'e-invariant elementary classical system in this 
way, will be called the \emph{centre-of-mass line} of the system. 

If $\mu=0$, then $\tilde M_a=0$ can be solved for $\xi^a$ \emph{precisely} 
when $M_a=kp_a$, which, as we noted above, is equivalent to $S_a=\chi p_a$. 
Thus, in particular, if $\mu=0$ and the Pauli--Lubanski spin vector is 
\emph{spacelike}, then \emph{no} translation $\xi^a$ yielding $\tilde M_a
=0$ exists. 

If $\mu=0$, $S_a=\chi p_a$ and $M_a=kp_a$, then the translations $\xi^a$ that 
yield $\tilde M_a=0$ satisfy $p_a\xi^a=k$, and hence these translations form 
a \emph{null hyperplane} with $p^a$ as its null normal, rather than only a 
single null straight line with tangent $p^a$. This set of translations 
can be reduced further in a natural way to form only a 1-parameter family 
if $\chi=0$ also holds. Therefore, an elementary system with $\mu=0$ and $S
_a=\chi p_a$ with $\chi\not=0$ can be `localized' only to a null hyperplane 
${\cal N}$, and it can be `localized' further in a natural way to a null
straight line $\gamma^a$ precisely when $\chi=0$. This result is analogous 
to the result that a relativistic \emph{quantum mechanical} particle with 
zero rest mass can be localized (with respect to the Newton--Wigner 
position operator \cite{NeWign}) only when its helicity is vanishing 
\cite{Wigh,Lunn,Ange}. 

Thus, to summarize, in the zero rest mass case (\ref{eq:2.1.3}) does 
\emph{not} provide a decomposition of the total angular momentum into its 
spin and orbital parts, and such elementary systems would have well defined 
centre-of-mass lines only if their helicity were zero. But since in the 
present paper we intend to recover the metric structure of the Minkowski 
space from the orbital part of the angular momentum by expressing the 
distance between the centre-of-mass straight lines in terms of the spins 
(as in the $E(3)$-invariant case in \cite{Sz22a}), in (\ref{eq:2.1.3}) we 
\emph{must} assume that the rest mass $\mu$ is strictly positive. The 
elementary systems with zero rest mass do not seem to provide an 
appropriate framework for the present project. Thus, in the rest of the 
paper, we concentrate only on the $\mu>0$ case. 

The (future) mass shell ${\cal M}^+_\mu:=\{p^a\in\mathbb{R}^{1,3}\vert\,\,
p^0>0,\, p^ap^b\eta_{ab}=\mu^2>0\}$ is a spacelike hypersurface in the 
momentum space $\mathbb{R}^{1,3}$. (We discuss the geometry of ${\cal M}
^+_\mu$ further in Appendix \ref{sub-A.1}.) Its cotangent bundle, $T^*
{\cal M}^+_\mu$, is homeomorphic to the manifold of the future directed 
timelike straight lines $\gamma^a$: the unit tangent of $\gamma^a$ is 
$p^a/\mu$, while a point on $\gamma^a$ can be chosen to be the intersection 
point of $\gamma^a$ and the spacelike hyperplane through the origin of 
$\mathbb{R}^{1,3}$ with normal proportional to $p^a$. This point of 
intersection is given by the spatial vector $M^a/\mu^2$. 

Clearly, if $\gamma^a$ is any timelike straight line, then this can always 
be obtained from the special one $\gamma^a_0(u):=u\,\delta^a_0$, $u\in 
\mathbb{R}$, by an appropriate Lorentz boost $\Lambda^a{}_b$ and a 
translation $\xi^a$: $\gamma^a(u)=\Lambda^a{}_b\gamma^b_0(u)+\xi^a$. The 
ambiguity in $\xi^a$ is that we can add to it any term proportional with 
$p^a$. Thus, the Cartesian frame in $\mathbb{R}^{1,3}$ is a special 
co-moving, centre-of-mass frame for the system whose centre-of-mass world 
line is $\gamma^a_0$; and whose centre-of-mass vector is zero, its 
energy-momentum has only time component, and the Pauli--Lubanski spin 
vector points e.g. in the $z$-direction.


\subsection{Classical two-particle systems and the empirical distance}
\label{sub-2.2}

In this subsection, we show that the Lorentzian distance between any two 
non-parallel timelike straight lines, considered to be the centre-of-mass 
lines of Poincar\'e-invariant elementary classical mechanical systems with 
positive rest masses, can be expressed in terms of \emph{Poincar\'e-invariant 
observables} of the two-particle system and its constituent elementary 
subsystems. 

Let $(p^a_{\bi},J^{ab}_{\bi})$, ${\bi}=1,2$, characterize two 
Poincar\'e-invariant elementary classical mechanical systems, and let us form 
their formal union characterized by $p^a:=p^a_1+p^a_2$ and $J^{ab}:=J^{ab}_1+
J^{ab}_2$. Then the rest mass, the Pauli--Lubanski spin and the centre-of-mass 
vectors of the composite system are defined in terms of $p^a$ and $J^{ab}$ 
according to the general rules. As a consequence of the definitions, $S^a_{12}
:=S^a-S^a_1-S^a_2=\frac{1}{2}\varepsilon^a{}_{bcd}(J^{bc}_1p^d_2+J^{bc}_2p^d_1)$; 
and, in a similar way, $P^2_{12}:=\frac{1}{2}(\eta_{ab}p^ap^b-\mu^2_1-\mu^2_2)=
\eta_{ab}p^a_1p^b_2$. Since both $p^a_1$ and $p^a_2$ are future pointing and 
timelike, $P^2_{12}\geq\mu_1\mu_2>0$ holds, in which the equality holds 
precisely when $p^a_1$ and $p^a_2$ are parallel with each other. As we will 
see, $S^a_{12}$ and $P^2_{12}$ play fundamental role in the subsequent analyses 
because they characterize the relationship between the two elementary 
subsystems in the composite system. 

Since both $p^a_1$ and $p^a_2$ are \emph{timelike}, the definitions and 
equation (\ref{eq:2.1.3}) yield 
\begin{equation}
S^a_{12}=\bigl(\frac{S^a_1}{\mu^2_1}+\frac{S^a_2}{\mu^2_2}\bigr)P^2_{12}-
\frac{p^a_1p^b_2}{\mu^2_1}S_{1b}-\frac{p^a_2p^b_1}{\mu^2_2}S_{2b}+\varepsilon^a
{}_{bcd}\bigl(\frac{M^b_1}{\mu^2_1}-\frac{M^b_2}{\mu^2_2}\bigr)p^c_1p^d_2.
\label{eq:2.2.1}
\end{equation}
If the two 4-momenta are not parallel, then $P^4_{12}>\mu^2_1\mu^2_2$, and the 
last term on the right is not identically zero. Then (\ref{eq:2.2.1}) can be 
solved for $M^a_1/\mu^2_1-M^a_2/\mu^2_2$: 
\begin{equation}
\frac{M^a_1}{\mu^2_1}-\frac{M^a_2}{\mu^2_2}=\frac{1}{\mu^2_1\mu^2_2-P^4_{12}}
\varepsilon^a{}_{bcd}\Bigl(S^b_{12}-\bigl(\frac{S^b_1}{\mu^2_1}+\frac{S^b_2}
{\mu^2_2}\bigr)P^2_{12}\Bigr)p^c_1p^d_2+u_1\frac{p^a_1}{\mu_1}+u_2\frac{p^a_2}
{\mu_2}, \label{eq:2.2.2}
\end{equation}
where $u_1,u_2\in\mathbb{R}$ are arbitrary. Hence, although its component 
in the timelike 2-plane spanned by $p^a_1$ and $p^a_2$ is ambiguous, its 
component in the orthogonal spacelike 2-plane, viz. 
\begin{equation}
d^a_{12}:=\Pi^a_b\Bigl(\frac{M^b_1}{\mu^2_1}-\frac{M^b_2}{\mu^2_2}\Bigr)=
-\frac{1}{P^4_{12}-\mu^2_1\mu^2_2}\varepsilon^a{}_{bcd}p^b_1p^c_2\Bigl(S^d_{12}
-P^2_{12}\bigl(\frac{S^d_1}{\mu^2_1}+\frac{S^d_2}{\mu^2_2}\bigr)\Bigr),
\label{eq:2.2.3}
\end{equation}
is well defined. Here 
\begin{eqnarray}
\Pi^a_b:=\!\!\!\!&{}\!\!\!\!&\delta^a_b+\frac{1}{P^4_{12}-\mu^2_1\mu^2_2}
  \Bigl(\mu^2_2p^a_1p_{1b}+\mu^2_1p^a_2p_{2b}-P^2_{12}\bigl(p^a_1p_{2b}+p^a_2
  p_{1b}\bigr)\Bigr)= \nonumber \\
=\!\!\!\!&{}\!\!\!\!&-\frac{1}{P^4_{12}-\mu^2_1\mu^2_2}\bigl(\varepsilon
  ^a{}_{ecd}p^c_1p^d_2\bigr)\bigl(\varepsilon^e{}_{bgh}p^g_1p^h_2\bigr)
  \label{eq:2.2.4}
\end{eqnarray}
is the projection to the spacelike 2-plane orthogonal to $p^a_1$ and $p^a_2$. 

$d^a_{12}$ is a translation invariant Lorentzian spacelike 4-vector. It 
points from a well defined point $\nu_{21}$ of the centre-of-mass line of 
the second system to a point $\nu_{12}$ of the the centre-of-mass line of 
the first system, and it is orthogonal to these two straight lines. Thus, 
this is the \emph{relative position vector} of the first system with 
respect to the second at a particular instant. Its length, defined by 
$d_{12}:=\sqrt{-\eta_{ab}d^a_{12}d^b_{12}}$ and whose physical dimension is, 
indeed, $cm$, is just the \emph{spatial Lorentzian distance} between the 
points $\nu_{21}$ and $\nu_{12}$ in the spatial 2-plane orthogonal to the 
timelike 2-plane spanned by $p^a_1$ and $p^a_2$. Note that \emph{it is 
given exclusively by Poincar\'e invariant observables of the composite 
system and its constituent subsystems}; and it is analogous to the 
empirical distance between the Euclidean-invariant elementary classical 
mechanical systems \cite{Sz22a}. Thus, we call it the \emph{empirical 
distance} between the two straight lines. $d^a_{12}$ is zero precisely 
when the two centre-of-mass lines intersect each other; and it is 
invariant with respect to the rescalings $(p^a_1,J^{ab}_1)\mapsto(\alpha p
^a_1,\alpha J^{ab}_1)$ and $(p^a_2,J^{ab}_2)\mapsto(\beta p^a_2,\beta J^{ab}
_2)$ for any $\alpha,\beta>0$. 

Let us represent the timelike straight lines as $\gamma^a_{\bi}(u)=\Lambda
^a_{\bi}{}_b\gamma^b_0(u)+\xi^a_{\bi}$, ${\bi}=1,2$, i.e. by a pair $(\Lambda
^a_{\bi}{}_b,\xi^a_{\bi})$ of Poincar\'e transformations (in fact, by a pair 
of Lorentz boosts and translations) and the straight line $\gamma^a_0$. 
Then with the choice $\gamma^a_0(u):=u\,\delta^a_0$ (as at the end of the 
previous subsection) the 4-momenta are given by $p^a_1=\mu_1\Lambda^a_1{}_0$ 
and $p^a_2=\mu_2\Lambda^a_2{}_0$, and hence the explicit form of the 
(non-negative) square of the spatial Lorentzian distance between $\gamma^a
_1$ and $\gamma^a_2$ is 
\begin{eqnarray}
\bigl(D_{12}\bigr)^2\!\!\!\!&:=\!\!\!\!&-\Pi_{ab}\Bigl(\xi^a_1-\xi^a_2\Bigr)
  \Bigl(\xi^b_1-\xi^b_2\Bigr)=\frac{1}{P^4_{12}-\mu^2_1\mu^2_2}\Bigl(\xi^a_1-
  \xi^a_2\Bigr)\Bigl(\xi^b_1-\xi^b_2\Bigr)\bigl(\varepsilon_{aecd}p^c_1p^d_2
  \bigr)\bigl(\varepsilon^e{}_{bgh}p^g_1p^h_2\bigr)= \nonumber \\
\!\!\!\!&=\!\!\!\!&-\frac{1}{((\Lambda^{-1}_1\Lambda_2)_{00})^2-1}\Bigl(
  \xi^a_1-\xi^a_2\Bigr)\Bigl(\xi^b_1-\xi^b_2\Bigr)\varepsilon_{acde}\Lambda
  ^c_1{}_0\Lambda^d_2{}_0\,\varepsilon_{bgh}{}^e\Lambda^g_1{}_0\Lambda^h_2
  {}_0. \label{eq:2.2.5}
\end{eqnarray}
This is invariant with respect to the rotations about $\gamma^a_0$ in the 
Lorentz transformations $\Lambda^a_{\bi}{}_b$, it is free of the 
ambiguities $\xi^a_{\bi}\mapsto\xi^a_{\bi}+up^a_{\bi}$ in the translations, and 
it is independent of the rest masses $\mu_1$ and $\mu_2$. The expression 
$(d_{12})^2=-\eta_{ab}d^a_{12}d^b_{12}$ above is an alternative form of 
(\ref{eq:2.2.5}), given by classical observables of a Poincar\'e-invariant 
composite system and its elementary subsystems. As we will see in subsection 
\ref{sub-4.3}, $(D_{12})^2$ can be recovered in the classical limit of the 
square of the empirical distance between $E(1,3)$-invariant elementary 
\emph{quantum mechanical} systems, where the latter is built from the 
basic \emph{quantum observables} analogously to how the alternative 
expression $(d_{12})^2$ is built form the classical observables. 

If the two 4-momenta, $p^a_1$ and $p^a_2$, are parallel, then the above 
strategy to recover the distance between the corresponding centre-of-mass 
lines does not work. Nevertheless, following an argumentation analogous to 
that in \cite{Sz22a}, one can show that this distance can be recovered as 
the \emph{limit} of distances between elementary systems with 
\emph{non-parallel 4-momenta}; but the points analogous to $\nu_{21}$ and 
$\nu_{12}$ above are not well defined. Therefore, it is enough to consider 
timelike straight lines with non-parallel tangents. 


\subsection{The Minkowski metric from empirical distances}
\label{sub-2.3}

In Minkowski space the metric determines the Lorentzian distance between 
any two (e.g. non-parallel) timelike straight lines, and now we show that 
the converse is also true: if the distance between any two non-parallel 
timelike straight lines is known, then the Lorentzian distance function on 
the Minkowski space is completely determined. 

Let the points $x^a_1,x^a_2\in\mathbb{R}^{1,3}$ be spacelike separated. Then 
there is a uniquely determined timelike 3-plane through $x^a_1$, and another 
timelike 3-plane through $x^a_2$ which are orthogonal to the direction of 
$x^a_1-x^a_2$. Then let us choose any timelike straight line $\gamma^a_1$ 
through $x^a_1$ lying in the timelike 3-plane above; and, in a similar way, 
let $\gamma^a_2$ be any timelike straight line through $x^a_2$ in the 
corresponding timelike 3-plane through $x^a_2$. Clearly, $\gamma^a_1$ and 
$\gamma^a_2$ can be chosen to be non-parallel. Then the Lorentzian distance 
between the points $x^a_1$ and $x^a_2$ is just the empirical distance 
between $\gamma^a_1$ and $\gamma^a_2$; i.e. in this case the Lorentzian 
distance is realized directly by the empirical distance. 

Next, let us suppose that the points $x^a_1$ and $x^a_2$ are \emph{timelike} 
separated and $x^a_1$ is in the chronological future of $x^a_2$. These two 
points determine uniquely a timelike straight line $\gamma^a$ through them, 
and the point with the coordinates $x^a:=(x^a_1+x^a_2)/2$ is on this straight 
line. Let $T^2:=\eta_{ab}(x^a_1-x^a_2)(x^b_1-x^b_2)$, the square of the 
Lorentzian distance between $x^a_1$ and $x^a_2$. Then let $\Sigma$ be the 
spacelike hyperplane through $x^a$ whose timelike normal is proportional 
to $x^a_1-x^a_2$, i.e. that $\Sigma$ is the `instantaneous spacelike 
3-space' with respect to $\gamma^a$ through $x^a$, and let $\tilde x^a$ be 
any point in $\Sigma$ whose spatial distance from $x^a$ in $\Sigma$ is 
$T/2$. Finally, let $\tilde\gamma^a$ be any timelike straight line through 
$\tilde x^a$ which is orthogonal to the spacelike straight line through 
$x^a$ and $\tilde x^a$. Then the empirical distance between the timelike 
straight lines $\gamma^a$ and $\tilde\gamma^a$ is just $T/2$. Hence, there 
are timelike straight lines such that the empirical distance between them 
reproduces the Lorentzian distance between any two timelike separated 
points. 

To summarize, considering the timelike straight lines of the Minkowski 
space $\mathbb{R}^{1,3}$ to be centre-of-mass lines of elementary, 
Poincar\'e-invariant classical mechanical systems with positive rest mass, 
the Lorentzian distance between these straight lines can be expressed in 
terms of Poincar\'e-invariant classical observables, or at least the limit 
of such observables. Since the Lorentzian distance between any two points 
of the Minkowski space can be written as the Lorentzian distance between 
appropriate timelike straight lines, we conclude that \emph{the metric 
structure of the Minkowski space can be determined by Poincar\'e-invariant 
classical observables of Poincar\'e-invariant elementary systems}. It is 
this characterization of the Minkowski metric by observables of elementary 
classical \emph{physical systems} that makes it possible to derive the 
Minkowski metric from observables of \emph{quantum} systems, too, in the 
classical limit.


\section{Poincar\'e invariant elementary quantum mechanical 
systems}
\label{sec-3}

\subsection{The definition of the elementary systems}
\label{sub-3.1}

A \emph{Poincar\'e-invariant elementary quantum mechanical system} is 
defined to be a system whose states belong to the representation space 
of a \emph{unitary, irreducible} representation of the \emph{quantum 
mechanical} Poincar\'e group $E(1,3)$; and, in this representation, the 
momentum and angular momentum tensor operators, ${\bf p}^a$ and ${\bf J}
^{ab}$, are the self-adjoint generators of the translations and 
boost-rotations (see \cite{NeWign}). In what follows, the letters in 
boldface will denote quantum mechanical operators. Such a representation 
is characterized by a fixed value $\mu^2\geq0$ and $w$, respectively, of 
the two Casimir operators 
\begin{equation}
{\bf P}^2:=\eta_{ab}{\bf p}^a{\bf p}^b, \hskip 20pt
{\bf S}^2:=\eta_{ab}{\bf S}^a{\bf S}^b. \label{eq:3.1.1}
\end{equation}
Here ${\bf S}_a:=\frac{1}{2}\varepsilon_{abcd}{\bf J}^{bc}{\bf p}^d$ is the 
Pauli--Lubanski spin operator. The commutators of ${\bf p}_a$ and ${\bf J}
_{ab}$ can be obtained formally from (\ref{eq:2.1.1a})-(\ref{eq:2.1.1c}) 
with the $p_a\mapsto{\bf p}_a$, $J_{ab}\mapsto-({\rm i}/\hbar){\bf J}_{ab}$ 
substitution: 
\begin{eqnarray}
&{}&\bigl[{\bf p}_a,{\bf p}_b\bigr]=0, \hskip 20pt
  \bigl[{\bf p}_a,{\bf J}_{bc}\bigr]={\rm i}\hbar\bigl(\eta_{ab}{\bf p}_c-\eta
  _{ac}{\bf p}_b\bigr), \label{eq:3.1.2a} \\
&{}&\bigl[{\bf J}_{ab},{\bf J}_{cd}\bigr]={\rm i}\hbar\bigl(\eta_{ac}{\bf J}
  _{db}-\eta_{ad}{\bf J}_{cb}+\eta_{bd}{\bf J}_{ca}-\eta_{bc}{\bf J}_{da}\bigr).
  \label{eq:3.1.2c}
\end{eqnarray}
Moreover, under the action of the translations with $\xi^a$, these transform 
as ${\bf p}^a\mapsto\tilde{\bf p}^a={\bf p}^a$ and ${\bf J}^{ab}\mapsto\tilde
{\bf J}^{ab}={\bf J}^{ab}+2\xi^{[a}{\bf p}^{b]}$, while, under the action of $SL
(2,\mathbb{C})$, they transform as Lorentzian vector and anti-symmetric 
tensor operators. For the sake of brevity, we call such a system a single 
particle, though, as Newton and Wigner stress \cite{NeWign}, such a system 
is \emph{not} necessarily `elementary' in the usual sense that it does not 
have any internal structure: it may have such a structure, but, in the 
given physical context, it is considered to be irrelevant. 

All the operators above are formally self-adjoint. However, ${\bf M}_a:=
{\bf J}_{ab}{\bf p}^b$ is \emph{not} self-adjoint, because ${\bf M}_a^\dagger
={\bf p}^b{\bf J}_{ab}={\bf M}_a+[{\bf p}^b,{\bf J}_{ab}]={\bf M}_a-3{\rm i}
\hbar{\bf p}_a$. Thus we form ${\bf C}_a:=\frac{1}{2}({\bf M}_a+{\bf M}
^\dagger_a)={\bf M}_a-\frac{3}{2}{\rm i}\hbar{\bf p}_a$, which is formally 
self-adjoint by definition, and we consider this to be the 
\emph{centre-of-mass operator}. The commutators of these operators can be 
derived from those for ${\bf p}_a$ and ${\bf J}_{ab}$ above. In particular, 
in any unitary, irreducible representation labeled by $\mu$ they are 
\begin{eqnarray}
&{}&\bigl[{\bf S}_a,{\bf S}_b\bigr]=-{\rm i}\hbar\,\varepsilon_{abcd}{\bf S}^c
  {\bf p}^d, \hskip 15pt
\bigl[{\bf S}_a,{\bf p}_b\bigr]=0, \label{eq:3.1.3a} \\
&{}&[{\bf C}_a,{\bf C}_b\bigr]={\rm i}\hbar\mu^2{\bf J}_{ab}, \hskip 15pt
  \bigl[{\bf C}_a,{\bf p}_b\bigr]={\rm i}\hbar\bigl({\bf p}_a{\bf p}_b-\mu^2
  \eta_{ab}{\bf I}\bigr), \hskip 15pt
\bigl[{\bf C}_a,{\bf S}_b\bigr]={\rm i}\hbar{\bf S}_a{\bf p}_b, \,\,\,\,\,\,
\,\,\,\, \label{eq:3.1.3b}
\end{eqnarray}
where ${\bf I}$ is the identity operator. As a consequence of the 
definitions, the analog of (\ref{eq:2.1.3}) holds for the operators too: 
\begin{equation}
\mu^2{\bf J}_{ab}=-\varepsilon_{abcd}{\bf S}^c{\bf p}^d+{\bf C}_a{\bf p}_b-
{\bf C}_b{\bf p}_a. \label{eq:3.1.4}
\end{equation}
Under translations by $\xi^a$ the operators ${\bf p}^a$ and ${\bf S}_a$ do 
not change, but ${\bf C}_a\mapsto\tilde{\bf C}_a={\bf C}_a+\xi^b(\mu^2\eta_{ab}
-{\bf p}_a{\bf p}_b)$. 

Several different forms of unitary, irreducible representations of $E(1,3)$ 
are known. Our approach is based on the use of square integrable completely 
symmetric spinor fields with index $2s$, say $\phi^{A_1...A_{2s}}$, on the 
future mass-shell ${\cal M}^+_\mu$ with rest mass $\mu$ in the classical 
momentum space. Here, $s$ and $\mu$ are determined by the two Casimir 
operators. The key points of this representation are sketched in 
\cite{StWi,Haag}, and we summarize them in Appendix \ref{sub-A.2}: in this 
representation the operators ${\bf p}^a$ are multiplication, while ${\bf J}
_{ab}$ are differential operators acting on the spinor fields, and they are 
given explicitly by equations (\ref{eq:A.2.2a}) and (\ref{eq:A.2.2b}), 
respectively. 

However, to prove our main result, it will be enough to consider only 
states $\phi$, or in the bra-ket notation $\vert\phi\rangle$, given by 
spinor fields $\phi^{A_1...A_{2s}}$ that are obtained by 
$E(1,3)$-transformations from special states $\vert\psi\rangle$, specified 
by special spinor fields $\psi^{A_1...A_{2s}}$, which are some sort of 
\emph{centre-of-mass states}. It is well known (see e.g. \cite{PR,HT}) 
that, in general, any totally symmetric spinor field with $2s$ indices, say 
$\psi^{A_1...A_{2s}}$, can be completely specified by $2s+1$ complex scalars, 
$\psi_r$, $r=0,1,...,2s$. The centre-of-mass states, as we will see in the 
next subsection, are those that can be specified by \emph{two} such scalars, 
viz. $\psi_{2s}$ and $\psi_0$, which, in addition, are holomorphic and 
anti-holomorphic, respectively, with respect to a certain complex structure. 
Thus, in the calculation of the empirical distance, we use the form of the 
quantum mechanical operators that fits naturally to this representation of 
the quantum states. This Newman--Penrose (NP) form of the operators is 
determined in Appendix \ref{sub-A.3}. The technical, geometric background 
to the appendices \ref{sub-A.2} and \ref{sub-A.3} is summarized in Appendix 
\ref{sub-A.1}.


\subsection{The co-moving, centre-of-mass states}
\label{sub-3.2}

According to subsection \ref{sub-2.1}, every state of a Poincar\'e-invariant 
elementary classical mechanical system with positive rest mass can be 
obtained by an appropriate Poincar\'e transformation from the special state 
in which $M^a=0$, $p^a=\mu\delta^a_0$ and $S^a=S\delta^a_3$, where $\mu>0$ 
and $S\geq0$ are fixed by the two Casimir invariants. (Thus, this state may 
be called a `special co-moving, centre-of-mass state'.) 

In the case of $E(3)$-invariant elementary quantum mechanical systems, the 
centre-of-mass states were introduced as the states that minimized the 
expectation value of the square of the centre-of-mass vector operator, and 
the expectation value of the Euclidean centre-of-mass vector operator itself 
turned, in fact, out to be zero in all these states \cite{Sz22a}. These 
states are given by spinor fields of the form $\psi_{A_1...A_{2s}} =({}_{-s}
Y_{s,m}/p)o_{A_1}\cdots o_{A_{2s}}$ or $(-)^{2s}({}_sY_{s,m}/p)$ $\iota_{A_1}\cdots
\iota_{A_{2s}}$ on the 2-sphere ${\cal S}_p$ of radius $p$ (the analog of the 
`mass shell'), where $p$ is the magnitude of the linear momentum, ${}_{\pm s}
Y_{s,m}$, $m=-s,-s+1,...,s$, are special spin weighted spherical harmonics 
on ${\cal S}_p$, and $\{o^A,\iota^A\}$ is a Newman--Penrose normalized 
spinor basis adapted to ${\cal S}_p$. The value of $p$ and $s$ is fixed by 
the two Casimir operators of $E(3)$. (For the notations and the technical 
background, see Appendix \ref{sub-A.1} and \ref{sub-A.2}.) 

Unfortunately, in the actual $E(1,3)$-invariant case, the mass shell is 
\emph{non-compact} and the metric $\eta_{ab}$ by means of which the square 
of ${\bf C}^a$ is defined is \emph{indefinite}, yielding that the analysis 
done in \cite{Sz22a} cannot be successfully repeated in the present case. 
Nevertheless, in this subsection, we specify certain special states, 
$\vert\psi_{s,m}\rangle$, motivated by the centre-of-mass states of the 
$E(3)$-invariant systems above, and we justify \emph{a posteriori} that 
these are, in fact, analogous to the special co-moving, centre-of-mass 
states of the $E(1,3)$-invariant elementary classical mechanical systems, 
viz. that $\langle\psi_{s,m}\vert{\bf C}^a\vert\psi_{s,m}\rangle=0$, $\langle
\psi_{s,m}\vert{\bf p}^a\vert\psi_{s,m}\rangle\sim\delta^a_0$ and $\langle
\psi_{s,m}\vert{\bf S}^a\vert\psi_{s,m}\rangle\sim\delta^a_3$ hold in all of 
these states. The factor of proportionality in the last two expressions 
will also be fixed here, but the justification of that choice will be 
given only in the next subsection. 

Since $E(3)\subset E(1,3)$, the structure of the centre-of-mass states in 
the $E(3)$-invariant case suggests to consider the special states $\vert
\psi^\pm_{s,m}\rangle$ given, respectively, by the completely symmetric 
spinor fields 
\begin{equation}
\psi^+_{A_1...A_{2s}}=f\,{}_{-s}Y_{s,m}\,o_{A_1}\cdots o_{A_{2s}}
\hskip 20pt {\rm and} \hskip 20pt 
\psi^-_{A_1...A_{2s}}=(-)^{2s}f\,{}_{s}Y_{s,m}\,\iota_{A_1}\cdots\iota_{A_{2s}} 
\label{eq:3.2.1}
\end{equation}
on ${\cal M}^+_\mu$. Here $f$ is a \emph{real} valued function of $p$ 
satisfying the normalization condition $1=\langle\psi^\pm_{s,m}\vert\psi^\pm
_{s,m}\rangle:=\langle\psi^\pm_{A_1...A_{2s}}\vert\psi^\pm_{A_1...A_{2s}}\rangle=
\int^\infty_0f^2(p)^2(p^0)^{-1}{\rm d}p$ (see equation (\ref{eq:A.1.2})). For 
$s>0$, $\psi^+_{A_1...A_{2s}}$ and $\psi^-_{A_1...A_{2s}}$ are orthogonal to each 
other with respect to the $L_2$ scalar product (\ref{eq:A.2.1}), but they 
coincide for $s=0$. The normalization condition requires the fall-off 
$f=o(1/p)$, i.e. that $f$ should fall-off \emph{slightly faster} than $1/p$ 
for large $p$. However, the various basic quantum observables have well 
defined action on these states only if $f$ has some faster fall-off (see 
Appendix \ref{sub-A.2}). Thus, temporarily, we require the faster fall-off 
$f=o(p^{-3/2})$, but the ultimate fall-off that we choose for the function 
$f$ at the end of this subsection will be the much faster exponential decay. 

However, the states $\vert\psi^\pm_{s,m}\rangle$ in themselves cannot be 
expected to be co-moving,  centre-of-mass states of $E(1,3)$-invariant 
elementary systems. In fact, while in the $E(1,3)$-invariant case what the 
Pauli--Lubanski spin operator determines is the \emph{non-negative} $s$, 
in the $E(3)$-invariant case the analogous `spin' operator (which is, in 
fact, a \emph{helicity} operator) gives the same $s$ but with \emph{either 
sign $\pm$}. Thus the sign of the `spin' in the centre-of-mass states of 
the $E(3)$-invariant systems is also fixed by the Casimir invariants, and 
hence these states have some definite chirality. Such a chirality is not 
provided by the Casimir invariants of the $E(1,3)$-invariant systems. 
Therefore, the co-moving centre-of-mass states of these systems can be 
expected to be only those \emph{combinations} of the states $\vert\psi
^\pm_{s,m}\rangle$ that are symmetric in the two chiralities. We will see 
that this expectation is correct and is justified by the explicit 
calculations of the expectation values of the basic quantum observables. 

First, let us compute the expectation value of the centre-of-mass operator. 
Recalling that by (\ref{eq:A.1.6a}) the only non-zero component $\psi_r$ 
of the spinor fields of the form $\psi^+_{A_1...A_{2s}}$ is $\psi_{2s}$, by 
(\ref{eq:A.1.11a}) equation (\ref{eq:A.3.3}) reduces to 
\begin{eqnarray}
\bigl({\bf C}^a\psi^+_{s,m}\bigr)_r&\!\!\!\!=&\!\!\!\!{\rm i}\hbar\delta_{r,2s}
  \Bigl(\frac{3}{2}p^af\,{}_{-s}Y_{s,m}+\mu p^0\,v^a\frac{{\rm d}f}{{\rm d}p}
  {}_{-s}Y_{s,m}+\mu^2\bar m^af\bigl({\edth}{}_{-s}Y_{s,m}\bigr)\Bigr)-
  \nonumber \\
&\!\!\!\!-&\!\!\!\!\frac{\rm i}{\sqrt{2}}\hbar\delta_{r,2s-1}\frac{\mu p^0}{p}
  \,m^af{}_{-s}Y_{s,m}. \label{eq:3.2.2}
\end{eqnarray}
(To avoid potential confusion, we inserted a comma between the two indices 
of the Kronecker delta.) Using the explicit form of the components of $v^a$ 
and $m^a$ given in Appendix \ref{sub-A.1}, equations (\ref{eq:A.1.4b}), 
(\ref{eq:A.1.6b}), (\ref{eq:A.2.1}) and the fact that the spherical 
harmonics have unit $L_2$-norm on the unit sphere, from (\ref{eq:3.2.2}) we 
obtain 
\begin{equation}
\langle\psi^+_{s,m}\vert{\bf C}^0\vert\psi^+_{s,m}\rangle={\rm i}\hbar\int
_{{\cal M}^+_\mu}p^0f\Bigl(\frac{3}{2}f+p\frac{{\rm d}f}{{\rm d}p}\Bigr)\vert
{}_{-s}Y_{s,m}\vert^2{\rm d}v_\mu=\frac{\rm i}{2}\hbar\int^\infty_0\frac{\rm d}
{{\rm d}p}\Bigl((p)^3f^2\Bigr){\rm d}p=0. \label{eq:3.2.2a}
\end{equation}
In the last step we used that, for large $p$, $f=o(p^{-3/2})$. Also, for $a
=i=1,2,3$, (\ref{eq:3.2.2}) gives 
\begin{eqnarray}
\langle\psi^+_{s,m}\vert{\bf C}^i\vert\psi^+_{s,m}\rangle\!\!\!\!&=\!\!\!\!&
  {\rm i}\hbar\int_{{\cal M}^+_\mu}\Bigl(\bigl(\frac{3}{2}p^if^2+(p^0)^2
  \frac{p^i}{p}f\frac{{\rm d}f}{{\rm d}p}-\mu^2({\edth}\bar m^i)f^2\bigr)
  \vert{}_{-s}Y_{s,m}\vert^2+ \nonumber \\
\!\!\!\!&{}\!\!\!\!&\hskip 30pt +\mu^2{\edth}\bigl(\bar m^if^2\vert{}_{-s}
  Y_{s,m}\vert^2\bigr)\Bigr){\rm d}v_\mu= \nonumber \\
\!\!\!\!&=\!\!\!\!&{\rm i}\hbar\langle{}_{-s}Y_{s,m}\vert\frac{{\bf p}^i}{p}
  \vert{}_{-s}Y_{s,m}\rangle_1\int^\infty_0\Bigl(\frac{1}{2}pf^2+\frac{(p^0)
  ^2}{p}f^2+(p^0)^2f\frac{{\rm d}f}{{\rm d}p}\Bigr)\frac{(p)^2}{p^0}{\rm d}
  p=\nonumber \\
\!\!\!\!&=\!\!\!\!&-\frac{\rm i}{2}\hbar\frac{m}{s+1}\delta^i_3\int
  ^\infty_0\frac{\rm d}{{\rm d}p}\Bigl(p^0(p)^2f^2\Bigr){\rm d}p=0, 
  \label{eq:3.2.2b}
\end{eqnarray}
where, according to Appendix \ref{sub-A.4}, $\langle\,.\,\vert\,.\,\rangle
_1$ denotes the $L_2$ scalar product on the \emph{unit} 2-sphere. 
Here, to get the first expression, we formed a total ${\edth}$-derivative 
and used the explicit form of $v^i$ and equation (\ref{eq:A.1.6b}) given 
in Appendix \ref{sub-A.1}; then we used ${\edth}\bar m^i=p^i/\mu^2-p^0v^i/
(\mu p)=-p^i/(p)^2$; and, finally, we used equations (\ref{eq:A.4.1b}), 
(\ref{eq:A.4.2b}) and (\ref{eq:A.4.3b}), we formed a total derivative and 
used the fall-off $f=o(p^{-3/2})$. In a similar way, one can show that 
\begin{eqnarray}
&{}&\langle\psi^-_{s,m}\vert{\bf C}^a\vert\psi^-_{s,m}\rangle=0,
  \label{eq:3.2.2c} \\
&{}&\langle\psi^-_{s,m}\vert{\bf C}^a\vert\psi^+_{s,m}\rangle=
  \overline{\langle\psi^+_{s,m}\vert{\bf C}^a\vert\psi^-_{s,m}\rangle}=
  -\frac{\rm i}{\sqrt{2}}\hbar\mu\delta^a_3\delta_{2s,1}\frac{4m}{3}\int
  ^\infty_0f^2p{\rm d}p. \label{eq:3.2.2d}
\end{eqnarray}
Therefore, $\vert\psi^\pm_{s,m}\rangle$ and $(\vert\psi^+_{s,m}\rangle+\vert
\psi^-_{s,m}\rangle)/\sqrt{2}$ could be centre-of-mass states. 

Using (\ref{eq:A.3.0}), similar calculations yield that 
\begin{eqnarray}
&{}&\langle\psi^+_{s,m}\vert{\bf p}^a\vert\psi^+_{s,m}\rangle=
  \delta^a_0\int^\infty_0f^2(p)^2{\rm d}p-\frac{m}{s+1}\delta^a_3\int^\infty
  _0f^2\frac{(p)^3}{p^0}{\rm d}p, \label{eq:3.2.3a} \\
&{}&\langle\psi^-_{s,m}\vert{\bf p}^a\vert\psi^-_{s,m}\rangle=\delta^a_0\int
  ^\infty_0f^2(p)^2{\rm d}p+\frac{m}{s+1}\delta^a_3\int^\infty_0f^2\frac{(p)^3}
  {p^0}{\rm d}p, \label{eq:3.2.3b} \\
&{}&\langle\psi^-_{s,m}\vert{\bf p}^a\vert\psi^+_{s,m}\rangle=
  \langle\psi^+_{s,m}\vert{\bf p}^a\vert\psi^-_{s,m}\rangle=\delta^a_0
  \delta_{s,0}\int^\infty_0f^2(p)^2{\rm d}p. \label{eq:3.2.3c}
\end{eqnarray}
Hence, $\vert\psi^\pm_{s,m}\rangle$ are \emph{not} co-moving states, i.e. 
$\langle\psi^\pm_{s,m}\vert{\bf p}^a\vert\psi^\pm_{s,m}\rangle$ are not 
proportional to $\delta^a_0$, but their sum, $(\vert\psi^+_{s,m}\rangle+
\vert\psi^-_{s,m}\rangle)/\sqrt{2}$, are. In an analogous way, using 
(\ref{eq:A.3.1}), we obtain 
\begin{eqnarray}
&{}&\langle\psi^+_{s,m}\vert{\bf S}^a\vert\psi^+_{s,m}\rangle=
  s\hbar\Bigl(-\delta^a_0\int^\infty_0f^2\frac{(p)^3}{p^0}{\rm d}p+\delta^a_3
  \frac{m}{s+1}\int^\infty_0f^2(p)^2{\rm d}p\Bigr), \label{eq:3.2.4a} \\
&{}&\langle\psi^-_{s,m}\vert{\bf S}^a\vert\psi^-_{s,m}\rangle=
  s\hbar\Bigl(\delta^a_0\int^\infty_0f^2\frac{(p)^3}{p^0}{\rm d}p+\delta^a_3
  \frac{m}{s+1}\int^\infty_0f^2(p)^2{\rm d}p\Bigr), \label{eq:3.2.4b} \\
&{}&\langle\psi^-_{s,m}\vert{\bf S}^a\vert\psi^+_{s,m}\rangle=
  \langle\psi^+_{s,m}\vert{\bf S}^a\vert\psi^-_{s,m}\rangle=\mu\hbar
  \frac{2m}{3}\delta_{2s,1}\delta^a_3. \label{eq:3.2.4c}
\end{eqnarray}
Hence, the expectation value 4-vectors $\langle\psi^\pm_{s,m}\vert{\bf S}^a
\vert\psi^\pm_{s,m}\rangle$ are not proportional to $\delta^a_3$, but $\langle
\psi^+_{s,m}+\psi^-_{s,m}\vert{\bf S}^a\vert\psi^+_{s,m}+\psi^-_{s,m}\rangle$ are. 

Therefore, to summarize the moral of equations 
(\ref{eq:3.2.2a})-(\ref{eq:3.2.4c}), the states $\vert\psi_{s,m}\rangle$, 
given by the spinor fields 
\begin{equation}
\psi_{A_1...A_{2s}}:=\frac{1}{\sqrt{2}}f\Bigl({}_{-s}Y_{s,m}o_{A_1}\cdots o_{A_{2s}}
+(-)^{2s}{}_sY_{s,m}\iota_{A_1}\cdots\iota_{A_{2s}}\Bigr) \label{eq:3.2.5}
\end{equation}
on ${\cal M}^+_\mu$ for $2s=0,1,2,...$, $m=-s,-s+1,...,s$ and for some real 
functions $f$ of $p$ with the $f=o(p^{-3/2})$ fall-off and normalization 
condition $\int^\infty_0f^2(p)^2(p^0)^{-1}{\rm d}p=1$, are special states of 
$E(1,3)$-invariant elementary quantum mechanical systems with positive rest 
mass $\mu$ and spin $s$ in which the expectation value of ${\bf C}^a$, 
${\bf p}^a$ and ${\bf S}^a$ is zero, proportional to $\delta^a_0$ and to 
$\delta^a_3$, respectively. In particular, $\langle\psi_{s,m}\vert{\bf p}^a
\vert\psi_{s,m}\rangle\eta_{ab}\langle\psi_{s,m}\vert{\bf S}^b\vert\psi_{s,m}
\rangle=0$ holds. Note that $\langle\psi_{s,m}\vert\psi_{s,m}\rangle=1+\delta
_{s,0}$, i.e. these states are normalized only for $s>0$. 

Later, we will need to know the action of ${\bf J}^{ab}$ on these special 
states, as well as their expectation values. In particular, for the action 
of ${\bf J}^{ab}$ on $\vert\psi^+_{s,m}\rangle$ (\ref{eq:A.3.4}) gives 
\begin{eqnarray}
\bigl({\bf J}^{ab}\psi^+_{s,m}\bigr)_r={\rm i}\hbar\delta_{r,2s}\Bigl(
  \!\!\!\!\!&{}\!\!\!\!\!&-\bigl(p^av^b-p^bv^a\bigr)\frac{p^0}{\mu}
  \frac{{\rm d}f}{{\rm d}p}\,{}_{-s}Y_{s,m}+\bigl(p^a\bar m^b-p^b\bar m^a
  \bigr)\frac{\sqrt{s}}{p}f{}_{-s+1}Y_{s,m}+ \nonumber \\
\!\!\!\!\!&{}\!\!\!\!\!&+\bigl(m^a\bar m^b-m^b\bar m^a\bigr)sf\,{}_{-s}
  Y_{s,m}\Bigr)+ \nonumber \\
+\frac{\rm i}{\sqrt{2}}\hbar\delta_{r,2s-1}\Bigl(\!\!\!\!\!&{}\!\!\!\!\!&
  \bigl(p^am^b-p^bm^a\bigr)\frac{p^0}{\mu p}-\bigl(v^am^b-v^bm^a\bigr)
  \Bigr)f\,{}_{-s}Y_{s,m}; \label{eq:3.2.6}
\end{eqnarray}
and (\ref{eq:A.3.4}) gives a similar expression for $({\bf J}^{ab}\psi^-_{s,m})
_r$, too. Then, using $m^i={\edth}p^i$, (\ref{eq:A.4.7a}), (\ref{eq:A.4.7b}) 
and the techniques of the calculations above we find that 
\begin{eqnarray}
\langle\psi^\pm_{s,m}\vert{\bf J}^{ab}\vert\psi^\pm_{s,m}\rangle\!\!\!\!&=
  \!\!\!\!& m\hbar\bigl(\delta^a_1\delta^b_2-\delta^a_2\delta^b_1\bigr), 
  \label{eq:3.2.7a}\\
\langle\psi^-_{s,m}\vert{\bf J}^{ab}\vert\psi^+_{s,m}\rangle\!\!\!\!&=\!\!\!\!&
  \overline{\langle\psi^+_{s,m}\vert{\bf J}^{ab}\vert\psi^-_{s,m}\rangle}=
  {\rm i}\mu\hbar\frac{2m}{3}\bigl(\delta^a_0\delta^b_3-\delta^a_3\delta^b_0
  \bigr)\delta_{2s,1}\int^\infty_0f^2\frac{p}{p^0}{\rm d}p; \label{eq:3.2.7b}
\end{eqnarray}
and hence that 
\begin{equation}
\langle\psi_{s,m}\vert{\bf J}^{ab}\vert\psi_{s,m}\rangle=m\hbar\bigl(\delta
^a_1\delta^b_2-\delta^a_2\delta^b_1\bigr). \label{eq:3.2.7c}
\end{equation}
Therefore, while the expectation value of ${\bf p}^a$ and ${\bf S}^a$ in 
the special co-moving, centre-of-mass states still depends on the specific 
choice for the function $f$, the expectation value of ${\bf J}^{ab}$ does 
not. To have a definite value of $\langle\psi_{s,m}\vert{\bf p}^a\vert\psi
_{s,m}\rangle$ and $\langle\psi_{s,m}\vert{\bf S}^a\vert\psi_{s,m}\rangle$ 
too, we should specify $f$. 

Our choice for $f$ is the square root of the Gaussian distribution function 
on the \emph{hyperboloidal} ${\cal M}^+_\mu$, which is concentrated on the 
point $p^a=(\mu,0,0,0)$. Namely, with the formal substitution $x^i:=p^i/\mu$ 
and for any $\epsilon>0$, the normalization condition for the standard 
three-dimensional Gaussian distribution function $g_\epsilon=g_\epsilon(x^i)$ 
on $\mathbb{R}^3$ can be written as 
\begin{eqnarray*}
1\!\!\!\!&=\!\!\!\!&\int_{\mathbb{R}^3}\Bigl(\frac{1}{\sqrt{2\pi}\epsilon}
  \Bigr)^3\exp\bigl(-\frac{1}{2\epsilon^2}\delta_{ij}x^ix^j\bigr){\rm d}^3x
  =\Bigl(\frac{1}{\sqrt{2\pi}\epsilon\mu}\Bigr)^3\int_{{\cal M}^+_\mu}\exp
  \bigl(-\frac{(p)^2}{2\epsilon^2\mu^2}\bigr)p^0{\rm d}v_\mu= \\
\!\!\!\!&=\!\!\!\!&4\pi\Bigl(\frac{1}{\sqrt{2\pi}\epsilon\mu}\Bigr)^3
  \int^\infty_0(p)^2\exp\bigl(-\frac{(p)^2}{2\epsilon^2\mu^2}\bigr){\rm d}p,
\end{eqnarray*}
where we used (\ref{eq:A.1.2}). Comparing this equation with the 
normalization condition for $f$, we find that 
\begin{equation}
f=f_\epsilon:=\sqrt[4]{\frac{2}{\pi}}\sqrt{\frac{p^0}{\epsilon^3\mu^3}}\exp
\Bigl(-\frac{1}{4}\frac{(p)^2}{\epsilon^2\mu^2}\Bigr) \label{eq:3.2.8}
\end{equation}
could be a possible choice. Clearly, this satisfies even the strongest 
fall-off condition. In what follows, we use this $f_\epsilon$ in $\vert
\psi_{s,m}\rangle$. We justify this choice in the next subsection, though 
many other choices could be equally good. Thus, in contrast to the 
centre-of-mass states of the $E(3)$-invariant systems, the present 
co-moving, centre-of-mass states are \emph{not} canonically determined by 
the physical system itself: they depend on our choice for $f$, too. 

With this particular choice for $f$, the expectation values of ${\bf p}^a$ 
and ${\bf S}^a$ are 
\begin{eqnarray}
\langle\psi_{s,m}\vert{\bf p}^a\vert\psi_{s,m}\rangle\!\!\!\!&=\!\!\!\!&
  \delta^a_0\bigl(1+\delta_{s,0}\bigr)\int^\infty_0f_\epsilon^2(p)^2{\rm d}p,
  \label{eq:3.2.9a}\\
\langle\psi_{s,m}\vert{\bf S}^a\vert\psi_{s,m}\rangle\!\!\!\!&=\!\!\!\!&
  m\hbar\delta^a_3\Bigl(\frac{s}{s+1}\int^\infty_0f_\epsilon^2(p)^2{\rm d}p+
  \frac{2}{3}\mu\delta_{2s,1}\Bigr). \label{eq:3.2.9b}
\end{eqnarray}
Since 
\begin{equation}
\int^\infty_0f_\epsilon^2(p)^2{\rm d}p=\frac{4}{\sqrt{\pi}}\mu\int^\infty_0
x^2\sqrt{1+2\epsilon^2x^2}\,\exp\bigl(-x^2\bigr){\rm d}x, \label{eq:3.2.9c}
\end{equation}
for this integral we have the estimates 
\begin{eqnarray}
\mu\!\!\!\!&=\!\!\!\!&\frac{4}{\sqrt{\pi}}\mu\int^\infty_0\,x^2\exp\bigl(
  -x^2\bigr){\rm d}x\leq\int^\infty_0f_\epsilon^2(p)^2{\rm d}p\leq \nonumber \\
\!\!\!\!&\leq\!\!\!\!&\frac{4}{\sqrt{\pi}}\mu\int^\infty_0\bigl(1+2\epsilon^2
  x^2\bigr)\,x^2\exp\bigl(-x^2\bigr){\rm d}x=\mu\bigl(1+3\epsilon^2\bigr),
  \label{eq:3.2.9d}
\end{eqnarray}
and hence we find that, in the $\epsilon\to0$ limit, this integral tends 
to $\mu$. Here we used the known expression 
\begin{equation}
\int^\infty_0x^n\exp(-x^2){\rm d}x=\frac{1}{2}\Gamma(\frac{n+1}{2}), 
\hskip 20pt n=0,1,2,3,... \label{eq:3.2.10}
\end{equation}
for the definite integrals, where the Gamma function is known to be 
$\Gamma(k+1/2)=(2k)!\sqrt{\pi}/(4^kk!)$ and $\Gamma(k+1)=k!$ for $k=
0,1,2,...$. Therefore, for $s>1/2$, the expectation values 
(\ref{eq:3.2.9a}) and (\ref{eq:3.2.9b}) tend in the $\epsilon\to0$ limit 
to $\mu\delta^a_0$ and $\hbar\mu ms(s+1)^{-1}\delta^a_3$, respectively,. 

The sequence of states by means of which the classical limit of 
$E(1,3)$-invariant systems is defined in the next subsection will be based 
on $\vert\psi_{s,m}\rangle$, given by (\ref{eq:3.2.5}) with (\ref{eq:3.2.8}), 
but in which we should link $\epsilon$ to $s$. 


\subsection{The definition of the classical limit}
\label{sub-3.3}

Intuitively, the classical limit of a quantum system with basic observables 
${\bf O}_\alpha$, $\alpha=1,2,...$, can be expected to be defined by a 
sequence of normalized states, $\vert\phi_k\rangle$, $k\in\mathbb{N}$, in 
which the expectation values $\langle\phi_k\vert{\bf O}_\alpha\vert\phi_k
\rangle$ tend in the $k\to\infty$ limit to their large classical value, 
formally to infinity, such that the rate of growth of the standard 
deviations $\Delta_{\phi_k}{\bf O}_\alpha$, given by $(\Delta_\phi{\bf O}_\alpha
)^2=\langle\phi\vert{\bf O}_\alpha^2\vert\phi\rangle-(\langle\phi\vert{\bf O}
_\alpha\vert\phi\rangle)^2$, is \emph{definitely smaller} than the typical 
rate of growth of the expectation values. Thus, in this limit, the standard 
deviations should tend to zero \emph{relative} to the typical expectation 
values. 

In this subsection, we construct such a sequence of states 
\emph{explicitly} for $E(1,3)$-invariant systems, which states are based 
on the co-moving, centre-of-mass states introduced in the previous 
subsection. 

The notion of classical limit of $SU(2)$-invariant quantum mechanical 
systems was introduced by Wigner \cite{Wi} in the canonical angular 
momentum basis. This notion was extended in a natural way in \cite{Sz22a} 
to $E(3)$-invariant quantum mechanical systems. This limit was defined by 
a sequence of states belonging to unitary irreducible representations of 
$E(3)$ such that, in each of these states the quantum number $\vert m
\vert$ took its maximal value, and the two Casimir invariants, the 
magnitude of the linear momentum and of the spin (more precisely, of the 
helicity), viz. $p$ and $s$, respectively, tended to infinity in the 
\emph{same} order, say with $s$. In this limit, the expectation value of 
the basic observables ${\bf p}^i$ and ${\bf J}^i$ also grow typically with 
$s$, but the standard deviations grow only with $\sqrt{s}$, which is 
\emph{slower} than that of the typical expectation values. 

Since $E(3)\subset E(1,3)$, the classical limit of $E(1,3)$-invariant 
quantum mechanical systems should be compatible with this notion. Thus, 
we will say that a sequence of special co-moving, centre-of-mass states 
defines a classical limit if $\vert m\vert=s\to\infty$ and $\mu\to\infty$ 
such that, in this limit, $\mu=Ms+O(1/s)$ for some positive constant $M$. 
Hence, the Casimir invariants $s$ and $\mu$ are required to tend to their 
large classical value in the \emph{same order}\footnote{
For example, the mass and spin of the electron in cgs units are $\mu_e=9.1
\times10^{-29}\,gram$ and $s_e=\hbar/2=3.3\times10^{-28}gram\,cm^2\,sec^{-1}$, 
respectively. Thus, if the classical scale is that in which the relevant 
quantities are of (or maybe a few orders less than the) order one in these 
units, then \emph{both} the spin and the mass should be enlarged roughly by 
the same factor. Hence, accepting that enlarging the spin by a factor of 
$10^{28}$ can be approximated in the given circumstances by tending with $s$ 
to infinity, then it seems reasonable to approximate enlarging the mass by 
the factor $10^{29}$ in the same way. In the $c=1$ units (that we use in the 
present paper), $\mu_e$ remains the same but $s_e=1.1\times10^{-38}gram\,cm^3$. 
This raises the possibility that in the classical limit it is only the spin 
$s$ that should tend to infinity, but the rest mass $\mu$ only to a large 
but \emph{finite} value. Another possibility is that both $s$ and $\mu$ tend 
to infinity, but \emph{in different orders}, e.g. when $\mu$ grows only like 
$\sqrt{s}$. Although in these cases $\epsilon$ would not be forced to be 
linked to $s$ and it could be kept finite (see below), but these asymptotic 
properties would not guarantee the finiteness of the empirical distance.
}. Then by (\ref{eq:3.2.7c}), (\ref{eq:3.2.9a}) and (\ref{eq:3.2.9c}) the 
expectation value of the basic observables ${\bf p}^a$ and ${\bf J}^{ab}$ of 
the $E(1,3)$-invariant system also diverge, typically with $s$. 

To see whether or not this limit is physically well established, we should 
check that the rate of growth of the standard deviation of the basic 
observables in this limit is less than that of the expectation values. In 
this subsection, we show that this is the case \emph{provided} the parameter 
$\epsilon$ of the Gaussian distribution function $f^2_\epsilon$ tends to zero 
in an appropriate way as $s\to\infty$. Thus the parameter $\epsilon$ should 
be linked to $s$. 

A straightforward calculation yields that the square of the standard 
deviation of the various components of ${\bf p}^a$ in the special co-moving, 
centre-of-mass states $\vert\psi_{s,m}\rangle$ is given by 
\begin{eqnarray}
\Bigl(\Delta_{\psi_{s,m}}{\bf p}^0\Bigr)^2\!\!\!\!&=\!\!\!\!&\mu^2\Bigl(1-
  \frac{16}{\pi}\bigl(\int^\infty_0\sqrt{1+2\epsilon^2x^2}\,x^2\exp(-x^2)
  {\rm d}x\bigr)^2+6\epsilon^2\Bigr)\leq \nonumber \\
\!\!\!\!&\leq\!\!\!\!&\mu^2\Bigl(1-\frac{16}{\pi}\bigl(\int^\infty_0x^2
  \exp(-x^2){\rm d}x\bigr)^2+6\epsilon^2\Bigr)=6\mu^2\epsilon^2,
  \label{eq:3.3.1a} \\
\Bigl(\Delta_{\psi_{s,m}}{\bf p}^1\Bigr)^2\!\!\!\!&=\!\!\!\!&\Bigl(\Delta
  _{\psi_{s,m}}{\bf p}^2\Bigr)^2=\frac{(s+1)^2-m^2}{(s+1)(2s+3)}6\mu^2
  \epsilon^2, \label{eq:3.3.1b} \\
\Bigl(\Delta_{\psi_{s,m}}{\bf p}^3\Bigr)^2\!\!\!\!&=\!\!\!\!&\frac{s+1+2m^2}
  {(s+1)(2s+3)}6\mu^2\epsilon^2, \label{eq:3.3.1c}
\end{eqnarray}
where we used (\ref{eq:3.2.9a}), (\ref{eq:3.2.9d}), (\ref{eq:3.2.10}), 
(\ref{eq:A.4.0}), (\ref{eq:A.4.4a}) and (\ref{eq:A.4.4b}). Thus, although 
in the $\vert m\vert=s\to\infty$ limit $\Delta_{\psi_{s,m}}{\bf p}^1$ and 
$\Delta_{\psi_{s,m}}{\bf p}^2$ diverge like $\sqrt{s}$ even with finite 
$\epsilon$, the rate of growth of $\Delta_{\psi_{s,m}}{\bf p}^3$ is $s$ 
\emph{unless $\epsilon$ tends to zero}, say as $\epsilon=1/s^\alpha$ for some 
$\alpha>0$. Under this condition, by (\ref{eq:3.3.1a}), $\Delta_{\psi_{s,m}}
{\bf p}^0$ grows also not faster than $s^{1-\alpha}$. 

Using the same techniques, in particular $\langle\psi_{s,m}\vert({\bf J}
^{ab})^2\vert\psi_{s,m}\rangle=\langle{\bf J}^{ab}\psi_{s,m}\vert{\bf J}^{ab}
\psi_{s,m}\rangle$ and equations (\ref{eq:3.2.5}), (\ref{eq:3.2.6}) and 
(\ref{eq:A.1.6b}), $\langle\psi_{s,m}\vert({\bf J}^{ab})^2\vert\psi_{s,m}
\rangle$ can be calculated in a straightforward way. Since by 
(\ref{eq:A.1.6b}) the terms of the form $\langle\psi^-_{s,m}\vert({\bf J}
^{ab})^2\vert\psi^+_{s,m}\rangle$ are non-zero only for $s\leq1$, and since 
basically we are interested only in the $s\to\infty$ limit, we need to 
calculate only $\langle\psi^\pm_{s,m}\vert({\bf J}^{ab})^2\vert\psi^\pm_{s,m}
\rangle$. 

In particular, since by (\ref{eq:A.1.3}) and (\ref{eq:A.1.4a}) $m^i\bar m
^j-m^j\bar m^i={\rm i}\delta^{ik}\delta^{jl}\varepsilon_{klr}p^r/p$ and $p^im^j-
p^jm^i={\rm i}\delta^{ik}\delta^{jl}\varepsilon_{klr}pm^r$ hold, using $2p^im^i
={\edth}((p^i)^2)$, equations (\ref{eq:A.1.11a}), (\ref{eq:A.1.11b}) and 
the normalization condition for $f$, by integration by parts we obtain 
\begin{eqnarray*}
\langle\psi^\pm_{s,m}\vert({\bf J}^{ij})^2\vert\psi^\pm_{s,m}\rangle\!\!\!\!&=
  \!\!\!\!&\delta^{ik}\delta^{jl}\varepsilon_{klr}\hbar^2\Bigl(2s^2\langle{}
  _{\mp s}Y_{s,m}\vert(\frac{p^r}{p})^2\vert{}_{\mp s}Y_{s,m}\rangle_1- \\
\!\!\!\!&-\!\!\!\!&s^2\langle{}_{\mp(s-1)}Y_{s,m}\vert(\frac{p^r}{p})^2\vert
  {}_{\mp(s-1)}Y_{s,m}\rangle_1+s\langle{}_{\mp(s-1)}Y_{s,m}\vert m^r\bar m^r
  \vert{}_{\mp(s-1)}Y_{s,m}\rangle_1\Bigr). 
\end{eqnarray*}
Here $\varepsilon_{ijk}$ is the three dimensional anti-symmetric Levi-Civita 
symbol. Using the explicit form (\ref{eq:A.4.5a}), (\ref{eq:A.4.5b}), 
(\ref{eq:A.4.7a}) and (\ref{eq:A.4.7b}) of the various matrix elements and 
equation (\ref{eq:3.2.7c}), finally we obtain that, for $s>1$, 
\begin{equation}
\bigl(\Delta_{\psi_{s,m}}{\bf J}^{23}\bigr)^2=\bigl(\Delta_{\psi_{s,m}}{\bf J}
^{31}\bigr)^2=\frac{1}{2}\hbar^2\bigl(s^2-m^2+s\bigr), \hskip 30pt
\Delta_{\psi_{s,m}}{\bf J}^{12}=0. \label{eq:3.3.2a}
\end{equation}
These expressions are precisely those obtained in the case of $SU(2)$ and 
$E(3)$-invariant systems, these are independent of the specific form of 
the function $f$, and they have the correct and expected $\sqrt{s}$ growth 
for the standard deviation of ${\bf J}^{ij}$ in the $\vert m\vert=s\to
\infty$ limit. 

In a similar way, by (\ref{eq:3.2.7c}), (\ref{eq:3.2.8}), (\ref{eq:3.2.10}) 
and (\ref{eq:A.4.7a}), for $s>1$ we obtain 
\begin{eqnarray}
&{}&\Bigl(\Delta_{\psi_{s,m}}{\bf J}^{0i}\Bigr)^2=s\hbar^2\bigl(\frac{1}{2}+
  \frac{1}{\epsilon^2}\bigr)+ \nonumber \\
&{}&+\hbar^2\langle{}_{\pm s}Y_{s,m}\vert(\frac{{\bf p}^i}{p})^2\vert{}_{\pm s}
  Y_{s,m}\rangle_1\Bigl(\frac{2}{\sqrt{\pi}}\epsilon^2\int^\infty_0\frac{x^4}
  {1+2\epsilon^2x^2}e^{-x^2}{\rm d}x+\frac{9}{4}-s+\frac{1}{\epsilon^2}
  \bigl(\frac{3}{4}-s\bigr)\Bigr)+ \nonumber \\
&{}&+\frac{1}{2}\hbar^2s\langle{}_{\pm(s-1)}Y_{s,m}\vert(\frac{{\bf p}^i}{p})
  ^2\vert{}_{\pm(s-1)}Y_{s,m}\rangle_1. \label{eq:3.3.3}
\end{eqnarray}
Then by (\ref{eq:A.4.4a}) and (\ref{eq:A.4.4b}) this yields 
\begin{eqnarray}
&{}& \hskip -20pt\Bigl(\Delta_{\psi_{s,m}}{\bf J}^{01}\Bigr)^2=\Bigl(
  \Delta_{\psi_{s,m}}{\bf J}^{02}\Bigr)^2= \nonumber \\
&{}&=\frac{\hbar^2}{(s+1)(2s+3)}\Bigl\{(s^2-m^2)\Bigl(\frac{2}{\sqrt{\pi}}
  \epsilon^2\int^\infty_0\frac{x^4}{1+2\epsilon^2x^2}e^{-x^2}{\rm d}x+
  \frac{3}{4}-\frac{1}{2}s+\frac{1}{\epsilon^2}\bigl(\frac{3}{4}-s\bigr)
  \Bigr)+ \nonumber \\
&{}& \hskip 80pt +(2s+1)\frac{2}{\sqrt{\pi}}\epsilon^2\int^\infty_0
  \frac{x^4}{1+2\epsilon^2x^2}e^{-x^2}{\rm d}x+\frac{9}{4}+5s+\frac{5}{2}s^2
  +s^3+ \nonumber \\
&{}& \hskip 80pt +\frac{1}{\epsilon^2}\Bigl(\frac{3}{4}+\frac{7}{2}s+3s^2
  +2s^3\Bigr)\Bigr\}, \label{eq:3.3.3a}
\end{eqnarray}
and 
\begin{eqnarray}
&{}& \hskip -20pt\Bigl(\Delta_{\psi_{s,m}}{\bf J}^{03}\Bigr)^2= \nonumber \\
&{}&=\frac{\hbar^2}{(s+1)(2s+3)}\Bigl\{(m^2-s^2)\Bigl(\frac{2}{\sqrt{\pi}}
  \epsilon^2\int^\infty_0\frac{x^4}{1+2\epsilon^2x^2}e^{-x^2}{\rm d}x+
  \frac{3}{4}-\frac{1}{2}s+\frac{1}{\epsilon^2}\bigl(\frac{3}{4}-s\bigr)
  \Bigr)+ \nonumber \\
&{}& \hskip 80pt +(2s^2+s+1)\frac{2}{\sqrt{\pi}}\epsilon^2\int^\infty_0
  \frac{x^4}{1+2\epsilon^2x^2}e^{-x^2}{\rm d}x+\frac{9}{4}+\frac{17}{4}s+
  \frac{9}{2}s^2+ \nonumber \\
&{}& \hskip 80pt +\frac{1}{\epsilon^2}\Bigl(\frac{3}{4}+\frac{11}{4}s+
  \frac{11}{2}s^2\Bigr)\Bigr\}. \label{eq:3.3.3b}
\end{eqnarray}
Thus, for finite $\epsilon$ the variances $(\Delta_{\psi_{s,m}}{\bf J}^{0i})
^2$ grow like $s$ in the $s\to\infty$ limit even if $m\not=s$. However, as 
we concluded from (\ref{eq:3.3.1a}) and (\ref{eq:3.3.1c}), we had to assume 
that $\epsilon\to0$ when $s\to\infty$. Hence, with the assumption $\epsilon
=s^{-\alpha}$ for some $\alpha>0$, by (\ref{eq:3.3.3a}) and (\ref{eq:3.3.3b}) 
even in the $m=s$ case, $(\Delta_{\psi_{s,m}}{\bf J}^{01})^2$ and $(\Delta
_{\psi_{s,m}}{\bf J}^{02})^2$ grow like $s^{1+2\alpha}$ and $(\Delta_{\psi_{s,m}}
{\bf J}^{03})^2$ grows like $s^{2\alpha}$ as $s\to\infty$. Comparing these 
rates with those obtained for $(\Delta_{\psi_{s,m}}{\bf p}^a)^2$ above, the 
optimal choice for $\epsilon$ seems to be $\epsilon=1/\sqrt[4]{s}$, 
yielding $s\sqrt{s}$ or $\sqrt{s}$ divergences for the variances. 
Therefore, the sequence $\vert\psi_{s,s}\rangle$ of states for $2s=3,4,...$ 
with $\epsilon=1/\sqrt[4]{s}$ can be used to define the classical limit 
of $E(1,3)$-invariant quantum systems. 

Clearly, the sequence of states of the form ${\bf U}_{(A,\xi)}\vert\psi_{s,s}
\rangle$ defines the same kind of classical limit, where ${\bf U}_{(A,\xi)}$ 
is the unitary operator representing the Poincar\'e transformation $(A^A{}_B,
\xi^a)\in E(1,3)$. The general, large classical values $p^a$ and $J^{ab}$ 
of ${\bf p}^a$ and ${\bf J}^{ab}$, respectively, can be recovered by 
sequences of states of this form. The states that we use in the proof of 
our main result will have this structure.


\section{Two-particle systems}
\label{sec-4}

\subsection{The relative position and distance operators}
\label{sub-4.1}

On the tensor product space ${\cal H}_1\otimes{\cal H}_2$ of two irreducible 
representation spaces, labeled, respectively, by $(\mu_1,s_1)$ and $(\mu_2,
s_2)$, equation (\ref{eq:3.1.4}) yields 
\begin{eqnarray}
&{}&\frac{1}{2}\mu^2_1\varepsilon^a{}_{bcd}{\bf J}^{bc}_1\otimes{\bf p}^d_2=
  {\bf S}^a_1{\bf p}_{1b}\otimes{\bf p}^b_2-{\bf S}_{1b}{\bf p}^a_1\otimes
  {\bf p}^b_2+\varepsilon^a{}_{bcd}{\bf C}^c_1{\bf p}^d_1\otimes{\bf p}^b_2, 
\label{eq:4.1.1a} \\
&{}&\frac{1}{2}\mu^2_2\varepsilon^a{}_{bcd}{\bf p}^d_1\otimes{\bf J}^{bc}_2=
  {\bf p}^b_1\otimes{\bf S}^a_2{\bf p}_{2b}-{\bf p}^b_1\otimes{\bf S}_{2b}
  {\bf p}^a_2+\varepsilon^a{}_{bcd}{\bf p}^b_1\otimes{\bf C}^c_2{\bf p}^d_2.
\label{eq:4.1.1b}
\end{eqnarray}
Let the 4-momentum and Pauli--Lubanski spin operators of the composite system 
be denoted by ${\bf p}^a$ and ${\bf S}^a$, respectively, and let us introduce 
the notations 
\begin{eqnarray}
&{}&{\bf P}^2_{12}:=\frac{1}{2}\bigl(\eta_{ab}{\bf p}^a{\bf p}^b-(\mu^2_1+\mu^2
  _2){\bf I}_1\otimes{\bf I}_2\bigr)=\eta_{ab}{\bf p}^a_1\otimes{\bf p}^b_2,
  \label{eq:4.1.2a}\\
&{}&{\bf S}^a_{12}:={\bf S}^a-{\bf S}^a_1\otimes{\bf I}_2-{\bf I}_1\otimes
  {\bf S}^a_2=\frac{1}{2}\varepsilon^a{}_{bcd}\bigl({\bf J}^{bc}_1\otimes
  {\bf p}^d_2+{\bf p}^d_1\otimes{\bf J}^{bc}_2\bigr). \label{eq:4.1.2b}
\end{eqnarray}
Here ${\bf I}_{\bi}$, ${\bi}=1,2$, denote the identity operators on the 
respective Hilbert spaces. The operators ${\bf P}^2_{12}$ and ${\bf S}^a_{12}$ 
characterize the relationship between the two subsystems in the composite 
system. For later use, it seems useful to calculate the commutator of these 
with some other operators. By the second of (\ref{eq:3.1.3a}), it follows 
that $[{\bf S}^a_1\otimes{\bf I}_2,{\bf P}^2_{12}]=[{\bf I}_1\otimes{\bf S}
^a_2,{\bf P}^2_{12}]=0$; and, by the second of (\ref{eq:3.1.2a}), it follows
that $[{\bf S}^a_{12},{\bf p}^b_1\otimes{\bf I}_2]=-[{\bf S}^a_{12},{\bf I}_1
\otimes{\bf p}^b_2]=-{\rm i}\hbar\varepsilon^{ab}{}_{cd}{\bf p}^c_1\otimes
{\bf p}^d_2$. The latter implies that $[{\bf S}^a_{12},{\bf P}^2_{12}]=0$. 

Using (\ref{eq:4.1.1a})-(\ref{eq:4.1.1b}) and the commutators 
(\ref{eq:3.1.3a})-(\ref{eq:3.1.3b}), from (\ref{eq:4.1.2b}) we obtain 
\begin{eqnarray*}
{\bf S}^a_{12}=\!\!\!\!&-\!\!\!\!&\frac{1}{\mu^2_1}{\bf p}^a_1{\bf S}_{1b}
  \otimes{\bf p}^b_2-\frac{1}{\mu^2_2}{\bf p}^b_1\otimes{\bf p}^a_2{\bf S}
  _{2b}+ \\
\!\!\!\!&+\!\!\!\!&\Bigl(\frac{1}{\mu^2_1}{\bf S}^a_1\otimes{\bf I}_2+
  \frac{1}{\mu^2_2}{\bf I}_1\otimes{\bf S}^a_2\Bigr){\bf P}^2_{12}+\varepsilon
  ^a{}_{bcd}{\bf p}^b_1\otimes{\bf p}^c_2\Bigl(\frac{{\bf C}^d_1}{\mu^2_1}\otimes
  {\bf I}_2-{\bf I}_1\otimes\frac{{\bf C}^d_2}{\mu^2_2}\Bigr)= \\
=\!\!\!\!&-\!\!\!\!&\frac{1}{\mu^2_1}p^a_1p^b_2{\bf S}_{1b}
  \otimes{\bf I}_2-\frac{1}{\mu^2_2}p^a_2p^b_1{\bf I}_1\otimes{\bf S}_{2b}+ \\
\!\!\!\!&+\!\!\!\!&P^2_{12}\Bigl(\frac{1}{\mu^2_1}{\bf S}^a_1\otimes{\bf I}_2+
  \frac{1}{\mu^2_2}{\bf I}_1\otimes{\bf S}^a_2\Bigr)+\varepsilon^a{}_{bcd}
  p^b_1p^c_2\Bigl(\frac{{\bf C}^d_1}{\mu^2_1}\otimes{\bf I}_2-{\bf I}_1\otimes
  \frac{{\bf C}^d_2}{\mu^2_2}\Bigr).
\end{eqnarray*}
In the second step, we used that in the actual representation ${\bf p}^a_1$, 
${\bf p}^a_2$ and ${\bf P}^2_{12}$ are multiplication operators (see Appendix 
\ref{sub-A.2}). Multiplying it (from the left) by $\varepsilon^e{}_{cda}p^c_1
p^d_2$ and recalling that $(\varepsilon^a{}_{cde}p^c_1p^d_2)(\varepsilon^e{}
_{fgb}p^f_1p^g_2)=-(P^4_{12}-\mu^2_1\mu^2_2)\Pi^a_b$, where $\Pi^a_b$ is the 
projection to the spacelike 2-plane orthogonal to $p^a_1$ and $p^a_2$ (see 
equation (\ref{eq:2.2.5})), we obtain 
\begin{equation}
{\bf d}^a_{12}:=\Pi^a_b\Bigl(\frac{{\bf C}^b_1}{\mu^2_1}\otimes{\bf I}_2-
{\bf I}_1\otimes\frac{{\bf C}^b_2}{\mu^2_2}\Bigr)=-\frac{1}{P^4_{12}-\mu^2_1
\mu^2_2}\varepsilon^a{}_{bcd}p^b_1p^c_2\Bigl({\bf S}^d_{12}-P^2_{12}\bigl(
\frac{{\bf S}^d_1}{\mu^2_1}\otimes{\bf I}_2+{\bf I}_1\otimes\frac{{\bf S}
^d_2}{\mu^2_2}\bigr)\Bigr). \label{eq:4.1.3}
\end{equation}
Since ${\bf S}^e_{12}$ does not commute with $\varepsilon^a{}_{bcd}p^b_1p^c_2$, 
the operator ${\bf d}^a_{12}$ cannot be expected to be self-adjoint. In fact, 
\begin{equation}
\bigl({\bf d}^a_{12}\bigr)^\dagger-{\bf d}^a_{12}=-\frac{1}{P^4_{12}-\mu^2_1\mu
^2_2}\Bigl[\varepsilon^a{}_{bcd}p^b_1p^c_2,{\bf S}^d_{12}\Bigr]=
\frac{2{\rm i}\hbar}{P^4_{12}-\mu^2_1\mu^2_2}\Bigl(\bigl(P^2_{12}+\mu^2_1
\bigr)p^a_2-\bigl(P^2_{12}+\mu^2_2\bigr)p^a_1\Bigr). \label{eq:4.1.4}
\end{equation}
Here we used that ${\bf S}^e_1$ and ${\bf S}^e_2$ commute with both 
$\varepsilon^a{}_{bcd}p^b_1p^c_2$ and $1/(P^4_{12}-\mu^2_1\mu^2_2)$, and 
${\bf S}^e_{12}$ commutes with $1/(P^4_{12}-\mu^2_1\mu^2_2)$, which can be 
verified directly using e.g. expression (\ref{eq:A.2.3b}) of the 
centre-of-mass operator. Thus the self-adjoint part of ${\bf d}^a_{12}$ 
can be written into the form 
\begin{equation}
{\bf D}^a_{12}:=\frac{1}{2}\bigl({\bf d}^a_{12}+\bigl({\bf d}^a_{12}\bigr)
^\dagger\bigr)={\bf d}^a_{12}+\frac{{\rm i}\hbar}{P^4_{12}-\mu^2_1\mu^2_2}
\Bigl(\bigl(P^2_{12}+\mu^2_1\bigr)p^a_2-\bigl(P^2_{12}+\mu^2_2\bigr)p^a_1
\Bigr). \label{eq:4.1.5}
\end{equation}
${\bf D}^a_{12}$ can be interpreted as the `relative position operator' of 
the first elementary system with respect to the second. However, the 
`correction term' in (\ref{eq:4.1.5}) that makes ${\bf d}^a_{12}$ self-adjoint 
is rather trivial: since $p_{1a}{\bf D}^a_{12}=-p_{2a}{\bf D}^a_{12}={\rm i}
\hbar{\bf I}_1\otimes{\bf I_2}$, its only non-zero component is proportional 
to $p^a_1-p^a_2$, which is a universal expression: $(p_{1a}-p_{a2}){\bf D}^a
_{12}=2{\rm i}\hbar{\bf I}_1\otimes{\bf I}_2$. This ${\bf D}^a_{12}$ is the 
quantum mechanical analog of (\ref{eq:2.2.3}). 

Nevertheless, despite the non-self-adjointness of ${\bf d}^a_{12}$, 
\emph{its square, $\eta_{ab}{\bf d}^a_{12}{\bf d}^b_{12}$, is self-adjoint}. 
To see this, in $(\eta_{ab}{\bf d}^a_{12}{\bf d}^b_{12})^\dagger=\eta_{ab}
({\bf d}^a_{12})^\dagger({\bf d}^b_{12})^\dagger$ it is enough to use 
(\ref{eq:4.1.3})--(\ref{eq:4.1.4}) and the commutators above to obtain that 
it is, in fact, $\eta_{ab}{\bf d}^a_{12}{\bf d}^b_{12}$. Its manifestly 
self-adjoint form is given explicitly by 
\begin{eqnarray}
\eta_{ab}\!\!\!\!\!&{}\!\!\!\!\!&{\bf d}^a_{12}{\bf d}^b_{12}=\eta_{ab}{\bf D}
  ^a_{12}{\bf D}^b_{12}+\hbar^2\frac{2P^2_{12}+\mu^2_1+\mu^2_2}{P^4_{12}-\mu^2_1
  \mu^2_2}{\bf I}_1\otimes{\bf I}_2= \label{eq:4.1.6} \\
\!\!\!\!\!&{}\!\!\!\!\!&=\Bigl(\frac{{\bf C}^a_1}{\mu^2_1}\otimes{\bf I}_2-
  {\bf I}_1\otimes\frac{{\bf C}^a_2}{\mu^2_2}\Bigr)\Pi_{ab}\Bigl(
  \frac{{\bf C}^b_1}{\mu^2_1}\otimes{\bf I}_2-{\bf I}_1\otimes\frac{{\bf C}
  ^b_2}{\mu^2_2}\Bigr)= \nonumber \\
\!\!\!\!\!&{}\!\!\!\!\!&=-\frac{1}{P^4_{12}-\mu^2_1\mu^2_2}\Bigl(\frac{{\bf C}
  ^a_1}{\mu^2_1}\otimes{\bf I}_2-{\bf I}_1\otimes\frac{{\bf C}^a_2}{\mu^2_2}
  \Bigr)\varepsilon_{acde}p^c_1p^d_2\,\varepsilon^e{}_{ghb}p^g_1p^h_2\Bigl(
  \frac{{\bf C}^b_1}{\mu^2_1}\otimes{\bf I}_2-{\bf I}_1\otimes\frac{{\bf C}
  ^b_2}{\mu^2_2}\Bigr). \nonumber
\end{eqnarray}
Here, to derive the expression in the second line from the first 
expression in (\ref{eq:4.1.3}) we used $[{\bf C}^a_1\otimes{\bf I}_2,
\Pi^b_a]\Pi_{bc}=[{\bf I}_1\otimes{\bf C}^a_2,\Pi^b_a]\Pi_{bc}=0$; while to
get the third line we used (\ref{eq:2.2.5}) and that ${\bf S}^a_{12}$ 
commutes with $1/(P^4_{12}-\mu^2_1\mu^2_2)$. Thus, the operator for the 
square of the distance between the two constituent, elementary subsystems 
in the composite system can be \emph{defined} by either $\eta_{ab}{\bf d}^a
_{12}{\bf d}^b_{12}$ or $\eta_{ab}{\bf D}^a_{12}{\bf D}^b_{12}$; and they 
differ only in a term of order $\hbar^2$. (\ref{eq:4.1.6}) is a well 
defined operator, given by the centre-of-mass and the 4-momentum operators, 
or by ${\bf S}^a_{12}$, ${\bf P}^2_{12}$ and the 4-momentum and spin 
operators of the constituent systems \emph{in an $E(1,3)$-invariant manner}. 

However, in contrast to the classical expression (\ref{eq:2.2.5}) (see also 
(\ref{eq:2.2.4})), $P^4_{12}$ in the denominator is \emph{not} constant: 
this is $P^4_{12}=(\eta_{ab}p^a_1p^b_2)^2$, an expression of the variables 
$p^a_1$ and $p^a_2$ of the wave functions representing the states of the 
subsystems. Hence, it could be difficult to evaluate the expectation value 
or to calculate the standard deviation of (\ref{eq:4.1.6}). Nevertheless, 
the role of the factor $1/(P^4_{12}-\mu^2_1\mu^2_2)$ in (\ref{eq:4.1.6}) is 
only the normalization of the middle factor $\varepsilon_{acde}p^c_1p^d_2\,
\varepsilon^e{}_{ghb}p^g_1p^h_2$ to be the projection $\Pi_{ab}$. Hence, its 
role is analogous to the magnitude of the angular momentum 3-vectors in the 
definition of the empirical angle $\delta_{ab}J^a_1J^b_2/\vert J_1\vert
\vert J_2\vert$ in $SU(2)$-invariant systems \cite{Sz21c}, and to the 
similar normalization factor in the definition of the empirical distance 
in $E(3)$-invariant systems \cite{Sz22a}. Thus, although $\eta_{ab}{\bf d}
^a_{12}{\bf d}^b_{12}$ might yield a well defined \emph{operator} for the 
distance between the subsystems, and certainly it would be worth studying 
this, in the present paper we follow the strategy of \cite{Sz21c} for the 
$SU(2)$ and of \cite{Sz22a} for the $E(3)$-invariant systems, and we 
calculate only the `empirical distance' based on (\ref{eq:4.1.6}) and 
introduced in the next subsection. As we will see, this approximation of 
the expectation value of the operator (\ref{eq:4.1.6}) is enough to prove 
our key result in subsection \ref{sub-4.3}. 


\subsection{The empirical distance}
\label{sub-4.2}

Let ${\cal S}_1$ and ${\cal S}_2$ be two $E(1,3)$-invariant elementary 
quantum mechanical systems, characterized by the Casimir invariants $(\mu_1,
s_1)$ and $(\mu_2,s_2)$, respectively. Then, motivated by (\ref{eq:4.1.6}) 
and introducing the notation 
\begin{equation}
\Sigma^a:=\frac{{\bf C}^a_1}{\mu^2_1}\otimes{\bf I}_2-{\bf I}_1\otimes
\frac{{\bf C}^a_2}{\mu^2_2}, \label{eq:4.2.s}
\end{equation}
their \emph{empirical distance} in their state $\phi_1$ and $\phi_2$, 
respectively, is defined to be the square root of 
\begin{equation}
d^2_{12}:=\frac{\langle\phi_1\otimes\phi_2\vert\Sigma^a\varepsilon_{acde}
p^c_1p^d_2\,\varepsilon^e{}_{ghb}p^g_1p^h_2\Sigma^b\vert\phi_1\otimes\phi_2
\rangle}{\langle\phi_1\otimes\phi_2\vert{\bf P}^4_{12}-\mu^2_1\mu^2_2
{\bf I}_1\otimes{\bf I}_2\vert\phi_1\otimes\phi_2\rangle}. \label{eq:4.2.1}
\end{equation}
Note that this is \emph{not} the expectation value of the \emph{operator} 
(\ref{eq:4.1.6}), it is the quotient of the expectation value of that's 
numerator and of that's denominator. Since, however, the denominator 
$1/(P^4_{12}-\mu^2_1\mu^2_2)$ in (\ref{eq:4.1.6}) is a multiplication rather 
than a differential operator, it is translation invariant. Hence, $d^2_{12}$ 
above can be expected not to deviate from the expectation value of 
(\ref{eq:4.1.6}) in the leading order in the classical limit. Since $d^2_{12}$ 
is not the expectation value of some operator, its standard deviation is not 
defined. Nevertheless, as we will see in subsection \ref{sub-4.3.2}, its 
\emph{uncertainty} can be defined. 

Clearly, just as in the $SU(2)$ and $E(3)$-invariant cases, this empirical 
geometrical quantity can be defined between any two constituent elementary 
subsystems of any composite system consisting of any finite number of 
subsystems, say ${\cal S}_1,...,{\cal S}_N$, in any of its state (not only 
in pure tensor product states of a bipartite system). In fact, in this more 
general case the Hilbert space of the pure states of the composite system is 
the tensor product ${\cal H}:={\cal H}_1\otimes\cdots\otimes{\cal H}_N$ of 
the Hilbert spaces of the constituent elementary subsystems, a general state 
is a density operator $\rho:{\cal H}\to{\cal H}$, and the square of the 
empirical distance between the ${\bf i}$'th and the ${\bf j}$'th subsystems, 
${\bi},{\bj}=1,...,N$, in this state is defined by 
\begin{equation}
d^2_{\bi\bj}:=\frac{{\rm tr}\Bigl(\rho\bigl(\Sigma^a_{\bi\bj}\varepsilon_{acde}
p^c_{\bi}p^d_{\bj}\,\varepsilon^e{}_{ghb}p^g_{\bi}p^h_{\bj}\Sigma^b_{\bi\bj}\bigr)
\Bigr)}{{\rm tr}\Bigl(\rho\bigl({\bf P}^4_{\bi\bj}-\mu^2_{\bi}\mu^2_{\bj}{\bf I}_1
\otimes\cdots\otimes{\bf I}_N\bigr)\Bigr)}, \label{eq:4.2.1a}
\end{equation}
where $\Sigma_{\bi\bj}$ and ${\bf P}^4_{\bi\bj}$ are the obvious generalizations 
of $\Sigma$ and ${\bf P}^4_{12}$, respectively. 
If, however, $\rho$ represents the tensor product $\phi:=\phi_1\otimes\cdots
\otimes\phi_N$ of pure states of the elementary subsystems, then 
(\ref{eq:4.2.1a}) reduces to an expression of the form (\ref{eq:4.2.1}) and 
the result depends only on the states $\phi_{\bi}$ and $\phi_{\bj}$, and will 
be independent of the states of the other subsystems. Since the states in 
the proof of our main result will be chosen to be such tensor product 
states, it is enough to carry out the subsequent analysis for a bipartite 
system. 

Unfortunately, the calculation of the empirical distance is considerably 
more complicated than in the $E(3)$-invariant case: while in the latter the 
relative position vector is proportional to the \emph{uniquely determined} 
direction orthogonal to the two linear momenta $p^i_1$ and $p^i_2$, $i=
1,2,3$, and hence the (signed) distance between the two subsystems is 
already given by the component of the relative position vector orthogonal 
to $p^i_1$ and $p^i_2$, in the present $E(1,3)$-invariant case the relative 
position vector has \emph{two} components orthogonal to $p^a_1$ and $p^a_2$. 
Thus, to get the distance between the subsystems in the $E(3)$-invariant 
case it was enough to consider the relative position vector itself, in the 
$E(1,3)$-invariant case we must calculate the \emph{square} of the relative 
position vector. This yields much more terms to evaluate in the calculation 
of the expectation values and variances. 

Since our aim is to recover the metric structure of the classical Minkowski 
space, we calculate these expectation values directly in the classical limit 
without evaluating them in general quantum states. This will be done in the 
next subsection. 


\subsection{The classical limit}
\label{sub-4.3}

The main result of the paper is the following theorem: 

\begin{theorem*}
Let $\gamma^a_1,\cdots,\gamma^a_N$ be timelike straight lines in the 
Minkowski space $\mathbb{R}^{1,3}$ such that no two of them are parallel. 
Then there are $E(1,3)$-invariant quantum mechanical systems ${\cal S}_1,
...,{\cal S}_N$ and a sequence of their pure quantum states $\phi_{1k},
...,\phi_{Nk}$, $k\in\mathbb{N}$, respectively, such that, in the $k\to
\infty$ limit, the empirical distances $d_{\bi\bj}$, calculated in the states 
$\phi_{1k}\otimes\cdots\otimes\phi_{Nk}$, tend with asymptotically vanishing 
uncertainty to the classical Lorentzian distances $D_{\bi\bj}$ between the 
straight lines $\gamma^a_{\bi}$ and $\gamma^a_{\bj}$, given by (\ref{eq:2.2.5}), 
for any ${\bi,\bj}=1,...,N$. 
\end{theorem*}

We split the proof of this theorem into two parts: the statement on the 
classical limit of the empirical distance will be proven in subsection 
\ref{sub-4.3.1}, while that on the uncertainty in subsection 
\ref{sub-4.3.2}. 


\subsubsection{The classical limit of the empirical distance}
\label{sub-4.3.1}

As we noted in connection with the general form (\ref{eq:4.2.1a}) of the 
empirical distance, if the state $\rho$ is the pure tensor product state 
$\phi_{1k}\otimes\cdots\otimes\phi_{Nk}$ as in the present case, then 
$d_{\bi\bj}$ depends only on the states $\phi_{\bi}$ and $\phi_{\bj}$ of 
${\cal S}_{\bi}$ and ${\cal S}_{\bj}$, respectively, and independent of the 
other factors in this tensor product. Hence, it is enough to prove the 
Theorem for $N=2$, i.e. when ${\bi}=1,2$. Then the operator in the 
numerator of (\ref{eq:4.2.1}) can be written as 
\begin{eqnarray}
{\bf A}:=\varepsilon_{acde}\varepsilon^e{}_{ghb}\Bigl\{\frac{1}{\mu^4_1}
  \!\!\!\!\!&{}\!\!\!\!\!&{\bf J}^a_1{}_{a_1}{\bf p}^{a_1}_1{\bf p}^c_1{\bf p}^g_1
  {\bf p}^{b_1}_1{\bf J}^b_1{}_{b_1}\otimes{\bf p}^d_2{\bf p}^h_2-\frac{1}{\mu
  ^2_1\mu^2_2}{\bf J}^a_1{}_{a_1}{\bf p}^{a_1}_1{\bf p}^c_1{\bf p}^g_1\otimes
  {\bf p}^d_2{\bf p}^h_2{\bf p}^{b_2}_2{\bf J}^b_2{}_{b_2}- \nonumber \\
-\frac{1}{\mu^2_1\mu^2_2}\!\!\!\!\!&{}\!\!\!\!\!&{\bf p}^c_1{\bf p}^g_1{\bf p}
  ^{b_1}_1{\bf J}^b_1{}_{b_1}\otimes{\bf J}^a_2{}_{a_2}{\bf p}^{a_2}_2{\bf p}^d
  _2{\bf p}^h_2+\frac{1}{\mu^4_2}{\bf p}^c_1{\bf p}^g_1\otimes{\bf J}^a_2
  {}_{a_2}{\bf p}^{a_2}_2{\bf p}^d_2{\bf p}^h_2{\bf p}^{b_2}_2{\bf J}^b_2{}_{b_2}
  \Bigr\}, \label{eq:4.3.1}
\end{eqnarray}
while that in the denominator is 
\begin{equation}
{\bf B}:=\eta_{ac}\eta_{bd}{\bf p}^a_1{\bf p}^b_1\otimes{\bf p}^c_2{\bf p}
^d_2-\mu^2_1\mu^2_2{\bf I}_1\otimes{\bf I}_2. \label{eq:4.3.2}
\end{equation}
We calculate the leading order terms in the expectation value of these 
operators in the states $\phi_1\otimes\phi_2$ specified below. 

Let us represent the timelike straight lines $\gamma^a_{\bi}$ by the 
Lorentz boosts $\Lambda^a_{\bi}{}_b$, the translations $\xi^a_{\bi}$ and a 
single timelike straight line $\gamma^a_0$ just as we did it at the end 
of subsection \ref{sub-2.1}: $\gamma^a_{\bi}(u)=\Lambda^a_{\bi}{}_b\gamma
^b_0(u)+\xi^a_{\bi}$; and let us consider the states $\vert\phi_{\bi}\rangle
=\exp({\rm i}p_e\xi^e_{\bi}/\hbar){\bf U}_{\bi}\vert\psi_{s_{\bi},s_{\bi}}\rangle$ 
of the $E(1,3)$-invariant elementary quantum mechanical system ${\cal S}
_{\bi}$. Here ${\bf U}_{\bi}={\bf U}_{(A_{\bi},0)}$, the unitary operator 
representing one of the two $SL(2,\mathbb{C})$ matrices $\pm A^A_{\bi}{}_B$ 
corresponding to the Lorentz boost $\Lambda^a_{\bi}{}_b$. Thus, the states 
$\vert\phi_{\bi}\rangle$ depend on $s_{\bi}$. However, to reduce the number 
of indices, we do not write out $s$ on the states, and we simply write e.g. 
$\vert\psi_{\bi}\rangle=\vert\psi_{s_{\bi},s_{\bi}}\rangle$. 

Next we evaluate the factors in the tensor products in the operators 
${\bf A}$ and ${\bf B}$. Using (\ref{eq:A.2.3a}) and (\ref{eq:A.2.3b}), 
for the first term in (\ref{eq:4.3.1}) we obtain that 
\begin{eqnarray}
\frac{1}{\mu^4_1}\!\!\!\!\!&{}\!\!\!\!\!&\varepsilon_{acde}\varepsilon^e
  {}_{ghb}\langle\phi_1\vert{\bf J}^a_{1a_1}{\bf p}^{a_1}_1{\bf p}^c_1{\bf p}
  ^g_1{\bf p}^{b_1}_1{\bf J}^b_{1b_1}\vert\phi_1\rangle\langle\phi_2\vert
  {\bf p}^d_2{\bf p}^h_2\vert\phi_2\rangle= \label{eq:4.3.3} \\
=\!\!\!\!\!&{}\!\!\!\!\!&-\xi^a_1\varepsilon_{acde}\xi^b_1\varepsilon_{bgh}
  {}^e\Lambda^c_1{}_{c_1}\Lambda^g_1{}_{g_1}\langle\psi_1\vert{\bf p}^{c_1}_1
  {\bf p}^{g_1}_1\vert\psi_1\rangle\Lambda^d_2{}_{d_2}\Lambda^h_2{}_{h_2}
  \langle\psi_2\vert{\bf p}^{d_2}_2{\bf p}^{h_2}_2\vert\psi_2\rangle+ 
  \nonumber \\
\!\!\!\!&{}\!\!\!\!&+\frac{1}{\mu^2_1}\xi^a_1\varepsilon_{acde}\varepsilon^e
  {}_{ghb}\Lambda^c_1{}_{c_1}\Lambda^g_1{}_{g_1}\Lambda^b_1{}_{b_1}\langle\psi_1
  \vert{\bf p}^{c_1}_1{\bf p}^{g_1}_1{\bf p}^f_1{\bf J}^{b_1}_{1f}\vert\psi_1
  \rangle\Lambda^d_2{}_{d_2}\Lambda^h_2{}_{h_2}\langle\psi_2\vert{\bf p}^{d_2}_2
  {\bf p}^{h_2}_2\vert\psi_2\rangle+ \nonumber \\
\!\!\!\!&{}\!\!\!\!&+\frac{1}{\mu^2_1}\varepsilon_{acde}\varepsilon^e{}_{ghb}
  \xi^b_1\Lambda^a_1{}_{a_1}\Lambda^c_1{}_{c_1}\Lambda^g_1{}_{g_1}\langle\psi_1
  \vert{\bf J}^{a_1}_{1f}{\bf p}^f_1{\bf p}^{c_1}_1{\bf p}^{g_1}_1\vert\psi_1
  \rangle\Lambda^d_2{}_{d_2}\Lambda^h_2{}_{h_2}\langle\psi_2\vert{\bf p}^{d_2}
  _2{\bf p}^{h_2}_2\vert\psi_2\rangle+ \nonumber \\
\!\!\!\!&{}\!\!\!\!&+\frac{1}{\mu^4_1}\varepsilon_{acde}\varepsilon^e{}_{ghb}
  \Lambda^a_1{}_{a_1}\Lambda^c_1{}_{c_1}\Lambda^g_1{}_{g_1}\Lambda^b_1{}_{b_1}
  \langle\psi_1\vert{\bf J}^{b_1}_{1f_1}{\bf p}^{f_1}_1{\bf p}^{c_1}_1{\bf p}
  ^{g_1}_1{\bf p}^{f_2}_1{\bf J}^{a_1}_{1f_2}\vert\psi_1\rangle\Lambda^d_2{}_{d_2}
  \Lambda^h_2{}_{h_2}\langle\psi_2\vert{\bf p}^{d_2}_2{\bf p}^{h_2}_2\vert\psi_2
  \rangle. \nonumber
\end{eqnarray}
Therefore, we should determine the expectation values $\langle\psi_{s,m}
\vert{\bf p}^a{\bf p}^b\vert\psi_{s,m}\rangle$, $\langle\psi_{s,m}\vert
{\bf p}^a{\bf p}^b{\bf p}^e{\bf J}^c{}_e\vert\psi_{s,m}\rangle$, $\langle
\psi_{s,m}\vert{\bf J}^a{}_e{\bf p}^e{\bf p}^b{\bf p}^c\vert\psi_{s,m}
\rangle$ and $\langle\psi_{s,m}\vert{\bf J}^a{}_e{\bf p}^e{\bf p}^b{\bf p}^c
{\bf p}^f{\bf J}^d{}_f\vert\psi_{s,m}\rangle$, at least for large $m=s$. 

By (\ref{eq:3.2.8}), (\ref{eq:3.2.9a}), (\ref{eq:3.2.10}), 
(\ref{eq:A.4.4a}) and (\ref{eq:A.4.4b}) it is easy to calculate $\langle
\psi_{s,m}\vert{\bf p}^a{\bf p}^b\vert\psi_{s,m}\rangle$: for $s>0$ it is 
\begin{eqnarray}
\langle\psi_{s,m}\vert{\bf p}^a{\bf p}^b\vert\psi_{s,m}\rangle\!\!\!\!&=
  \!\!\!\!&\mu^2\delta^a_0\delta^b_0\bigl(1+3\epsilon^2\bigr)+3\epsilon^2
  \mu^2\delta^a_i\delta^b_j\langle{}_{-s}Y_{s,m}\vert\frac{{\bf p}^i{\bf p}^j}
  {(p)^2}\vert{}_{-s}Y_{s,m}\rangle_1= \nonumber \\
\!\!\!\!&=\!\!\!\!&\mu^2\Bigl(\delta^a_0\delta^b_0+O(\frac{1}{\sqrt{s}})
  \Bigr). \label{eq:4.3.4}
\end{eqnarray}
This yields that, for $s_1,s_2>0$, the first term on the right hand side 
of (\ref{eq:4.3.3}) is 
\begin{eqnarray}
-\xi^a_1\varepsilon_{acde}\xi^b_1\varepsilon_{bgh}{}^e\!\!\!\!\!&{}
  \!\!\!\!\!&\Lambda^c_1{}_{c_1}\Lambda^g_1{}_{g_1}\langle\psi_1\vert
  {\bf p}^{c_1}_1{\bf p}^{g_1}_1\vert\psi_1\rangle\Lambda^d_2{}_{d_2}\Lambda
  ^h_2{}_{h_2}\langle\psi_2\vert{\bf p}^{d_2}_2{\bf p}^{h_2}_2\vert\psi_2
  \rangle= \nonumber \\
=\!\!\!\!\!&{}\!\!\!\!\!&\mu^2_1\mu^2_2\Bigl(-\xi^a_1\varepsilon_{acde}
  \Lambda^c_1{}_0\Lambda^d_2{}_0\,\xi^b_1\varepsilon_{bgh}{}^e\Lambda^g_1
  {}_0\Lambda^h_2{}_0+O(\frac{1}{\sqrt{s_1}})+O(\frac{1}{\sqrt{s_2}})\Bigr).
  \label{eq:4.3.5}
\end{eqnarray}
Note that, also for $s_1,s_2>0$, (\ref{eq:4.3.4}) gives the asymptotic 
form of 
\begin{eqnarray*}
\langle\phi_1\otimes\phi_2\vert{\bf P}^4_{12}\vert\phi_1\otimes\phi_2\rangle
  \!\!\!\!&=\!\!\!\!&\eta_{ac}\eta_{bd}\Lambda^a_{1a_1}\Lambda^b_{1b_1}\langle
  \psi_1\vert{\bf p}^{a_1}_1{\bf p}^{b_1}_1\vert\psi_1\rangle\Lambda^c_{2c_2}
  \Lambda^d_{2d_2}\langle\psi_2\vert{\bf p}^{c_2}_2{\bf p}^{d_2}_2\vert\psi_2
  \rangle= \nonumber \\
\!\!\!\!&=\!\!\!\!&(\Lambda^{-1}_1\Lambda_2)_{ac}(\Lambda^{-1}_1\Lambda_2)_{bd}
  \langle\psi_1\vert{\bf p}^a_1{\bf p}^b_1\vert\psi_1\rangle\langle\psi_2
  \vert{\bf p}^c_2{\bf p}^d_2\vert\psi_2\rangle;
\end{eqnarray*}
and hence of the operator ${\bf B}$ given by (\ref{eq:4.3.2}), too: 
\begin{equation}
\langle\phi_1\otimes\phi_2\vert{\bf B}\vert\phi_1\otimes\phi_2\rangle=
\mu^2_1\mu^2_2\Bigl\{\bigl((\Lambda^{-1}_1\Lambda_2)_{00}\bigr)^2-1+O\bigl(
\frac{1}{\sqrt{s_1}}\bigr)+O\bigl(\frac{1}{\sqrt{s_2}}\bigr)\Bigr\}.
\label{eq:4.3.6}
\end{equation}
Apart from the coefficient and the asymptotically vanishing terms between 
the brackets, this is just the denominator in the second line of the 
classical expression (\ref{eq:2.2.5}) for the Lorentzian distance between
$\gamma^a_1$ and $\gamma^a_2$. 

Since $\langle\psi_{s,m}\vert{\bf J}^a{}_e{\bf p}^e{\bf p}^b{\bf p}^c\vert
\psi_{s,m}\rangle=\overline{\langle\psi_{s,m}\vert{\bf p}^c{\bf p}^b{\bf p}^e
{\bf J}^a{}_e\vert\psi_{s,m}\rangle}$, it is enough to calculate only one 
of these. Using (\ref{eq:A.3.4}) and (\ref{eq:A.1.6b}), for $s>1/2$ we 
obtain 
\begin{eqnarray*}
\langle\psi_{s,m}\vert{\bf p}^a{\bf p}^b{\bf p}^e{\bf J}^c{}_e\vert\psi_{s,m}
  \rangle\!\!\!\!\!&{}\!\!\!\!\!&=\langle{\bf p}^b{\bf p}^a\psi_{s,m}\vert
  {\bf p}^e{\bf J}^c{}_e\psi_{s,m}\rangle= \\
=\frac{\rm i}{2}\hbar\mu^2\int_{{\cal M}^+_\mu}\Bigl(\!\!\!\!\!&{}\!\!\!\!\!&
  p^ap^bv^c\frac{p^0}{\mu}f\frac{{\rm d}f}{{\rm d}p}\bigl(\vert{}_{-s}Y_{s,m}
  \vert^2+\vert{}_sY_{s,m}\vert^2\bigr)+ \\
\!\!\!\!\!&{}\!\!\!\!\!&+p^ap^b\bigl(\bar m^c\overline{{}_{-s}Y_{s,m}}({\edth}
  {}_{-s}Y_{s,m})+m^c\overline{{}_sY_{s,m}}({\edth}'{}_sY_{s,m})\bigr)f^2\Bigr)
  {\rm d}v_\mu.
\end{eqnarray*}
Then by integration by parts and using $m^i\bar m^j+\bar m^im^j=\delta^{ij}
-p^ip^j/(p)^2$, $m^i\bar m^j-m^j\bar m^i={\rm i}\delta^{ik}\delta^{jl}
\varepsilon_{klr}p^r/p$ (see Appendix \ref{sub-A.4}) and equations 
(\ref{eq:A.4.7a}) and (\ref{eq:A.4.7b}), we obtain 
\begin{eqnarray*}
\langle\psi_{s,m}\vert\!\!\!\!\!&{}\!\!\!\!\!&{\bf p}^a{\bf p}^b{\bf p}^e
  {\bf J}^c{}_e\vert\psi_{s,m}\rangle={\rm i}\hbar\delta^a_0\delta^b_0\delta
  ^c_0\int^\infty_0(p^0)^2(p)^3f\frac{{\rm d}f}{{\rm d}p}{\rm d}p+ \\
+\!\!\!\!\!&{}\!\!\!\!\!&{\rm i}\hbar\delta^a_i\delta^b_j\delta^c_0\langle
  {}_{-s}Y_{s,m}\vert\frac{{\bf p}^i{\bf p}^j}{(p)^2}\vert{}_{-s}Y_{s,m}\rangle
  _1\int^\infty_0(p)^5f\frac{{\rm d}f}{{\rm d}p}{\rm d}p- \\
-\!\!\!\!\!&{}\!\!\!\!\!&\frac{\rm i}{2}\hbar\mu^2\bigl(\delta^a_0\delta^b_i
  +\delta^a_i\delta^b_0\bigr)\delta^c_k\Bigl({\rm i}(\delta^i_1\delta^k_2-
  \delta^i_2\delta^k_1)\frac{m}{s+1}+\delta^{ik}\Bigr)\int^\infty_0(p)^2f^2
  {\rm d}p+ \\
+\!\!\!\!\!&{}\!\!\!\!\!&{\rm i}\hbar\bigl(\delta^a_0\delta^b_i+\delta^a_i
  \delta^b_0\bigr)\delta^c_k\langle{}_{-s}Y_{s,m}\vert\frac{{\bf p}^i{\bf p}
  ^k}{(p)^2}\vert{}_{-s}Y_{s,m}\rangle_1\Bigl(\int^\infty_0(p^0)^2(p)^3f
  \frac{{\rm d}f}{{\rm d}p}{\rm d}p+\frac{3}{2}\mu^2\int^\infty_0(p)^2f^2
  {\rm d}p\Bigr). 
\end{eqnarray*}
Finally, using (\ref{eq:3.2.8}) and evaluating the integrals by 
(\ref{eq:3.2.10}), we find that 
\begin{equation}
\langle\psi_{s,m}\vert{\bf p}^a{\bf p}^b{\bf p}^e{\bf J}^c{}_e\vert\psi_{s,m}
\rangle=\mu^3\hbar\bigl(C^{abc}+O(\frac{1}{\sqrt{s}})\bigr), \label{eq:4.3.7a}
\end{equation}
where $C^{abc}$ is a constant. Therefore, the second and third lines on the 
right hand side of (\ref{eq:4.3.3}) have the structure 
\begin{equation}
\mu_1\mu^2_2\hbar\bigl(C+O(\frac{1}{\sqrt{s_1}})+O(\frac{1}{\sqrt{s_2}})
\bigr), \label{eq:4.3.7}
\end{equation}
where $C$ is a constant. We will not need the exact form of this term. 

Using the same techniques (i.e. forming total derivatives, using the 
explicit form of the various components of $v^a$, the product $m^i\bar m
^j$ expressed by $m^i$, $\bar m^i$ and $p^i$, the explicit form of $f$, the 
integral formula (\ref{eq:3.2.10}), the various matrix elements given in 
Appendix \ref{sub-A.4}, etc), the fourth term in (\ref{eq:4.3.3}) can be 
calculated in a straightforward way. However, this computation is 
considerably longer than the previous ones and we do not need its explicit 
form. We need to show only that the rate of the divergence of this term is 
smaller than the rate of the divergence of the denominator of 
(\ref{eq:4.3.6}). Next we show that this is the case. 

The last term on the right hand side of (\ref{eq:4.3.3}) can be written into 
the form 
\begin{equation}
-\frac{1}{\mu^4_1}(\Lambda^{-1}_1\Lambda_2)^h{}_{h_2}(\Lambda^{-1}_1\Lambda_2)
^g{}_{g_2}\varepsilon_{hbde}\varepsilon_{gac}{}^e\langle({\bf p}^b_1{\bf p}^{f_1}
_1{\bf J}^d_1{}_{f_1})\psi_1\vert({\bf p}^a_1{\bf p}^{f_2}_1{\bf J}^c_1{}_{f_2})
\psi_1\rangle\langle\psi_2\vert{\bf p}^{h_2}_2{\bf p}^{g_2}_2\vert\psi_2
\rangle, \label{eq:4.3.8a}
\end{equation}
in which the second expectation value is of the form $\mu^2_2(\delta^{h_2}_0
\delta^{g_2}_0+O(1/\sqrt{s}))$ by (\ref{eq:4.3.4}). Since also by 
(\ref{eq:4.3.4}) the order of the denominator in (\ref{eq:4.2.1}) is of 
$\mu^2_1\mu^2_2$, we should show only that the first expectation value in 
this expression tends to infinity more slowly than $\mu^6_1$, and hence 
this does not give any contribution to the numerator of (\ref{eq:4.2.1}) in 
the classical limit. 

Since by (\ref{eq:3.2.8}) 
\begin{equation*}
\frac{{\rm d}f}{{\rm d}p}=\frac{1}{2}p\bigl(\frac{1}{(p^0)^2}-\frac{1}
{\epsilon^2\mu^2}\bigr)f,
\end{equation*}
(\ref{eq:A.1.11a}), (\ref{eq:A.1.11b}) and (\ref{eq:A.3.4}) yield 
immediately that 
\begin{eqnarray}
\varepsilon_{aecd}({\bf p}^c{\bf p}^f{\bf J}^d{}_f\psi^+_{s,m})_r={\rm i}\hbar
  \mu^2\varepsilon_{aecd}\,p^c\!\!\!\!\!&{}\!\!\!\!\!&\Bigl(v^d\frac{p^0}{\mu}
  \frac{{\rm d}f}{{\rm d}p}{}_{-s}Y_{s,m}\delta_{r,2s}-\frac{1}{\sqrt{2}}
  \frac{p^0}{p\mu}m^df{}_{-s}Y_{s,m}\delta_{r,2s-1}+ \nonumber \\
+\!\!\!\!\!&{}\!\!\!\!\!&\bar m^df({\edth}{}_{-s}Y_{s,m})\delta_{r,2s}\Bigr)= 
  \nonumber \\
={\rm i}\hbar\mu^2\varepsilon_{aecd}\frac{p^c}{p}\!\!\!\!\!&{}\!\!\!\!\!&
  \Bigl(-\frac{1}{\sqrt{2}}\frac{p^0}{\mu}m^d{}_{-s}Y_{s,m}\delta_{r,2s-1}-
  \sqrt{s}\bar m^d{}_{-s+1}Y_{s,m}\delta_{r,2s}+ \nonumber \\
+\!\!\!\!\!&{}\!\!\!\!\!&\frac{1}{2}\frac{(p)^2}{p^0\mu}v^d{}_{-s}Y_{s,m}
  \delta_{r,2s}-\frac{1}{\epsilon^2}\frac{1}{2}\frac{(p)^2p^0}{\mu^3}v^d
  {}_{-s}Y_{s,m}\delta_{r,2s}\Bigr)f, \label{eq:4.3.8b}
\end{eqnarray}
and, in a similar way, that 
\begin{eqnarray}
\varepsilon_{aecd}({\bf p}^c{\bf p}^f{\bf J}^d{}_f\psi^-_{s,m})_r=(-)^{2s}
  {\rm i}\hbar\mu^2\!\!\!\!\!&{}\!\!\!\!\!&\varepsilon_{aecd}\frac{p^c}{p}
  \Bigl(\sqrt{s}m^d{}_{s-1}Y_{s,m}\delta_{r,0}+\frac{1}{2}\frac{(p)^2}{p^0\mu}
  v^d{}_sY_{s,m}\delta_{r,0}- \nonumber \\
-\!\!\!\!\!&{}\!\!\!\!\!&\frac{1}{\epsilon^2}\frac{1}{2}\frac{(p)^2p^0}
  {\mu^3}v^d{}_sY_{s,m}\delta_{r,0}+\frac{1}{\sqrt{2}}\frac{p^0}{\mu}\bar m^d
  {}_sY_{s,m}\delta_{r,1}\Bigr)f. \label{eq:4.3.8c}
\end{eqnarray}
Using $\varepsilon_{aecd}p^cm^d=-{\rm i}\mu(v_am_e-v_em_a)$ and $\varepsilon
_{aecd}p^cv^d=-{\rm i}\mu(m_a\bar m_e-m_e\bar m_a)$ (see Appendix 
\ref{sub-A.1}) and equations (\ref{eq:A.1.6b}) and (\ref{eq:A.2.1}), 
(\ref{eq:4.3.8b}) yields that 
\begin{eqnarray}
\langle\varepsilon^{he}{}_{gb}({\bf p}^g{\bf p}^{f_1}{\bf J}^b{}_{f_1})
  \!\!\!\!\!&{}\!\!\!\!\!&\psi^+_{s,m}\vert\varepsilon^d{}_{eac}({\bf p}^a
  {\bf p}^{f_2}{\bf J}^c{}_{f_2})\psi^+_{s,m}\rangle= \nonumber \\
=\hbar^4\mu^4\int_{{\cal M}^+_\mu}\Bigl\{\!\!\!\!\!&{}\!\!\!\!\!&-s\bigl(v^hv^d+
  \bar m^hm^d\bigr)\vert{}_{-s}Y_{s,m}\vert^2\bigl(\frac{p^0}{p}\bigr)^2-
  s\bigl(v^hv^d+m^h\bar m^d\bigr)\vert{}_{-s+1}Y_{s,m}\vert^2\bigl(\frac{\mu}{p}
  \bigr)^2- \nonumber \\
\!\!\!\!\!&{}\!\!\!\!\!&-\frac{1}{4}\bigl(m^h\bar m^d+\bar m^hm^d\bigr)\vert
  {}_{-s}Y_{s,m}\vert^2\Bigl((\frac{p}{p^0})^2-\frac{2}{\epsilon^2}(\frac{p}
  {\mu})^2+\frac{1}{\epsilon^4}(\frac{pp^0}{\mu^2})^2\Bigr)- \nonumber \\
\!\!\!\!\!&{}\!\!\!\!\!&-\frac{\sqrt{s}}{2}v^hm^d{}_{-s}Y_{s,m}\,
  \overline{{}_{-s+1}Y_{s,m}}\bigl(\frac{\mu}{p^0}-\frac{1}{\epsilon^2}
  \frac{p^0}{\mu}\bigr)- \nonumber \\
\!\!\!\!\!&{}\!\!\!\!\!&-\frac{\sqrt{s}}{2}\bar m^hv^d\overline{{}_{-s}Y_{s,m}}
  \,{}_{-s+1}Y_{s,m}\bigl(\frac{\mu}{p^0}-\frac{1}{\epsilon^2}\frac{p^0}{\mu}
  \bigr)\Bigr\}{\rm d}{\cal S}_1\frac{(p)^2}{p^0}f^2{\rm d}p. 
  \label{eq:4.3.8d}
\end{eqnarray}
Substituting the explicit form of $v^a$ here (see Appendix \ref{sub-A.1}), 
the integrals of the resulting expression on the unit sphere ${\cal S}_1$ 
can be evaluated using $m^i={\edth}p^i$, equations (\ref{eq:A.1.11a}) and 
(\ref{eq:A.1.11b}) and the matrix elements given in Appendix \ref{sub-A.4}. 
The evaluation of the integrals with respect to $p$ can be based on the 
integral 
\begin{equation}
\int^\infty_0(p)^n(p^0)^mf^2{\rm d}p=\frac{2}{\sqrt{\pi}}2^{\frac{n}{2}}(
\epsilon\mu)^{n-2}\mu^{m+1}\int^\infty_0x^n(1+2\epsilon^2x^2)^{\frac{m+1}{2}}
e^{-x^2}{\rm d}x, \hskip 20pt n,m\in\mathbb{Z} \label{eq:4.3.8e}
\end{equation}
where the integral on the right is a finite, bounded function of $\epsilon$. 
Thus, its behaviour in the classical limit depends only on the coefficient 
$(\epsilon\mu)^{n-2}\mu^{m+1}$. Using these, for the order of the leading term 
of the $00$-component of (\ref{eq:4.3.8d}) is $\mu^4s$, the order of the 
leading term of the $0i$-components is $\mu^4s\sqrt[4]{s}$, and that of the 
$ij$-components is $\mu^4s\sqrt{s}$. Using (\ref{eq:4.3.8c}), we obtain 
similar orders. Finally, by (\ref{eq:A.1.6b}) the `cross-term' $\langle
\varepsilon^{he}{}_{gb}({\bf p}^g{\bf p}^{f_1}{\bf J}^b{}_{f_1})\psi^+_{s,m}\vert
\varepsilon^d{}_{eac}({\bf p}^a{\bf p}^{f_2}{\bf J}^c{}_{f_2})\psi^-_{s,m}
\rangle$ is zero for $s>1$. Therefore, since $\psi_{s,m}=(\psi^+_{s,m}+\psi^-
_{s,m})/\sqrt{2}$, we obtain that the leading term in $\langle\varepsilon^{he}
{}_{gb}({\bf p}^g{\bf p}^{f_1}{\bf J}^b{}_{f_1})\psi_1\vert\varepsilon^d{}_{eac}
({\bf p}^a{\bf p}^{f_2}{\bf J}^c{}_{f_2})\psi_1\rangle$ is of order $\mu^4s
\sqrt{s}$, and hence, for $s>1$, the fourth term on the right hand side of 
(\ref{eq:4.3.3}) has the structure 
\begin{equation}
\mu^2_2\hbar^2\bigl(C's_1\sqrt{s_1}+O(s_1)+O(\frac{1}{\sqrt{s_2}})\bigr), 
\label{eq:4.3.8}
\end{equation}
where $C'$ is a constant. 

Since the first and fourth terms in (\ref{eq:4.3.1}) have the same 
structure after interchanging the indices $1$ and $2$, the asymptotic forms 
(\ref{eq:4.3.5}), (\ref{eq:4.3.7}) and (\ref{eq:4.3.8}) immediately yield 
the asymptotic form of the expectation value of the first and the last terms 
in (\ref{eq:4.3.1}): 
\begin{eqnarray}
\frac{1}{\mu^4_1}\!\!\!\!\!&{}\!\!\!\!\!&\varepsilon_{acde}\varepsilon^e
  {}_{ghb}\langle\phi_1\vert{\bf J}^a_{1a_1}{\bf p}^{a_1}_1{\bf p}^c_1{\bf p}
  ^g_1{\bf p}^{b_1}_1{\bf J}^b_{1b_1}\vert\phi_1\rangle\langle\phi_2\vert
  {\bf p}^d_2{\bf p}^h_2\vert\phi_2\rangle+ \nonumber \\
+\frac{1}{\mu^4_2}\!\!\!\!\!&{}\!\!\!\!\!&\varepsilon_{acde}\varepsilon^e
  {}_{ghb}\langle\phi_1\vert{\bf p}^c_1{\bf p}^g_1\vert\phi_1\rangle\langle
  \phi_2\vert{\bf J}^a_{2a_2}{\bf p}^{a_2}_2{\bf p}^d_2{\bf p}^h_2{\bf p}^{b_2}
  _2{\bf J}^b_{2b_2}\vert\phi_2\rangle= \nonumber \\
=\!\!\!\!\!&{}\!\!\!\!\!&\mu^2_1\mu^2_2\Bigl(-\xi^a_1\varepsilon_{acde}
  \Lambda^c_1{}_0\Lambda^d_2{}_0\,\xi^b_1\varepsilon_{bgh}{}^e\Lambda^g_1
  {}_0\Lambda^h_2{}_0+O(\frac{1}{\sqrt{s_1}})+O(\frac{1}{\sqrt{s_2}})\Bigr)
  +\nonumber \\
+\!\!\!\!\!&{}\!\!\!\!\!&\mu^2_1\mu^2_2\Bigl(-\xi^a_2\varepsilon_{acde}
  \Lambda^c_1{}_0\Lambda^d_2{}_0\,\xi^b_2\varepsilon_{bgh}{}^e\Lambda^g_1
  {}_0\Lambda^h_2{}_0+O(\frac{1}{\sqrt{s_1}})+O(\frac{1}{\sqrt{s_2}})\Bigr)
  +\nonumber \\
+\!\!\!\!\!&{}\!\!\!\!\!&\mu_1\mu^2_2\hbar\Bigl(C_1+O(\frac{1}{\sqrt{s_1}})
  +O(\frac{1}{\sqrt{s_2}})\Bigr)+\mu_2\mu^2_1\hbar\Bigl(C_2+O(\frac{1}
  {\sqrt{s_2}})+O(\frac{1}{\sqrt{s_1}})\Bigr)+
  \nonumber \\
+\!\!\!\!\!&{}\!\!\!\!\!&\mu^2_2\hbar^2\Bigl(C_3s_1\sqrt{s_1}+O(s_1)+
  O(\frac{1}{\sqrt{s_2}})\Bigr)+\mu^2_1\hbar^2\Bigl(C_4s_2\sqrt{s_2}+O(s_2)
  +O(\frac{1}{\sqrt{s_1}})\Bigr), \label{eq:4.3.9}
\end{eqnarray}
where $C_1$, $C_2$, $C_3$ and $C_4$ are constants. 

Using (\ref{eq:A.2.3a}) and (\ref{eq:A.2.3b}), the expectation value of 
the second term in (\ref{eq:4.3.1}) can be written as 
\begin{eqnarray}
-\frac{1}{\mu^2_1\mu^2_2}\!\!\!\!\!&{}\!\!\!\!\!&\varepsilon_{acde}\varepsilon
  ^e{}_{ghb}\langle\phi_1\vert{\bf J}^a_1{}_{a_1}{\bf p}^{a_1}_1{\bf p}^c_1
  {\bf p}^g_1\vert\phi_1\rangle\langle\phi_2\vert{\bf p}^d_2{\bf p}^h_2
  {\bf p}^{b_2}_2{\bf J}^b_2{}_{b_2}\vert\phi_2\rangle= \label{eq:4.3.10} \\
=\!\!\!\!\!&{}\!\!\!\!\!&\xi^a_1\varepsilon_{acde}\xi^b_2\varepsilon_{bgh}{}^e
  \Lambda^c_{1c_1}\Lambda^g_{1g_1}\langle\psi_1\vert{\bf p}^{c_1}_1{\bf p}^{g_1}_1
  \vert\psi_1\rangle\Lambda^d_{2d_2}\Lambda^h_{2h_2}\langle\psi_2\vert{\bf p}
  ^{d_2}_2{\bf p}^{h_2}_2\vert\psi_2\rangle+ \nonumber \\
+\!\!\!\!\!&{}\!\!\!\!\!&\frac{1}{\mu^2_2}\xi^a_1\varepsilon_{acde}\Lambda^c
  _{1c_1}\Lambda^g_{1g_1}\langle\psi_1\vert{\bf p}^{c_1}_1{\bf p}^{g_1}_1\vert
  \psi_1\rangle\varepsilon_{bgh}{}^e\Lambda^d_{2d_2}\Lambda^h_{2h_2}\Lambda^b
  _{2b_2}\langle\psi_2\vert{\bf p}^{d_2}_2{\bf p}^{h_2}_2{\bf p}^f_2{\bf J}^{b_2}
  _{2f}\vert\psi_2\rangle+ \nonumber \\
+\!\!\!\!\!&{}\!\!\!\!\!&\frac{1}{\mu^2_1}\varepsilon_{acde}\xi^b_2
  \varepsilon_{bgh}{}^e\Lambda^a_{1a_1}\Lambda^c_{1c_1}\Lambda^g_{1g_1}\langle
  \psi_1\vert{\bf J}^{a_1}_{1f}{\bf p}^f_1{\bf p}^{c_1}_1{\bf p}^{g_1}_1\vert
  \psi_1\rangle\Lambda^d_{2d_2}\Lambda^h_{2h_2}\langle\psi_2\vert{\bf p}^{d_2}
  _2{\bf p}^{h_2}_2\vert\psi_2\rangle+ \nonumber \\
+\!\!\!\!\!&{}\!\!\!\!\!&\frac{1}{\mu^2_1\mu^2_2}\varepsilon_{acde}
  \varepsilon_{bgh}{}^e\Lambda^a_{1a_1}\Lambda^c_{1c_1}\Lambda^g_{1g_1}\langle
  \psi_1\vert{\bf J}^{a_1}_{1f_1}{\bf p}^{f_1}_1{\bf p}^{c_1}_1{\bf p}^{g_1}_1
  \vert\psi_1\rangle\Lambda^d_{2d_2}\Lambda^h_{2h_2}\Lambda^b_{2b_2}\langle
  \psi_2\vert{\bf p}^{d_2}_2{\bf p}^{h_2}_2{\bf p}^{f_2}_2{\bf J}^{b_2}_{2f_2}
  \vert\psi_2\rangle. \nonumber
\end{eqnarray}
Using (\ref{eq:4.3.4}) and (\ref{eq:4.3.7a}), we obtain the asymptotic 
form of the second term in (\ref{eq:4.3.3}) immediately. But since the 
third term in (\ref{eq:4.3.3}) has the same structure; and also the second 
and third terms in (\ref{eq:4.3.1}) have the same structure, finally we have 
that the asymptotic form of these two terms in (\ref{eq:4.3.1}) is 
\begin{eqnarray}
&{}&\mu^2_1\mu^2_2\Bigl(2\xi^a_1\varepsilon_{acde}\Lambda^c_1{}_0\Lambda^d_2
  {}_0\,\xi^b_2\varepsilon_{bgh}{}^e\Lambda^g_1{}_0\Lambda^h_2{}_0+
  O(\frac{1}{\sqrt{s_1}})+O(\frac{1}{\sqrt{s_2}})\Bigr)+ \nonumber \\
&{}&+\mu^2_1\mu_2\hbar\Bigl(C_5+O(\frac{1}{\sqrt{s_1}})+O(\frac{1}
  {\sqrt{s_2}})\Bigr)+\mu_1\mu^2_2\hbar\Bigl(C_6+O(\frac{1}{\sqrt{s_1}})
  +O(\frac{1}{\sqrt{s_2}})\Bigr)+ \nonumber \\
&{}&+\mu_1\mu_2\hbar^2\Bigl(C_7+O(\frac{1}{\sqrt{s_1}})+O(\frac{1}
  {\sqrt{s_2}})\Bigr), \label{eq:4.3.11}
\end{eqnarray}
where $C_5$, $C_6$ and $C_7$ are constants.   

The sum of (\ref{eq:4.3.9}) and (\ref{eq:4.3.11}) gives the expectation 
value of ${\bf A}$, i.e. the numerator of (\ref{eq:4.2.1}), in the state 
$\phi:=\phi_1\otimes\phi_2$ for large $s_1$ and $s_2$, which is 
\begin{eqnarray}
\langle\phi\vert{\bf A}\vert\phi\rangle=\mu^2_1\mu^2_2
  \Bigl(\!\!\!\!\!&{}\!\!\!\!\!&-(\xi^a_1\varepsilon_{acde}\Lambda^c_1{}_0
  \Lambda^d_2{}_0)(\xi^b_1\varepsilon_{bgh}{}^e\Lambda^g_1{}_0\Lambda^h_2{}_0)
  -(\xi^a_2\varepsilon_{acde}\Lambda^c_1{}_0\Lambda^d_2{}_0)(\xi^b_2
  \varepsilon_{bgh}{}^e\Lambda^g_1{}_0\Lambda^h_2{}_0)+ \nonumber \\
\!\!\!\!\!&{}\!\!\!\!\!&+2(\xi^a_1\varepsilon_{acde}\Lambda^c_1{}_0\Lambda^d
  _2{}_0)(\xi^b_2\varepsilon_{bgh}{}^e\Lambda^g_1{}_0\Lambda^h_2{}_0)+
  O(\frac{1}{\sqrt{s_1}})+O(\frac{1}{\sqrt{s_2}})\Bigr)+ \nonumber \\
+\mu_1\mu^2_2\hbar\Bigl(\!\!\!\!\!&{}\!\!\!\!\!&c_1+O(\frac{1}{\sqrt{s_1}})
  +O(\frac{1}{\sqrt{s_2}})\Bigr)+\mu^2_1\mu_2\hbar\Bigl(c_2+O(\frac{1}
  {\sqrt{s_1}})+O(\frac{1}{\sqrt{s_2}})\Bigr)+ \nonumber \\
+\mu_1\mu_2\hbar^2\Bigl(\!\!\!\!\!&{}\!\!\!\!\!&c_3s_1\sqrt{s_1}+O(s_1)+
  c_4s_2\sqrt{s_2}+O(s_2)\Bigr), \label{eq:4.3.12}
\end{eqnarray}
where $c_1$, $c_2$, $c_3$ and $c_4$ are constants. Comparing this with the 
numerator of (\ref{eq:2.2.5}) and recalling the form (\ref{eq:4.3.6}) of 
the expectation value of ${\bf B}$ in the same state, we conclude that, 
in the $s_1=s_2\to\infty$ limit, the empirical distance $d_{12}$ tends to 
the classical Lorentzian distance $D_{12}$; and, apart from an additive 
constant, the index $k$ in the Theorem can be chosen to be $2s_1=2s_2$. 


\subsubsection{The uncertainty}
\label{sub-4.3.2}

As we already stressed at the end of subsection \ref{sub-4.1} and at the 
beginning of subsection \ref{sub-4.2}, the square of the empirical distance 
(\ref{eq:4.2.1}) is the quotient of two expectation values, $\langle\phi_1
\otimes\phi_2\vert{\bf A}\vert\phi_1\otimes\phi_2\rangle$ and $\langle\phi_1
\otimes\phi_2\vert{\bf B}\vert\phi_1\otimes\phi_2\rangle$, rather than the 
expectation value of a single operator, say of $\eta_{ab}{\bf d}^a_{12}{\bf d}
^b_{12}$ given by (\ref{eq:4.1.6}). Thus, its standard deviation is not 
defined. However, borrowing the idea from experimental physics how the error 
of a quantity, built from experimental data, is defined, in \cite{Sz22a} we 
introduced the notion of \emph{uncertainty} of the $E(3)$-invariant empirical 
distance in the state $\phi_1\otimes\phi_2$. Namely, if $Q=Q(q^1,...,q^n)$ 
is a differentiable function and in a series of experiments we obtain the 
mean values $\bar q^\alpha$ and errors $\delta q^\alpha$ of the quantities 
$q^\alpha$, $\alpha=1,...,n$, then the mean value and error of $Q$ are 
defined to be $\bar Q:=Q(\bar q^1,...,\bar q^n)$ and 
\begin{equation*}
\delta Q:=\vert\frac{\partial Q}{\partial q^1}(\bar q^\alpha)\vert\delta q
^1+\cdots+\vert\frac{\partial Q}{\partial q^n}(\bar q^\alpha)\vert\delta q^n,
\end{equation*}
respectively. 

Motivated by this formula, we define the \emph{uncertainty} of the square of 
the empirical distance (\ref{eq:4.2.1}) in the state $\phi:=\phi_1\otimes
\phi_2$ by 
\begin{equation}
\delta_\phi{d^2_{12}}:=\frac{\Delta_\phi{\bf A}}{\langle\phi\vert{\bf B}
\vert\phi\rangle}+\frac{\langle\phi\vert{\bf A}\vert\phi\rangle}{(\langle
\phi\vert{\bf B}\vert\phi\rangle)^2}\Delta_\phi{\bf B}
=\Bigl(\frac{\Delta_\phi{\bf A}}{\langle\phi\vert{\bf A}\vert\phi\rangle}+
\frac{\Delta_\phi{\bf B}}{\langle\phi\vert{\bf B}\vert\phi\rangle}\Bigr)
d^2_{12}, \label{eq:4.3.13}
\end{equation}
where $\Delta_\phi{\bf A}$ and $\Delta_\phi{\bf B}$ are the standard 
deviations in the state $\phi$. We show that both terms between the 
brackets tend to zero in the classical limit. Since in this limit $d^2
_{12}$ is bounded, this implies that the uncertainty $\delta_\phi{d^2_{12}}$ 
tends to zero. 

It is easy to show that the second term between the brackets is zero in 
this limit. In fact, the variance of ${\bf B}$ in the state $\phi:=\phi_1
\otimes\phi_2$ is 
\begin{eqnarray*}
\bigl(\Delta_\phi{\bf B}\bigr)^2\!\!\!\!&=\!\!\!\!&\langle\phi_1\otimes
  \phi_2\vert{\bf B}^2\vert\phi_1\otimes\phi_2\rangle-\bigl(\langle\phi_1
  \otimes\phi_2\vert{\bf B}\vert\phi_1\otimes\phi_2\rangle\bigr)^2= \\
\!\!\!\!&=\!\!\!\!&\langle\phi_1\otimes\phi_2\vert{\bf P}^8_{12}\vert\phi_1
  \otimes\phi_2\rangle-\mu^4_1\mu^4_2\Bigl(\bigl((\Lambda^{-1}_1\Lambda_2)
  _{00}\bigr)^4+O(\frac{1}{\sqrt{s_1}})+O(\frac{1}{\sqrt{s_2}})\Bigr),
\end{eqnarray*}
where, in the last step, we used (\ref{eq:4.3.6}). Since by 
(\ref{eq:A.4.6}) we have that 
\begin{eqnarray*}
\langle\phi_1\otimes\phi_2\!\!\!\!\!&{}\!\!\!\!\!&\vert{\bf P}^8_{12}\vert
  \phi_1\otimes\phi_2\rangle= \\
=\!\!\!\!\!&{}\!\!\!\!\!&(\Lambda^{-1}_1\Lambda_2)_{ab}(\Lambda^{-1}_1
  \Lambda_2)_{cd}(\Lambda^{-1}_1\Lambda_2)_{ef}(\Lambda^{-1}_1\Lambda_2)_{gh}
  \langle\psi_1\vert{\bf p}^a_1{\bf p}^c_1{\bf p}^e_1{\bf p}^g_1\vert\psi_1
  \rangle\langle\psi_2\vert{\bf p}^b_2{\bf p}^d_2{\bf p}^f_2{\bf p}^h_2\vert
  \psi_2\rangle= \\
=\!\!\!\!\!&{}\!\!\!\!\!&\mu^4_1\mu^4_2\Bigl(\bigl((\Lambda^{-1}_1\Lambda_2)
  _{00}\bigr)^4+O(\frac{1}{\sqrt{s_1}})+O(\frac{1}{\sqrt{s_2}})\Bigr),
\end{eqnarray*}
we conclude that the second term on the right in (\ref{eq:4.3.13}) tends to 
zero if $s_1,s_2\to\infty$. 

To show that the first term also tends to zero in this limit, we should 
show that, in the leading order, the expectation value $\langle\phi\vert
{\bf A}^2\vert\phi\rangle$ is just $(\langle\phi\vert{\bf A}\vert\phi
\rangle)^2$ and all the remaining terms grow slower than $\mu^4_1\mu^4_2$. 
By (\ref{eq:4.3.1}) the expectation value $\langle\phi\vert{\bf A}^2\vert
\phi\rangle$ can be written as 
\begin{eqnarray}
\langle\phi\vert{\bf A}^2\vert\phi\rangle\!\!\!\!\!&{}\!\!\!\!\!&=\varepsilon
   _{a_1c_1d_1e}\varepsilon^e{}_{g_1h_1b_1}\varepsilon_{a_2c_2d_2f}\varepsilon^f
   {}_{g_2h_2b_2}\times \nonumber \\
\times\Bigl\{\!\!\!\!\!&{}\!\!\!\!\!&\frac{1}{\mu^8_1}\langle\phi_1\vert
   ({\bf J}^{a_1}_{1a}{\bf p}^a_1{\bf p}^{c_1}_1)({\bf p}^{g_1}_1{\bf p}^b_1
   {\bf J}^{b_1}_{1b})({\bf J}^{a_2}_{1a'}{\bf p}^{a'}_1{\bf p}^{c_2}_1)({\bf p}
   ^{g_2}_1{\bf p}^{b'}_1{\bf J}^{b_2}_{1b'})\vert\phi_1\rangle\langle\phi_2
   \vert{\bf p}^{d_1}_2{\bf p}^{h_1}_2{\bf p}^{d_2}_2{\bf p}^{h_2}_2\vert\phi_2
   \rangle- \nonumber \\
-\!\!\!\!\!&{}\!\!\!\!\!&\frac{1}{\mu^6_1\mu^2_2}\langle\phi_1\vert
   ({\bf J}^{a_1}_{1a}{\bf p}^a_1{\bf p}^{c_1}_1)({\bf p}^{g_1}_1{\bf p}^b_1
   {\bf J}^{b_1}_{1b})({\bf J}^{a_2}_{1a'}{\bf p}^{a'}_1{\bf p}^{c_2}_1){\bf p}
   ^{g_2}_1\vert\phi_1\rangle\langle\phi_2\vert{\bf p}^{d_1}_2{\bf p}^{h_1}_2
   {\bf p}^{d_2}_2({\bf p}^{h_2}_2{\bf p}^{b'}_2{\bf J}^{b_2}_{2b'})\vert\phi_2
   \rangle- \nonumber \\ 
-\!\!\!\!\!&{}\!\!\!\!\!&\frac{1}{\mu^6_1\mu^2_2}\langle\phi_1\vert
   ({\bf J}^{a_1}_{1a}{\bf p}^a_1{\bf p}^{c_1}_1)({\bf p}^{g_1}_1{\bf p}^b_1
   {\bf J}^{b_1}_{1b}){\bf p}^{c_2}_1({\bf p}^{g_2}_1{\bf p}^{b'}_1{\bf J}^{b_2}
   _{1b'})\vert\phi_1\rangle\langle\phi_2\vert{\bf p}^{d_1}_2{\bf p}^{h_1}_2
   ({\bf J}^{a_2}_{2a'}{\bf p}^{a'}_2{\bf p}^{d_2}_2){\bf p}^{h_2}_2\vert\phi_2
   \rangle+ \nonumber \\ 
+\!\!\!\!\!&{}\!\!\!\!\!&\frac{1}{\mu^4_1\mu^4_2}\langle\phi_1\vert
   ({\bf J}^{a_1}_{1a}{\bf p}^a_1{\bf p}^{c_1}_1)({\bf p}^{g_1}_1{\bf p}^b_1
   {\bf J}^{b_1}_{1b}){\bf p}^{c_2}_1{\bf p}^{g_2}_1\vert\phi_1\rangle\langle
   \phi_2\vert{\bf p}^{d_1}_2{\bf p}^{h_1}_2({\bf J}^{a_2}_{2a'}{\bf p}^{a'}_2
   {\bf p}^{d_2}_2)({\bf p}^{h_2}_2{\bf p}^{b'}_2{\bf J}^{b_2}_{2b'})\vert\phi_2
   \rangle- \nonumber
\end{eqnarray}
\begin{eqnarray}
-\!\!\!\!\!&{}\!\!\!\!\!&\frac{1}{\mu^6_1\mu^2_2}\langle\phi_1\vert
   ({\bf J}^{a_1}_{1a}{\bf p}^a_1{\bf p}^{c_1}_1){\bf p}^{g_1}_1({\bf J}^{a_2}
   _{1a'}{\bf p}^{a'}_1{\bf p}^{c_2}_1)({\bf p}^{g_2}_1{\bf p}^b_1{\bf J}^{b_2}
   _{1b})\vert\phi_1\rangle\langle\phi_2\vert{\bf p}^{d_1}_2({\bf p}^{h_1}_2
   {\bf p}^{b'}_2{\bf J}^{b_1}_{2b'}){\bf p}^{d_2}_2{\bf p}^{h_2}_2\vert\phi_2
   \rangle+ \nonumber \\
+\!\!\!\!\!&{}\!\!\!\!\!&\frac{1}{\mu^4_1\mu^4_2}\langle\phi_1\vert({\bf J}
   ^{a_1}_{1a}{\bf p}^a_1{\bf p}^{c_1}_1){\bf p}^{g_1}_1({\bf J}^{a_2}_{1a'}
   {\bf p}^{a'}_1{\bf p}^{c_2}_1){\bf p}^{g_2}_1\vert\phi_1\rangle\langle\phi_2
   \vert{\bf p}^{d_1}_2({\bf p}^{h_1}_2{\bf p}^b_2{\bf J}^{b_1}_{2b}){\bf p}^{d_2}
   _2({\bf p}^{h_2}_2{\bf p}^{b'}_2{\bf J}^{b_2}_{2b'})\vert\phi_2\rangle+ 
   \nonumber \\
+\!\!\!\!\!&{}\!\!\!\!\!&\frac{1}{\mu^4_1\mu^4_2}\langle\phi_1\vert({\bf J}
   ^{a_1}_{1a}{\bf p}^a_1{\bf p}^{c_1}_1){\bf p}^{g_1}_1{\bf p}^{c_2}_1({\bf p}
   ^{g_2}_1{\bf p}^b_1{\bf J}^{b_2}_{1b})\vert\phi_1\rangle\langle\phi_2\vert
   {\bf p}^{d_1}_2({\bf p}^{h_1}_2{\bf p}^{b'}_2{\bf J}^{b_1}_{2b'})({\bf J}^{a_2}
   _{2a'}{\bf p}^{a'}_2{\bf p}^{d_2}_2){\bf p}^{h_2}_2\vert\phi_2\rangle- 
   \nonumber \\
-\!\!\!\!\!&{}\!\!\!\!\!&\frac{1}{\mu^2_1\mu^6_2}\langle\phi_1\vert({\bf J}
   ^{a_1}_{1a}{\bf p}^a_1{\bf p}^{c_1}_1){\bf p}^{g_1}_1{\bf p}^{c_2}_1{\bf p}
   ^{g_2}_1\vert\phi_2\rangle\langle\phi_2\vert{\bf p}^{d_1}_2({\bf p}^{h_1}_2
   {\bf p}^b_2{\bf J}^{b_1}_{2b})({\bf J}^{a_2}_{2a'}{\bf p}^{a'}_2{\bf p}^{d_2}
   _2)({\bf p}^{h_2}_2{\bf p}^{b'}_2{\bf J}^{b_2}_{2b'})\vert\phi_2\rangle- 
   \nonumber
\end{eqnarray}
\begin{eqnarray}
-\!\!\!\!\!&{}\!\!\!\!\!&\frac{1}{\mu^6_1\mu^2_2}\langle\phi_1\vert
   {\bf p}^{c_1}_1({\bf p}^{g_1}_1{\bf p}^b_1{\bf J}^{b_1}_{1b})({\bf J}^{a_2}
   _{1a}{\bf p}^a_1{\bf p}^{c_2}_1)({\bf p}^{g_2}_1{\bf p}^{b'}_1{\bf J}^{b_2}
   _{1b'})\vert\phi_1\rangle\langle\phi_2\vert({\bf J}^{a_1}_{2a'}{\bf p}^{a'}
   _2{\bf p}^{d_1}_2){\bf p}^{h_1}_2{\bf p}^{d_2}_2{\bf p}^{h_2}_2\vert\phi_2
   \rangle + \nonumber\\
+\!\!\!\!\!&{}\!\!\!\!\!&\frac{1}{\mu^4_1\mu^4_2}\langle\phi_1\vert
   {\bf p}^{c_1}_1({\bf p}^{g_1}_1{\bf p}^b_1{\bf J}^{b_1}_{1b})({\bf J}^{a_2}
   _{1a}{\bf p}^a_1{\bf p}^{c_2}_1){\bf p}^{g_2}_1\vert\phi_1\rangle\langle
   \phi_2\vert({\bf J}^{a_1}_{2a'}{\bf p}^{a'}_2{\bf p}^{d_1}_2){\bf p}^{h_1}_2
   {\bf p}^{d_2}_2({\bf p}^{h_2}_2{\bf p}^{b'}_2{\bf J}^{b_2}_{2b'})\vert\phi_2
   \rangle+ \nonumber \\
+\!\!\!\!\!&{}\!\!\!\!\!&\frac{1}{\mu^4_1\mu^4_2}\langle\phi_1\vert
   {\bf p}^{c_1}_1({\bf p}^{g_1}_1{\bf p}^b_1{\bf J}^{b_1}_{1b}){\bf p}^{c_2}_1
   ({\bf p}^{g_2}_1{\bf p}^{b'}_1{\bf J}^{b_2}_{1b'})\vert\phi_1\rangle\langle
   \phi_2\vert({\bf J}^{a_1}_{2a}{\bf p}^a_2{\bf p}^{d_1}_2){\bf p}^{h_1}_2
   ({\bf J}^{a_2}_{2a'}{\bf p}^{a'}_2{\bf p}^{d_2}_2){\bf p}^{h_2}_2\vert\phi
   _2\rangle- \nonumber \\
-\!\!\!\!\!&{}\!\!\!\!\!&\frac{1}{\mu^2_1\mu^6_2}\langle\phi_1\vert
   {\bf p}^{c_1}_1({\bf p}^{g_1}_1{\bf p}^b_1{\bf J}^{b_1}_{1b}){\bf p}^{c_2}_1
   {\bf p}^{g_2}_1\vert\phi_1\rangle\langle\phi_2\vert({\bf J}^{a_1}_{2a}
   {\bf p}^a_2{\bf p}^{d_1}_2){\bf p}^{h_1}_2({\bf J}^{a_2}_{2a'}{\bf p}^{a'}_2
   {\bf p}^{d_2}_2)({\bf p}^{h_2}_2{\bf p}^{b'}{\bf J}^{b_2}_{2b'})\vert\phi_2
   \rangle+ \nonumber
\end{eqnarray}
\begin{eqnarray}
+\!\!\!\!\!&{}\!\!\!\!\!&\frac{1}{\mu^4_1\mu^4_2}\langle\phi_1\vert
   {\bf p}^{c_1}_1{\bf p}^{g_1}_1({\bf J}^{a_2}_{1a}{\bf p}^a_1{\bf p}^{c_2}_1)
   ({\bf p}^{g_2}_1{\bf p}^b_1{\bf J}^{b_2}_{1b})\vert\phi_1\rangle\langle
   \phi_2\vert({\bf J}^{a_1}_{1a'}{\bf p}^{a'}_2{\bf p}^{d_1}_2)({\bf p}^{h_1}
   _2{\bf p}^{b'}_2{\bf J}^{b_1}_{2b'}){\bf p}^{d_2}_2{\bf p}^{h_2}_2\vert
   \phi_2\rangle- \nonumber \\
-\!\!\!\!\!&{}\!\!\!\!\!&\frac{1}{\mu^2_1\mu^6_2}\langle\phi_1\vert
   {\bf p}^{c_1}_1{\bf p}^{g_1}_1({\bf J}^{a_2}_{1a}{\bf p}^a_1{\bf p}^{c_2}_1)
   {\bf p}^{g_2}_1\vert\phi_1\rangle\langle\phi_2\vert({\bf J}^{a_1}_{1a'}
   {\bf p}^{a'}_2{\bf p}^{d_1}_2)({\bf p}^{h_1}_2{\bf p}^b_2{\bf J}^{b_1}_{2b})
   {\bf p}^{d_2}_2({\bf p}^{h_2}_2{\bf p}^{b'}{\bf J}^{b_2}_{2b'})\vert\phi_2
   \rangle- \nonumber \\
-\!\!\!\!\!&{}\!\!\!\!\!&\frac{1}{\mu^2_1\mu^6_2}\langle\phi_1\vert
   {\bf p}^{c_1}_1{\bf p}^{g_1}_1{\bf p}^{c_2}_1({\bf p}^{g_2}_1{\bf p}^b_1
   {\bf J}^{b_2}_{1b})\vert\phi_1\rangle\langle\phi_2\vert({\bf J}^{a_1}_{2a}
   {\bf p}^a_2{\bf p}^{d_1}_2)({\bf p}^{h_1}_2{\bf p}^{b'}_2{\bf J}^{b_1}_{2b'})
   ({\bf J}^{a_2}_{2a'}{\bf p}^{a'}_2{\bf p}^{d_2}_2){\bf p}^{h_2}_2\vert\phi_2
   \rangle+ \nonumber \\
+\!\!\!\!\!&{}\!\!\!\!\!&\frac{1}{\mu^8_2}\langle\phi_1\vert{\bf p}^{c_1}
   _1{\bf p}^{g_1}_1{\bf p}^{c_2}_1{\bf p}^{g_2}_1\vert\phi_1\rangle\langle
   \phi_2\vert({\bf J}^{a_1}_{2a}{\bf p}^a_2{\bf p}^{d_1}_2)({\bf p}^{h_1}_2
   {\bf p}^b_2{\bf J}^{b_1}_{2b})({\bf J}^{a_2}_{2a'}{\bf p}^{a'}_2{\bf p}^{d_2}
   _2)({\bf p}^{h_2}_2{\bf p}^{b'}_2{\bf J}^{b_2}_{2b'})\vert\phi_2\rangle
   \Bigr\}. \label{eq:4.3.14}
\end{eqnarray}
Thus, these terms are quartic expressions of `units' of the form 
$\varepsilon_{acde}{\bf p}^c{\bf p}^f{\bf J}^d{}_f$ between the round 
brackets. Note that, by the second commutator in (\ref{eq:3.1.2a}), the 
order of the factors in such a `unit' is irrelevant; moreover, $[{\bf p}_b,
\varepsilon_{acde}{\bf p}^c{\bf p}^h{\bf J}^d{}_h]={\rm i}\hbar\mu^2
\varepsilon_{abce}{\bf p}^c$ holds. 

As in subsection \ref{sub-4.3.1}, let the states be of the form $\vert\phi
_{\bi}\rangle=\exp({\rm i}p_e\xi^e_{\bi}/\hbar)\vert\chi_{\bi}\rangle:=\exp
({\rm i}p_e\xi^e_{\bi}/\hbar)$ ${\bf U}_{\bi}\vert\psi_{\bi}\rangle$. Then the 
action of a `unit' on such a state is 
\begin{equation*}
\varepsilon_{acde}{\bf p}^c{\bf p}^f{\bf J}^d{}_f\vert\phi\rangle=
\exp(\frac{\rm i}{\hbar}p_g\xi^g)\Bigl(\mu^2\varepsilon_{acde}{\bf p}^c
\xi^d+\varepsilon_{acde}{\bf p}^c{\bf p}^f{\bf J}^d{}_f\Bigr)\vert\chi
\rangle.
\end{equation*}
Hence, each of the 16 terms on the right hand side of (\ref{eq:4.3.14}) 
will be the sum of 16 terms. One of these 16 terms is a quartic expression 
of the translation(s) $\xi^a$, there are terms that are proportional to 
the first, others to the second and some to the third powers of the 
translation(s); and also there is one term which does not contain any 
$\xi^a$. 

The 16 terms being quartic in the translation(s) can be written as 
\begin{eqnarray}
\!\!\!\!\!&{}\!\!\!\!\!&\Bigl(\xi^{a_1}_1\varepsilon_{a_1c_1d_1e}\xi^{b_1}_1
  \varepsilon_{b_1g_1h_1}{}^e-\xi^{a_1}_1\varepsilon_{a_1c_1d_1e}\xi^{b_1}_2
  \varepsilon_{b_1g_1h_1}{}^e-\xi^{a_1}_2\varepsilon_{a_1c_1d_1e}\xi^{b_1}_1
  \varepsilon_{b_1g_1h_1}{}^e+ \label{eq:4.3.15} \\
\!\!\!\!\!&{}\!\!\!\!\!&+\xi^{a_1}_2\varepsilon_{a_1c_1d_1e}\xi^{b_1}_2
  \varepsilon_{b_1g_1h_1}{}^e\Bigr)\Bigl(\xi^{a_2}_1\varepsilon_{a_2c_2d_2f}\xi
  ^{b_2}_1\varepsilon_{b_2g_2h_2}{}^f-\xi^{a_2}_1\varepsilon_{a_2c_2d_2f}\xi
  ^{b_2}_2\varepsilon_{b_2g_2h_2}{}^f- \nonumber \\
\!\!\!\!\!&{}\!\!\!\!\!&-\xi^{a_2}_2\varepsilon_{a_2c_2d_2f}\xi^{b_2}_1
  \varepsilon_{b_2g_2h_2}{}^f+\xi^{a_2}_2\varepsilon_{a_2c_2d_2f}\xi^{b_2}_2
  \varepsilon_{b_2g_2h_2}{}^f\Bigr)\langle\chi_1\vert{\bf p}^{c_1}_1{\bf p}
  ^{g_1}_1{\bf p}^{c_2}_1{\bf p}^{g_2}_1\vert\chi_1\rangle\langle\chi_2\vert
  {\bf p}^{d_1}_2{\bf p}^{h_1}_2{\bf p}^{d_2}_2{\bf p}^{h_2}_2\vert\chi_2\rangle.
  \nonumber
\end{eqnarray}
Analogously to (\ref{eq:4.3.4}), it is a straightforward calculation to 
show that, for $s>0$, 
\begin{eqnarray*}
\langle\psi_{s,m}\!\!\!\!\!&{}\!\!\!\!\!&\vert{\bf p}^a{\bf p}^b{\bf p}^c
  {\bf p}^d\vert\psi_{s,m}\rangle=\delta^a_0\delta^b_0\delta^c_0\delta^d_0
  \langle\psi_{s,m}\vert({\bf p}^0)^4\vert\psi_{s,m}\rangle+\delta^a_i\delta
  ^b_j\delta^c_k\delta^d_l\langle\psi_{s,m}\vert{\bf p}^i{\bf p}^j{\bf p}^k
  {\bf p}^l\vert\psi_{s,m}\rangle+ \nonumber \\
+\!\!\!\!\!&{}\!\!\!\!\!&\bigl(\delta^a_0\delta^b_0\delta^c_i\delta^d_j+
  \delta^a_0\delta^b_i\delta^c_0\delta^d_j+\delta^a_0\delta^b_i\delta^c_j
  \delta^d_0+\delta^a_i\delta^b_0\delta^c_0\delta^d_j+\delta^a_i\delta^b_0
  \delta^c_j\delta^d_0+\delta^a_i\delta^b_j\delta^c_0\delta^d_0\bigr)\langle
  \psi_{s,m}\vert{\bf p}^0{\bf p}^0{\bf p}^i{\bf p}^j\vert\psi_{s,m}\rangle
  \nonumber \\
=\!\!\!\!\!&{}\!\!\!\!\!&\mu^4\Bigl(\delta^a_0\delta^b_0\delta^c_0\delta^d_0
  +O(\frac{1}{\sqrt{s}})\Bigr). \label{eq:4.3.16}
\end{eqnarray*}
Using this, expression (\ref{eq:4.3.15}) can be written into the form 
\begin{equation*}
\mu^4_1\mu^4_2\Bigl((\xi^a_1-\xi^a_2)\varepsilon_{acde}\Lambda^c_1{}_0
\Lambda^d_2{}_0(\xi^b_1-\xi^b_2)\varepsilon_{bgh}{}^e\Lambda^g_1{}_0\Lambda^h
_2{}_0+O(\frac{1}{\sqrt{s_1}})+O(\frac{1}{\sqrt{s_2}})\Bigr)^2,
\end{equation*}
which is precisely $(\langle\phi\vert{\bf A}\vert\phi\rangle)^2$. We show 
that the rate of growth of all the remaining terms in (\ref{eq:4.3.14}) 
is less than that of $\mu^4_1\mu^4_2$. 

Next, let us consider the terms in (\ref{eq:4.3.14}) that are cubic in the 
translation(s). There is only one `unit' in the expectation values of such 
a term. For example, one of the four such terms coming from the first term 
on the right of (\ref{eq:4.3.14}) is 
\begin{equation}
  \frac{1}{\mu^2_1}\xi^{a_1}_1\varepsilon_{a_1c_1d_1e}\xi^{b_1}_1\varepsilon
  _{b_1g_1h_1}{}^e\xi^{a_2}_1\varepsilon_{a_2c_2d_2f}\varepsilon_{b_2g_2h_2}{}^f
  \langle\chi_1\vert{\bf p}^{c_1}_1{\bf p}^{g_1}_1{\bf p}^{c_2}_1({\bf p}^{g_2}_1
  {\bf p}^b_1{\bf J}^{b_2}_1{}_b)\vert\chi_1\rangle
  \langle\chi_2\vert{\bf p}^{d_1}_2{\bf p}^{h_1}_2{\bf p}^{d_2}_2{\bf p}^{h_2}_2
  \vert\chi_2\rangle. \label{eq:4.3.17}
\end{equation}
By (\ref{eq:4.3.16}) the last factor grows in the classical limit as $\mu
^4_2$, and hence we should show only that the first expectation value grows 
more slowly than $\mu^6_1$. 

Since by (\ref{eq:A.2.3a}) and (\ref{eq:A.2.3b}) this expectation value can 
be written as 
\begin{equation*}
\Lambda^{c_1}{}_{d_1}\Lambda^{g_1}{}_{h_1}\Lambda^{c_2}{}_{d_2}\Lambda^f{}_e(\Lambda
^{-1})^g{}_h\varepsilon^e{}_{gg_2b_2}\langle\psi_{s,m}\vert{\bf p}^{d_1}{\bf p}
^{h_1}{\bf p}^{d_2}({\bf p}^{g_2}{\bf p}^b{\bf J}^{b_2}{}_b)\vert\psi_{s,m}\rangle
\end{equation*}
and $\psi_{s,m}=(\psi^+_{s,m}+\psi^-_{s,m})/\sqrt{2}$, it is enough to determine 
the order of the leading terms in the expectation values $\varepsilon^{ef}
{}_{gb}\langle\psi^\pm_{s,m}\vert{\bf p}^{d_1}{\bf p}^{h_1}{\bf p}^{d_2}({\bf p}^g
{\bf p}^c{\bf J}^b{}_c)\vert\psi^\pm_{s,m}\rangle$. (For large enough $s$ the 
`cross terms' $\varepsilon^{ef}{}_{gb}\langle\psi^-_{s,m}\vert{\bf p}^{d_1}
{\bf p}^{h_1}{\bf p}^{d_2}({\bf p}^g{\bf p}^c{\bf J}^b{}_c)\vert\psi^+_{s,m}
\rangle$ are vanishing.) Substituting (\ref{eq:4.3.8b}) here, we obtain 
\begin{eqnarray*}
\hbar\mu^2\int_{{\cal M}^+_\mu}p^{d_1}p^{h_1}p^{d_2}\Bigl\{\!\!\!\!\!&{}\!\!\!\!\!&
  -\mu\bigl(v^e\bar m^f-v^f\bar m^e\bigr){\edth}\bigl(\vert{}_{-s}Y_{s,m}\vert
  ^2\bigr)+\frac{1}{2}\frac{p}{p^0}\bigl(m^e\bar m^f-m^f\bar m^e\bigr)\vert
  {}_{-s}Y_{s,m}\vert^2- \\
\!\!\!\!\!&{}\!\!\!\!\!&-\frac{1}{2}\frac{1}{\epsilon^2}\frac{pp^0}{\mu^2}
  \bigl(m^e\bar m^f-m^f\bar m^e\bigr)\vert{}_{-s}Y_{s,m}\vert^2\Bigr\}{\rm d}
  {\cal S}_1\frac{(p)^2}{p^0}f^2{\rm d}p.
\end{eqnarray*}
Integrating by parts, using the explicit form of the components of $v^e$, the 
expressions $v^i\bar m^j-v^j\bar m^i=-{\rm i}p^0\eta^{ik}\eta^{jl}\varepsilon
_{klr}\bar m^r/\mu$ (Appendix \ref{sub-A.1}) and $m^i\bar m^j-m^j\bar m^i=
{\rm i}\eta^{ik}\eta^{jl}\varepsilon_{klr}p^r/p$ (Appendix \ref{sub-A.4}), the 
various matrix elements in Appendix \ref{sub-A.4} and equation 
(\ref{eq:4.3.8e}), we can determine the order of its leading term. For $d_1
=h_1=d_2=0$ it is $\mu^5\sqrt[4]{s}$, for $d_1=d_2=0$, $h_1=k$ it is $\mu^5$, 
for $d_1=k$, $d_2=l$, $h_1=0$ it is $\mu^5/\sqrt[4]{s}$, and for $d_1=k$, 
$d_2=l$, $h_1=m$ it is $\mu^5/\sqrt{s}$. 

Repeating the above analysis with $\psi^-_{s,m}$ and using (\ref{eq:4.3.8c}), 
we obtain the same leading orders. Finally, using the commutator $[{\bf p}
_b,\varepsilon_{acde}{\bf p}^c{\bf p}^h{\bf J}^d{}_h]={\rm i}\hbar\mu^2
\varepsilon_{abce}{\bf p}^c$, one can show that the position of the `unit' 
in (\ref{eq:4.3.17}) is irrelevant: the difference of any two such 
configurations give an even slowly growing term. This shows that \emph{any} 
term in (\ref{eq:4.3.14}) that is cubic in the translation(s) grow not 
faster than $\mu^3_1\mu^4_2\sqrt[4]{s_1}$ or $\mu^4_1\mu^3_2\sqrt[4]{s_2}$. 

The proof that the terms being quadratic or linear in or independent of 
the translation(s) in (\ref{eq:4.3.14}) grow more slowly than $\mu^4_1
\mu^4_2$ is similar.


\section{Final remarks}
\label{sec-5}

In the present investigations, the basic notion is a composite quantum 
system that is the union of $E(1,3)$-invariant \emph{elementary quantum 
mechanical systems}. Apart from their own intrinsic properties, fixed by 
their abstract algebra of observables and their representations, \emph{no} 
additional, extra structures, in particular, \emph{no} notion of 3-space 
or spacetime, is used. We found that, in addition to the familiar 
observables representing the usual physical quantities in quantum theory, 
one can construct observables that mimic \emph{geometrical notions}, e.g. 
angles between `directions', or distances between `centre-of-mass 
(world)lines'. Using these, genuine `quantum geometrical structures', 
\emph{defined by the quantum system itself}, can be introduced. The 
significance of these structures is that in the \emph{classical limit} 
these structures reproduce the well known, but \emph{a priori} given 
geometrical structures of the Minkowski space. Thus, the latter, like the 
geometrical structures in the case of the Euclidean 3-space 
\cite{Sz21c,Sz22a}, might be considered to be coming from quantum theory. 
This is the main message of the present paper. 

Nevertheless, different quantum observables may be used to define one kind 
of quantum geometry. For example, in the $E(3)$-invariant case, the linear 
momenta, the angular momenta, and the relative position vectors could be 
used to define some `quantum conformal structure', which are all different, 
though all three can be used to derive the one and the same conformal 
structure of the classical 3-space in the classical limit. In a similar 
way, in $E(1,3)$-invariant systems, the `quantum conformal structure' can 
be defined using the (timelike) 4-momenta, the (spacelike) Pauli--Lubanski 
spin, or the (spacelike) relative position vectors. The resulting `quantum 
conformal structures' are expected to yield the same conformal structure 
of the Minkowski space in the classical limit, though they are \emph{not} 
expected to be equivalent at the genuine quantum level. Thus, at the 
quantum level, there might be no uniquely defined geometric structure of 
a given kind: that may depend on the observables that that is based on. 

In the present (as well as in our previous \cite{Sz21c,Sz22a}) 
investigation(s) the observables representing geometric quantities were 
evaluated in pure tensor product states of the constituent elementary 
subsystems of the large composite system. Thus, the subsystems were assumed 
to be independent. The fact that non-trivial geometric quantities, e.g. 
the relative distance between them, could be derived is due to the 
structure of these geometric observables: all these are joint, or rather 
`entangled' observables. In particular, the empirical distance between the 
$E(1,3)$-invariant elementary system `1' and elementary system `2' is built 
from the quantum operators ${\bf P}^2_{12}:=\eta_{ab}{\bf p}^a_1\otimes{\bf p}
^b_2$ and ${\bf S}^a_{12}:=\frac{1}{2}\varepsilon^a{}_{bcd}({\bf J}^{bc}_1
\otimes{\bf p}^d_2+{\bf p}^d_1\otimes{\bf J}^{bc}_2)$. Since the observables 
are `entangled' in this sense, the states do not need to be entangled to 
get well defined, non-trivial results, e.g. actually the relative distance. 


\section{Acknowledgments}

The author wishes to thank the referees for their useful remarks and 
suggestions for the improvement of the presentation of the results, and 
also for the references \cite{Linn} and \cite{Feintzeig}. 
No funds, grants or support was received.


\appendix

\section{Appendix: The summary of the unitary, irreducible 
representations of $E(1,3)$}
\label{sec-A}

In the first of these appendices, mostly to fix the notations, we review 
the geometry of the mass-shell and the various structures thereon (for the 
general background, see \cite{PR,HT,NP}). Then we summarize the key results 
on the unitary, irreducible representations of the quantum mechanical 
Poincar\'e group $E(1,3)$ in a slightly more geometric form than it is 
usually given in standard textbooks (and which are only sketched e.g. in 
\cite{StWi,Haag}). In Appendix \ref{sub-A.3}, we rewrite the centre-of-mass 
and other relevant operators in the Newman--Penrose (NP) form that we use 
to calculate the empirical distant and its classical limit. Finally, in 
Appendix \ref{sub-A.4}, matrix elements of certain polynomials of the 
linear momentum ${\bf p}^i$ in the basis of spin weighted spherical 
harmonics are presented.

\subsection{The geometry of the mass-shell}
\label{sub-A.1}

The future mass shell ${\cal M}^+_\mu$ with rest mass $\mu>0$, defined in 
subsection \ref{sub-2.1}, is homeomorphic to $\mathbb{R}^3$. The induced 
(negative definite) metric on ${\cal M}^+_\mu$, $h_{ab}=\eta_{ab}-p_ap_b/\mu
^2$, is of constant curvature with curvature scalar ${\cal R}=-6/\mu^2$, 
and its extrinsic curvature is $\chi_{ab}=h_{ab}/\mu$. By definition, the 
Cartesian components $p^a=(p^0,p^i)$, $i=1,2,3$, of the position vectors on 
the mass shell satisfy $(p^0)^2=\mu^2+\delta_{ij}p^ip^j$. ${\cal M}^+_\mu$ can 
be foliated by the 2-spheres ${\cal S}_p:=\{p^a\in{\cal M}^+_\mu\vert\,p:=
\sqrt{\delta_{ij}p^ip^j}={\rm const}\,\}$. Clearly, $p\in[0,\infty)$, and 
for $p=0$ the `sphere' ${\cal S}_0$ is a single point. (To avoid confusion, 
the square of $p$ will be written as $(p)^2$, rather than $p^2$, which has 
already been used to denote the 2nd Cartesian component of $p^a$.) Let 
$(\zeta,\bar\zeta)$ be the standard complex stereographic coordinates on 
these 2-spheres, defined e.g. in terms of the more familiar spherical polar 
coordinates by $\zeta:=\exp({\rm i}\varphi)\cot(\theta/2)$. Then $(p,\zeta,
\bar\zeta)$ is an intrinsic coordinate system on ${\cal M}^+_\mu$; and 
$(\mu,p,\zeta,\bar\zeta)$ is a coordinate system inside the future null 
cone of the origin in $\mathbb{R}^{1,3}$. In these coordinates, the line 
element of the intrinsic metric on ${\cal M}^+_\mu$ is 
\begin{equation}
dh^2=-\frac{\mu^2}{\mu^2+(p)^2}\bigl(dp\bigr)^2-(p)^2\frac{4}{(1+\zeta\bar
\zeta)^2}d\zeta d\bar\zeta. \label{eq:A.1.1}
\end{equation}
Thus $\sqrt{-\det(h_{ab})}=\mu(p)^2/\sqrt{\mu^2+(p)^2}$, and hence the natural 
volume element on ${\cal M}^+_\mu$ divided by $\mu$ is given by 
\begin{equation}
{\rm d}v_\mu=\frac{1}{p^0}{\rm d}p^1{\rm d}p^2{\rm d}p^3=\frac{(p)^2}{\sqrt{\mu
^2+(p)^2}}{\rm d}p\,{\rm d}{\cal S}_1, \label{eq:A.1.2}
\end{equation}
where ${\rm d}{\cal S}_1$ is the area element on the \emph{unit} sphere. 
This ${\rm d}v_\mu$ is Lorentz-invariant, and it is well defined even on the 
$\mu=0$ mass shell, i.e. on the null cone of the origin in $\mathbb{R}^{1,3}$. 
It is this volume element that will be used in the definition of the Hilbert 
space of the square integrable spinor fields on ${\cal M}^+_\mu$. 

The Cartesian components of $p^a$ expressed by these coordinates are 
\begin{equation}
p^a=\Bigl(p^0,p\frac{\bar\zeta+\zeta}{1+\zeta\bar\zeta},{\rm i}p\frac{\bar
\zeta-\zeta}{1+\zeta\bar\zeta},p\frac{\zeta\bar\zeta-1}{1+\zeta\bar\zeta}
\Bigr), \label{eq:A.1.3}
\end{equation}
and $p^a/\mu$ is the future pointing unit timelike normal of ${\cal M}^+_\mu$ 
at the point $p^a\in{\cal M}^+_\mu$. Its spinor form, $p_{AA'}:=p_a\sigma^a
_{AA'}$, multiplied by $\sqrt{2}/\mu$ defines a positive definite Hermitean
pointwise scalar product on the spinor spaces, by means of which the primed 
spinors can be converted to unprimed ones. Here $\sigma^{AA'}_a$ are the four 
standard $SL(2,\mathbb{C})$ Pauli matrices (including $1/\sqrt{2}$), 
according to the conventions of \cite{PR}, and whose indices are 
lowered/raised by $\eta_{ab}$ and the symplectic spinor metrics $\varepsilon
_{AB}$, $\varepsilon_{A'B'}$ and the corresponding inverses. 

The outward directed unit normal of the 2-spheres ${\cal S}_p$ that is 
tangent to ${\cal M}^+_\mu$ is given by $v^a=(p/\mu,p^0p^i/\mu p)$. This $v^a$ 
is completed to be a frame field on ${\cal M}^+_\mu$ by the complex null 
vector field $m^a$ and its complex conjugate $\bar m^a$ such that $m^a$ and 
$\bar m^a$ are tangent to the 2-spheres ${\cal S}_p$ and that they satisfy 
$m^am_a=0$ and $m^a\bar m_a=-1$. They can be chosen to be linked to the 
complex stereographic coordinates so that, in particular, the Cartesian 
components of $m^a$ are 
\begin{equation}
m^a=\frac{1}{\sqrt{2}}\Bigl(0,\frac{1-\zeta^2}{1+\zeta\bar\zeta},{\rm i}
\frac{1+\zeta^2}{1+\zeta\bar\zeta},\frac{2\zeta}{1+\zeta\bar\zeta}\Bigr);
\label{eq:A.1.4a}
\end{equation}
while $v^a$ and $m^a$, as differential operators, are given by 
\begin{equation}
v^a\partial_a=\frac{p^0}{\mu}\frac{\partial}{\partial p},
  \hskip 20pt
m^a\partial_a=\frac{1}{\sqrt{2}p}\bigl(1+\zeta\bar\zeta\bigr)
  \frac{\partial}{\partial\bar\zeta}. \label{eq:A.1.4b}
\end{equation}
Thus, $\eta_{ab}=p_ap_b/\mu^2-v_av_b-m_a\bar m_b-\bar m_am_b$ and $p^av^bm^c
\bar m^d\varepsilon_{abcd}={\rm i}\mu$ hold, where $\varepsilon_{abcd}$ is the 
natural volume 4-form on $\mathbb{R}^{1,3}$. 

We associate a spinor basis $\{o^A,\iota^A\}$, normalized according to $o_A
\iota^A=1$, to the basis $\{p^a,v^a,m^a,\bar m^a\}$ such that the spinor 
form of these basis vectors has the form $p^{AA'}=\mu(o^A\bar o^{A'}+\iota^A
\bar\iota^{A'})/\sqrt{2}$, $v^{AA'}:=v^a\sigma^{AA'}_a=(o^A\bar o^{A'}-\iota^A
\bar\iota^{A'})/\sqrt{2}$ and $m^{AA'}:=m^a\sigma^{AA'}_a=o^A\bar\iota^{A'}$, 
respectively. Explicitly, this spinor dyad is given by 
\begin{equation}
o^A=-\sqrt{\frac{p^0+p}{\mu}}\frac{\rm i}{\sqrt{1+\zeta\bar\zeta}}
  \left(\begin{array}{c}
  \zeta \\
     1 \\
  \end{array}\right), 
\hskip 20pt
\iota^A=-\sqrt{\frac{p^0-p}{\mu}}\frac{\rm i}{\sqrt{1+\zeta\bar\zeta}}
     \left(\begin{array}{c}
     1 \\
     -\bar\zeta \\
     \end{array}\right). \label{eq:A.1.5}
\end{equation}
This basis is related to the Cartesian spinor basis $O^A=\delta^A_0$, $I^A=
\delta^A_1$ by a general, $p,\zeta,\bar\zeta$-dependent $SL(2,\mathbb{C})$ 
transformation. In particular, this is \emph{boosted} with respect to the 
Cartesian spinor basis, and the boost parameter is $\sqrt{(p^0+p)/\mu}=1/
\sqrt{(p^0-p)/\mu}$. Without this the bases $\{o^A,\iota^A\}$ and $\{O^A,
I^A\}$ would be linked only by a $\zeta,\bar\zeta$-dependent $SU(2)$, 
rather than a general $p,\zeta,\bar\zeta$-dependent $SL(2,\mathbb{C})$ 
transformation. The corresponding un-boosted dyad (i.e. without the boost 
parameter) will be denoted by $\{\tilde o{}^A,\tilde\iota{}^A\}$. 

By (\ref{eq:A.1.4b}) and (\ref{eq:A.1.5}) it is easy to check that 
\begin{eqnarray}
&{}&v^e\partial_eo^A=\frac{1}{2\mu}o^A, \hskip 15pt
  m^e\partial_eo^A=\beta o^A, \hskip 15pt
  \bar m^e\partial_eo^A=-\bar\beta o^A+\frac{1}{\sqrt{2}p}\frac{p^0+p}
  {\mu}\iota^A, \hskip 20pt \label{eq:A.1.5.a} \\
&{}&v^e\partial_e\iota^A=-\frac{1}{2\mu}\iota^A, \hskip 15pt
  m^e\partial_e\iota^A=-\beta\iota^A-\frac{1}{\sqrt{2}p}\frac{\mu}
  {p^0+p}o^A, \hskip 15pt
  \bar m^e\partial_e\iota^A=\bar\beta\iota^A, \hskip 20pt \label{eq:A.1.5.b}
\end{eqnarray}
where $\beta:=-(2\sqrt{2}p)^{-1}\zeta$. 

If $\phi_{A_1...A_{2s}}$ is a completely symmetric spinor field, then we form 
$\phi_r:=\phi_{A_1...A_{2s}}\iota^{A_1}\cdots\iota^{A_r}$ $o^{A_{r+1}}\cdots o
^{A_{2s}}$, $r=0,...,2s$ (see \cite{PR,HT}). The spinor field itself and the 
pointwise Hermitean scalar product of two such spinor fields can be given in 
terms of these functions: 
\begin{eqnarray}
&{}&\phi_{A_1...A_{2s}}=\sum^{2s}_{r=0}(-)^r{{2s}\choose{r}}\phi_{2s-r}\iota_{(A_1}
   \cdots\iota_{A_r}o_{A_{r+1}}\cdots o_{A_{2s})}, \label{eq:A.1.6a}\\
&{}&\bigl(\phi_{A_1...A_{2s}},\psi_{A_1...A_{2s}}\bigr):=2^s\frac{p^{A'_1A_1}}{\mu}
  \cdots\frac{p^{A'_{2s}A_{2s}}}{\mu}\bar\phi_{A'_1...A'_{2s}}\psi_{A_1...\psi_{2s}}=
  \sum^{2s}_{r=0}{{2s}\choose{r}}\bar\phi_r\psi_r. \hskip 20pt\label{eq:A.1.6b}
\end{eqnarray}
In particular, $\phi_r=0$ for $r=0,...,2s-1$ is equivalent to the 
proportionality of $\phi_{A_1...A_{2s}}$ to $o_{A_1}\cdots o_{A_{2s}}$, and $\phi_r
=0$ for $r=1,...,2s$ to its proportionality to $\iota_{A_1}\cdots\iota_{A_{2s}}$. 

Recalling that a scalar $\phi$ is said to have spin weight $\sigma$ if, 
under the transformation $\{o^A,\iota^A\}\mapsto\{\exp({\rm i}\alpha)o^A,
\exp(-{\rm i}\alpha)\iota^A\}$ of the spinor basis, it transforms as $\phi
\mapsto\exp({\rm i}\sigma\alpha)\phi$, the scalar $\phi_r$ has spin weight 
$s-r$ (see \cite{PR,HT}). Since the \emph{components} of the spin vectors 
of the NP basis can be considered to be the contractions with the Cartesian 
spinors, e.g. $o^0=-I_Ao^A$ and $o^1=O_Ao^A$, they are spin weighted scalars, 
and the spin weight of $o^A$ is $1/2$, while that of $\iota^A$ is $-1/2$. 
Hence the spin weight of $p^{AA'}$ and of $v^{AA'}$ is zero, while that of 
$m^{AA'}$ and $\bar m^{AA'}$ is $1$ and $-1$, respectively. 

Another particularly important examples for the spin weighted scalars are 
the spin weighted spherical harmonics, which can be defined by 
\begin{equation}
{}_\sigma Y_{j,m}:=N_{\sigma,j,m}O_{(A_1}\cdots O_{A_{j-m}}I_{A_{j-m+1}}\cdots I_{A_{2j})}
\tilde o^{A_1}\cdots\tilde o^{A_{j+\sigma}}\tilde\iota^{A_{j+\sigma+1}}\cdots
\tilde\iota^{A_{2j}}, \label{eq:A.1.7}
\end{equation}
where the factor of normalization is 
\begin{equation}
N_{\sigma,j,m}:=(-)^{j+m}\sqrt{\frac{2j+1}{4\pi}}\frac{(2j)!}{\sqrt{(j-m)!
(j+m)!(j-\sigma)!(j+\sigma)!}}, \label{eq:A.1.8}
\end{equation}
and $2\sigma\in\mathbb{Z}$, $j=\vert\sigma\vert,\vert\sigma\vert+1,\vert
\sigma\vert+2,...$ and $m=-j, -j+1,...,j$ (see \cite{PR}). Note that in the 
definition of ${}_\sigma Y_{j,m}$ above the \emph{un-boosted} spinor basis 
$\{\tilde o{}^A,\tilde\iota{}^A\}$ is used. Since, however, the boost 
parameter is constant on the $p={\rm const}$ 2-spheres, the notion of the 
spin weight defined with respect to $\{o^A,\iota^A\}$ and to $\{\tilde o^A,
\tilde\iota^A\}$ is the same. The advantage of using $\{\tilde o^A,\tilde
\iota^A\}$ rather than $\{o^A,\iota^A\}$ in the definition of ${}_\sigma Y
_{j,m}$ is that the resulting spherical harmonics depend only on $\zeta$ 
and $\bar\zeta$, but do not on the radial coordinate $p$. The spin weighted 
spherical harmonics form an orthonormal basis in the space of the square 
integrable spin weighted scalars on the \emph{unit sphere}. Note also that 
(after correction of a misprint in equation (4.15.104) of \cite{PR}) the 
complex conjugate of the spin weighted spherical harmonics is $\overline{
{}_\sigma Y_{j,m}}=(-)^{m-\sigma}{}_{-\sigma}Y_{j,-m}$. 

The edth and edth-prime operators of Newman and Penrose \cite{NP}, acting 
on spin weighted scalars with spin weight $\sigma$ on 2-spheres of radius 
$p$, can be defined by 
\begin{equation}
{\edth}\phi:=\frac{1}{\sqrt{2}p}\Bigl(\bigl(1+\zeta\bar\zeta\bigr)
\frac{\partial\phi}{\partial\bar\zeta}+\sigma\zeta\phi\Bigr), 
\hskip 20pt
{\edth}'\phi:=\frac{1}{\sqrt{2}p}\Bigl(\bigl(1+\zeta\bar\zeta\bigr)
\frac{\partial\phi}{\partial\zeta}-\sigma\bar\zeta\phi\Bigr). 
\label{eq:A.1.9}
\end{equation}
${\edth}$ increases, while ${\edth}'$ decreases the spin weight by one. These 
definitions yield, in particular, that 
\begin{equation}
{\edth}o^A=0, \hskip 20pt {\edth}'o^A=\frac{1}{\sqrt{2}p}\sqrt{
  \frac{p^0+p}{p^0-p}}\iota^A, \hskip 20pt 
{\edth}\iota^A=-\frac{1}{\sqrt{2}p}\sqrt{\frac{p^0-p}{p^0+p}}o^A, 
  \hskip 20pt {\edth}'\iota^A=0;  \label{eq:A.1.10}
\end{equation}
which imply that ${\edth}p^{AA'}=m^{AA'}$, ${\edth}v^{AA'}=(p^0/\mu p)m^{AA'}$, 
${\edth}m^{AA'}=0$ and ${\edth}'m^{AA'}=p^{AA'}/\mu^2-(p^0/\mu p)v^{AA'}$. Also, 
the action of these operators on the spin weighted spherical harmonics is 
\begin{eqnarray}
&{}&{\edth}\,{}_\sigma Y_{j,m}=-\frac{1}{\sqrt{2}p}\sqrt{(j+\sigma+1)(j-\sigma)}
  \,{}_{\sigma+1}Y_{j,m}, \label{eq:A.1.11a} \\
&{}&{\edth}'\,{}_\sigma Y_{j,m}=\frac{1}{\sqrt{2}p}\sqrt{(j-\sigma+1)(j+\sigma)}
  \,{}_{\sigma-1}Y_{j,m}. \label{eq:A.1.11b}
\end{eqnarray}
These imply that $({\edth}{\edth}'+{\edth}'{\edth})\,{}_\sigma Y_{j,m}=-([j
(j+1)-\sigma^2]/p^2)\,{}_\sigma Y_{j,m}$. A spin weighted scalar $\phi$ is 
said to be \emph{holomorphic} if ${\edth}'\phi=0$, and it is 
\emph{anti-holomorphic} if ${\edth}\phi=0$. 
(\ref{eq:A.1.11a})-(\ref{eq:A.1.11b}) imply that the space of the 
holomorphic scalars of weight $\sigma$ is spanned by the harmonics ${}
_\sigma Y_{\vert\sigma\vert,m}$ with $\sigma\leq0$ and that of the 
anti-holomorphic ones by the harmonics ${}_\sigma Y_{\sigma,m}$ with $\sigma
\geq0$. Both these spaces are $2\vert\sigma\vert+1$ complex dimensional. 


\subsection{The representation}
\label{sub-A.2}

We \emph{a priori} assume that the spectrum of the operators ${\bf p}^a$ 
form the solid future null cone in the classical momentum space $\mathbb{R}
^{1,3}$ (without its vertex), i.e. the set of the non-zero future pointing 
non-spacelike vectors. In the present paper, we consider only the case when 
the spectrum elements $p^a$ are \emph{strictly timelike}, i.e. when the 
corresponding rest mass $\mu$ is strictly positive. 

The representation space is the Hilbert space ${\cal H}_{\mu,s}$ of the square 
integrable \emph{totally symmetric} unprimed spinor fields $\phi^{A_1...A_{2s}}$, 
$2s=0,1,2,...$, on the future mass shell ${\cal M}^+_\mu$ of mass $\mu>0$, 
and the scalar product of $\phi^{A_1...A_{2s}}$ and $\psi^{A_1...A_{2s}}$ in 
${\cal H}_{\mu,s}$ is defined by 
\begin{equation}
\langle\phi^{A_1...A_{2s}}\vert\psi^{A_1...A_{2s}}\rangle:=2^s\int_{{\cal M}^+_\mu}
\frac{p_{A'_1A_1}}{\mu}...\frac{p_{A'_{2s}A_{2s}}}{\mu}\bar\phi^{A'_1...A'_{2s}}\psi
^{A_1...A_{2s}}{\rm d}v_\mu. \label{eq:A.2.1}
\end{equation}
This is the integral of $(\phi_{A_1...A_{2s}},\psi_{A_1...A_{2s}})$ given by 
(\ref{eq:A.1.6b}) on the future mass shell ${\cal M}^+_{\mu}$. This implies 
that $\phi^{A_1...A_{2s}}$ is normalizable precisely when its pointwise norm, 
$\vert\phi^{A_1...A_{2s}}\vert$, defined by the pointwise scalar product 
(\ref{eq:A.1.6b}), falls off as $o(1/p)$, i.e. $\vert\phi^{A_1...A_{2s}}\vert$ 
must fall off \emph{slightly faster} than $1/p$ for large $p$. Since 
(\ref{eq:A.1.6b}) defines a $\mathbb{C}$-linear isomorphism between the 
space of the primed and the unprimed spinors, it is enough to consider only 
e.g. the unprimed ones. 

Clearly, the rotation-boost Killing fields of $\mathbb{R}^{1,3}$ that are 
vanishing at the origin $p^a=0$ are tangent to the mass-shell and generate 
the isometries of ${\cal M}^+_\mu$, too. Then the action of $SL(2,\mathbb{C})$ 
on any $\phi^{A_1...A_{2s}}\in{\cal H}_{\mu,s}$ is defined to be just the action 
of the corresponding isometries, while the action of the translation with 
$\xi^a$ is a multiplication of $\phi^{A_1...A_{2s}}$ by the phase factor $\exp
({\rm i}p_a\xi^a/\hbar)$. Explicitly, for any $A^A{}_B\in SL(2,\mathbb{C})$ 
and $\xi^a\in\mathbb{R}^4$, this action is given by $({\bf U}_{(A,\xi)}\phi)
^{A_1...A_{2s}}(p^e):=\exp({\rm i}p_a\xi^a/\hbar)A^{A_1}{}_{B_1}\cdots A^{A_{2s}}{}
_{B_{2s}}\phi^{B_1...B_{2s}}((\Lambda^{-1})^e{}_bp^b)$. Here, the Lorentz matrix 
in the argument on the right is built from the $SL(2,\mathbb{C})$ matrix 
$A^A{}_B$ according to $\Lambda^a{}_b:=\sigma^a_{AA'}A^A{}_B\bar A^{A'}{}_{B'}
\sigma^{BB'}_b$. These transformations are unitary with respect to the 
scalar product (\ref{eq:A.2.1}). In this representation, the operators 
${\bf p}_a$ and ${\bf J}_{ab}$ are defined via 1-parameter families of 
$E(1,3)$ transformations, $\xi^a(u)=T^au$ and $A^A{}_B(u)=({\rm Exp}
(\lambda u))^A{}_B:=\delta^A_B+\lambda^A{}_Bu+\frac{1}{2}\lambda^A{}_C
\lambda^C{}_Bu^2+...$ for any $T^a\in\mathbb{R}^4$ and $\lambda^A{}_B\in sl
(2,\mathbb{C})$, respectively, and $u\in\mathbb{R}$. Explicitly, $({\rm i}
/\hbar)(T^a{\bf p}_a+M^{ab}{\bf J}_{ab})\phi^{A_1...A_{2s}}:=\frac{\rm d}
{{\rm d}u}({\bf U}_{(A(u),\xi(u))}\phi)^{A_1...A_{2s}}\vert_{u=0}$ with $2M^a{}_b
:=-\sigma^a_{AA'}(\lambda^A{}_B\delta^{A'}_{B'}+\bar\lambda^{A'}{}_{B'}\delta
^A_B)\sigma^{BB'}_b$. The resulting operators are self-adjoint on some 
dense subspaces of ${\cal H}_{\mu,s}$ and are given explicitly by 
\begin{eqnarray}
{\bf p}_a\phi^{A_1...A_{2s}}\!\!\!\!&=\!\!\!\!&p_a\phi^{A_1...A_{2s}},
  \label{eq:A.2.2a} \\
{\bf J}_{ab}\phi^{A_1...A_{2s}}\!\!\!\!&=\!\!\!\!&{\rm i}\hbar\Bigl(p_a
  \frac{\partial}{\partial p^b}-p_b\frac{\partial}{\partial p^a}\Bigr)
  \phi^{A_1...A_{2s}}+ \nonumber \\
\!\!\!\!&{}\!\!\!\!&+2{\rm i}\hbar s\,\sigma^{D'(A_1}_{[a}\sigma_{b]D'(B_1}
  \delta^{A_2}_{B_2}\cdots\delta^{A_{2s})}_{B_{2s})}\phi^{B_1...B_{2s}}.
  \label{eq:A.2.2b}
\end{eqnarray}
Thus, ${\bf p}_a$ acts as a multiplication operator, while ${\bf J}_{ab}$ is 
just ${\rm i}\hbar$-times the Lie derivative of the smooth spinor fields 
along the rotation-boost Killing vector $p_a(\partial/\partial p^b)-p_b
(\partial/\partial p^a)$ on ${\cal M}^+_\mu$. 
${\bf p}^a$ and ${\bf J}^{ab}$ are \emph{not} bounded, and hence they are 
defined only on some appropriate dense subspaces of ${\cal H}_{\mu,s}$: to 
ensure the square integrability of ${\bf p}_a\phi^{A_1...A_{2s}}$, the spinor 
field should fall off slightly faster than $1/(p)^2$ for large $p$, and 
${\bf J}_{ab}\phi^{A_1...A_{2s}}$ to be well defined, the spinor field must also 
be smooth. It is straightforward to check that, on an appropriate common 
domain, these operators satisfy the commutation relations 
(\ref{eq:3.1.2a})-(\ref{eq:3.1.2c}). 

(\ref{eq:A.2.2a}) and (\ref{eq:A.2.2b}) yield that, under the action of 
$E(1,3)$, the operators ${\bf p}^a$ and ${\bf J}^{ab}$ transform according to 
\begin{eqnarray}
&{}&{\bf U}^\dagger_{(A,\xi)}{\bf p}^a{\bf U}_{(A,\xi)}=\Lambda^a{}_b{\bf p}^b, 
  \label{eq:A.2.3a} \\
&{}&{\bf U}^\dagger_{(A,\xi)}{\bf J}^{ab}{\bf U}_{(A,\xi)}=\bigl(\xi^a\delta^b_c
  -\xi^b\delta^a_c\bigr)\Lambda^c{}_d{\bf p}^d+\Lambda^a{}_c\Lambda^b{}_d
  {\bf J}^{cd}. \label{eq:A.2.3b}
\end{eqnarray}
Thus, their transformation properties postulated in subsection \ref{sub-3.1} 
have been implemented successfully by the unitary operators ${\bf U}_{(A,\xi)}$. 

Equations (\ref{eq:A.2.2a}) and (\ref{eq:A.2.2b}) yield 
\begin{eqnarray}
{\bf S}_a\phi^{A_1...A_{2s}}\!\!\!\!&=\!\!\!\!&2s\hbar\,\sigma^{AA'}_ap_{A'}{}^{B}
  \delta^{(A_1}_{(A}\varepsilon_{B)(B_1}\delta^{A_2}_{B_2}\cdots\delta^{A_{2s})}
  _{B_{2s})}\phi^{B_1...B_{2s}}, \label{eq:A.2.4a} \\
{\bf C}_a\phi^{A_1...A_{2s}}\!\!\!\!&=\!\!\!\!&{\rm i}\hbar(p_ap^b-\mu^2\delta
  ^b_a)\frac{\partial\phi^{A_1...A_{2s}}}{\partial p^b}+\frac{3}{2}{\rm i}
  \hbar p_a\phi^{A_1...A_{2s}}\nonumber \\
\!\!\!\!&+\!\!\!\!&{\rm i}2s\hbar\,\sigma_a^{AA'}p_{A'}{}^{B}\delta^{(A_1}
  _{(A}\varepsilon_{B)(B_1}\delta^{A_2}_{B_2}\cdots\delta^{A_{2s})}_{B_{2s})}
  \phi^{B_1...B_{2s}}. \label{eq:A.2.4b}
\end{eqnarray}
By (\ref{eq:A.2.2a}), in this representation, ${\bf p}_a{\bf p}^a$ is $\mu
^2$-times the identity operator; and by (\ref{eq:A.2.4a}) ${\bf S}_a{\bf S}
^a$ is $w=-\hbar^2\mu^2s(s+1)$-times the identity operator. Therefore, 
${\cal H}_{\mu,s}$ is a subspace of, and, in fact, it coincides with the 
carrier space of the unitary, irreducible representation of $E(1,3)$ 
labeled by $\mu$ and $s$. 


\subsection{The NP form of operators}
\label{sub-A.3}

By (\ref{eq:A.2.2a}) and the definition of the spinor components $\phi_r$, 
$r=0,...,2s$, for the Newman--Penrose (NP) form of the energy-momentum 
vector operator we obtain 
\begin{equation}
\bigl({\bf p}_a\phi\bigr)_r:=\bigl({\bf p}_a\phi_{A_1...A_{2s}}\bigr)\iota^{A_1}
\cdots\iota^{A_r}o^{A_{r+1}}\cdots o^{A_{2s}}=p_a\phi_r \label{eq:A.3.0}
\end{equation}
Using the spinor identity $\delta^A_B=\iota^Ao_B-o^A\iota_B$, the spinorial 
expression of the vectors $p_a$, $v_a$, $m_a$ and $\bar m_a$ given in 
Appendix \ref{sub-A.1} and the notations therein, it is also straightforward 
to derive from (\ref{eq:A.2.4a}) the NP form of the Pauli--Lubanski spin 
vector operator: 
\begin{eqnarray}
\bigl({\bf S}_a\phi\bigr)_r:=\!\!\!\!&{}\!\!\!\!&\bigl({\bf S}_a\phi
  _{A_1...A_{2s}}\bigr)\iota^{A_1}\cdots\iota^{A_r}o^{A_{r+1}}\cdots o^{A_{2s}}= 
  \nonumber \\
=\!\!\!\!&{}\!\!\!\!&\mu\hbar\Bigl((s-r)v_a\phi_r+\frac{2s-r}{\sqrt{2}}m_a
  \phi_{r+1}+\frac{r}{\sqrt{2}}\bar m_a\phi_{r-1}\Bigr) \label{eq:A.3.1}
\end{eqnarray}
for $r=0,1,...,2s$. 

Contracting (\ref{eq:A.2.4b}) with the vectors of the basis $\{p^a,v^a,m^a,
\bar m^a\}$, we obtain that 
\begin{eqnarray}
p^a{\bf C}_a\phi^{A_1...A_{2s}}&\!\!\!\!=&\!\!\!\!\frac{3}{2}{\rm i}\hbar\mu^2
  \phi^{A_1...A_{2s}}, \label{eq:A.3.2a} \\
v^a{\bf C}_a\phi^{A_1...A_{2s}}&\!\!\!\!=&\!\!\!\!-{\rm i}\hbar\mu^2v^e\bigl(
  \partial_e\phi^{A_1...A_{2s}}\bigr)- \nonumber \\
&\!\!\!\!{}&\!\!\!\!-{\rm i}\hbar s\mu\,o^{(A_1}\phi^{A_2...A_{2s})B}\iota_B-{\rm i}
  \hbar s\mu\,\iota^{(A_1}\phi^{A_2...A_{2s})B}o_B, \label{eq:A.3.2b} \\
m^a{\bf C}_a\phi^{A_1...A_{2s}}&\!\!\!\!=&\!\!\!\!-{\rm i}\hbar\mu^2m^e\bigl(
  \partial_e\phi^{A_1...A_{2s}}\bigr)+{\rm i}\sqrt{2}\hbar s\mu\,o^{(A_1}\phi
  ^{A_2...A_{2s})B}o_B, \label{eq:A.3.2c} \\
\bar m^a{\bf C}_a\phi^{A_1...A_{2s}}&\!\!\!\!=&\!\!\!\!-{\rm i}\hbar\mu^2\bar 
  m^e\bigl(\partial_e\phi^{A_1...A_{2s}}\bigr)-{\rm i}\sqrt{2}\hbar s\mu\,\iota
  ^{(A_1}\psi^{A_2...A_{2s})B}\iota_B. \label{eq:A.3.2d}
\end{eqnarray}
To find their components in the basis $\{o^A,\iota^A\}$, let us use $\delta
^a_b=p^ap_b/\mu^2-v^av_b-m^a\bar m_b-\bar m^am_b$, equations (\ref{eq:A.1.4b}), 
(\ref{eq:A.1.5.a}), (\ref{eq:A.1.5.b}) and the definitions (\ref{eq:A.1.9}) 
of the edth operators. We obtain that, for any $r=0,1,...,2s$, 
\begin{eqnarray}
\bigl({\bf C}_a\phi\bigr)_r&\!\!\!\!:=&\!\!\!\!\bigl({\bf C}_a\phi
  _{A_1...A_{2s}}\bigr)\iota^{A_1}\cdots\iota^{A_r}o^{A_{r+1}}\cdots o^{A_{2s}}= 
  \nonumber \\
&\!\!\!\!=&\!\!\!\!\Bigl(\frac{1}{\mu^2}p_ap^b-v_av^b-m_a\bar m^b-\bar m_am^b
  \Bigr)\bigl({\bf C}_b\phi_{A_1...A_{2s}}\bigr)\iota^{A_1}\cdots\iota^{A_r}o
  ^{A_{r+1}}\cdots o^{A_{2s}}= \nonumber \\
&\!\!\!\!=&\!\!\!\!\frac{3}{2}{\rm i}\hbar p_a\phi_r+{\rm i}\hbar\mu p^0\,
  v_a\frac{\partial\phi_r}{\partial p}+ \nonumber \\
&\!\!\!\!{}&\!\!\!\!+{\rm i}\hbar\mu^2\,m_a\Bigl({\edth}'\phi_r-\frac{2s-r}
  {\sqrt{2}p}\frac{p^0}{\mu}\phi_{r+1}\Bigr)+{\rm i}\hbar\mu^2\,\bar m_a\Bigl(
  {\edth}\phi_r+\frac{r}{\sqrt{2}p}\frac{p^0}{\mu}\phi_{r-1}\Bigr) 
\label{eq:A.3.3}
\end{eqnarray}
holds. Using (\ref{eq:3.1.4}), the definitions yield that $v^ap^b{\bf J}_{ab}
=v^a{\bf C}_a$, $m^ap^b{\bf J}_{ab}=m^a{\bf C}_a$, $\bar m^ap^b{\bf J}_{ab}=
\bar m^a{\bf C}_a$, ${\rm i}\mu m^av^b{\bf J}_{ab}=-m^a{\bf S}_a$, ${\rm i}
\mu\bar m^av^b{\bf J}_{ab}=\bar m^a{\bf S}_a$ and ${\rm i}\mu m^a\bar m^b
{\bf J}_{ab}=v^a{\bf S}_a$ hold. Then by (\ref{eq:A.3.1}) and 
(\ref{eq:A.3.3}) these imply that 
\begin{eqnarray}
\bigl({\bf J}_{ab}\phi\bigr)_r\!\!\!\!&=\!\!\!\!&-{\rm i}\hbar\bigl(p_av_b-
  p_bv_a\bigr)\frac{p^0}{\mu}\frac{\partial\phi_r}{\partial p}-{\rm i}\hbar
  \bigl(p_am_b-p_bm_a\bigr)\Bigl({\edth}'\phi_r-\frac{2s-r}{\sqrt{2}}
  \frac{p^0}{\mu p}\phi_{r+1}\Bigr)- \nonumber\\
\!\!\!\!&{}\!\!\!\!&-{\rm i}\hbar\bigl(p_a\bar m_b-p_b\bar m_a\bigr)\Bigl(
  {\edth}\phi_r+\frac{r}{\sqrt{2}}\frac{p^0}{\mu p}\phi_{r-1}\Bigr)-{\rm i}
  \hbar\bigl(v_am_b-v_bm_a\bigr)\frac{2s-r}{\sqrt{2}}\phi_{r+1}+ \nonumber\\
\!\!\!\!&{}\!\!\!\!&+{\rm i}\hbar\bigl(v_a\bar m_b-v_b\bar m_a\bigr)
  \frac{r}{\sqrt{2}}\phi_{r-1}-{\rm i}\hbar\bigl(m_a\bar m_b-m_b\bar m_a
  \bigr)(s-r)\phi_r, \label{eq:A.3.4}
\end{eqnarray}
which is the NP form of the angular momentum operator ${\bf J}_{ab}$. 


\subsection{Matrix elements of homogeneous polynomials of ${\bf p}^i$}
\label{sub-A.4}

Denoting the $L_2$ scalar product on the \emph{unit 2-sphere} ${\cal S}_1$ 
by $\langle\, .\,\vert\, .\,\rangle_1$, by the orthonormality of the spin 
weighted spherical harmonics we have that 
\begin{equation}
\langle{}_\sigma Y_{k,n}\vert{\bf p}^0\vert{}_\sigma Y_{j,m}\rangle_1:=
\oint_{{\cal S}_1}p^0\,\overline{{}_\sigma Y_{k,n}}\,{}_\sigma Y_{j,m}{\rm d}
{\cal S}_1=p^0\delta_{k,j}\delta_{n,m}. \label{eq:A.4.0}
\end{equation}
In the appendix of \cite{Sz22a} we calculated the matrix elements of ${\bf 
p}^i$ in the basis of spin weighted spherical harmonics. There we found that 
the only non-zero matrix elements are
\begin{eqnarray}
\langle{}_\sigma Y_{j+1,n}\vert{\bf p}^1\vert{}_\sigma Y_{j,m}\rangle_1\!\!\!\!&{}
  \!\!\!\!&=\frac{p}{2(j+1)}\sqrt{\frac{(j+\sigma+1)(j-\sigma+1)}{(2j+1)(2j
  +3)}}\label{eq:A.4.1a} \\
\times\Bigl(\!\!\!\!&{}\!\!\!\!&\sqrt{(j-m+1)(j-m+2)}\delta_{n,m-1}-
  \sqrt{(j+m+1)(j+m+2)}\delta_{n,m+1}\Bigr), \nonumber\\
\langle{}_\sigma Y_{j,n}\vert{\bf p}^1\vert{}_\sigma Y_{j,m}\rangle_1\!\!\!\!&{}
  \!\!\!\!&=\frac{p\sigma}{2j(j+1)} \label{eq:A.4.1b}\\
\times\Bigl(\!\!\!\!&{}\!\!\!\!&\sqrt{(j+m)(j-m+1)}\delta_{n,m-1}+\sqrt{(j-m)
  (j+m+1)}\delta_{n,m+1}\Bigr), \nonumber\\
\langle{}_\sigma Y_{j-1,n}\vert{\bf p}^1\vert{}_\sigma Y_{j,m}\rangle_1\!\!\!\!&{}
  \!\!\!\!&=\frac{p}{2j}\sqrt{\frac{(j+\sigma)(j-\sigma)}{(2j-1)(2j+1)}}
  \label{eq:A.4.1c}\\
\times\Bigl(\!\!\!\!&{}\!\!\!\!&\sqrt{(j-m)(j-m-1)}\delta_{n,m+1}-\sqrt{(j+m)
  (j+m-1)}\delta_{n,m-1}\Bigr); \nonumber
\end{eqnarray}
\begin{eqnarray}
\langle{}_\sigma Y_{j+1,n}\vert{\bf p}^2\vert{}_\sigma Y_{j,m}\rangle_1\!\!\!\!&{}
  \!\!\!\!&={\rm i}\frac{p}{2(j+1)}\sqrt{\frac{(j+\sigma+1)(j-\sigma+1)}{(2j
  +1)(2j+3)}}\label{eq:A.4.2a} \\
\times\Bigl(\!\!\!\!&{}\!\!\!\!&\sqrt{(j+m+1)(j+m+2)}\delta_{n,m+1}+
  \sqrt{(j-m+1)(j-m+2)}\delta_{n,m-1}\Bigr), \nonumber\\
\langle{}_\sigma Y_{j,n}\vert{\bf p}^2\vert{}_\sigma Y_{j,m}\rangle_1\!\!\!\!&{}
  \!\!\!\!&={\rm i}\frac{p\sigma}{2j(j+1)} \label{eq:A.4.2b}\\
\times\Bigl(\!\!\!\!&{}\!\!\!\!&\sqrt{(j+m)(j-m+1)}\delta_{n,m-1}-
  \sqrt{(j-m)(j+m+1)}\delta_{n,m+1}\Bigr), \nonumber\\
\langle{}_\sigma Y_{j-1,n}\vert{\bf p}^2\vert{}_\sigma Y_{j,m}\rangle_1\!\!\!\!&{}
  \!\!\!\!&=-{\rm i}\frac{p}{2j}\sqrt{\frac{(j+\sigma)(j-\sigma)}{(2j-1)
  (2j+1)}}\label{eq:A.4.2c}\\
\times\Bigl(\!\!\!\!&{}\!\!\!\!&\sqrt{(j+m)(j+m-1)}\delta_{n,m-1}+
  \sqrt{(j-m)(j-m-1)}\delta_{n,m+1}\Bigr), \nonumber
\end{eqnarray}
\begin{eqnarray}
\langle{}_\sigma Y_{j+1,n}\vert{\bf p}^3\vert{}_\sigma Y_{j,m}\rangle_1\!\!\!\!&=
  \!\!\!\!&\frac{p}{j+1}\sqrt{\frac{(j+\sigma+1)(j-\sigma+1)(j+m+1)(j-m+1)}
  {(2j+1)(2j+3)}}\delta_{n,m}, \,\,\,\,\,\,\,\,\,\,\label{eq:A.4.3a}\\
\langle{}_\sigma Y_{j,n}\vert{\bf p}^3\vert{}_\sigma Y_{j,m}\rangle_1\!\!\!\!&=
  \!\!\!\!&\frac{pm\sigma}{j(j+1)}\delta_{n,m}, \label{eq:A.4.3b}\\
\langle{}_\sigma Y_{j-1,n}\vert{\bf p}^3\vert{}_\sigma Y_{j,m}\rangle_1\!\!\!\!&=
  \!\!\!\!&\frac{p}{j}\sqrt{\frac{(j+\sigma)(j-\sigma)(j+m)(j-m)}{(2j-1)
  (2j+1)}}\delta_{n,m}. \label{eq:A.4.3c}
\end{eqnarray}
Thus, the subspaces spanned by ${}_\sigma Y_{j,m}$ with given $\sigma$ and 
$j$ are not invariant under the action of the momentum operators; and while 
${\bf p}^3$ does not change the index $m$, ${\bf p}^1\pm{\rm i}{\bf p}^2$ 
increases/decreases the value of $m$. 

Also, using (\ref{eq:A.4.1a})-(\ref{eq:A.4.3c}), we found that 
\begin{eqnarray}
\langle{}_\sigma Y_{j,m}\vert{\bf p}^1{\bf p}^1\vert{}_\sigma Y_{j,m}\rangle_1
  \!\!\!\!&{}\!\!\!\!&=\langle{}_\sigma Y_{j,m}\vert{\bf p}^2{\bf p}^2\vert
  {}_\sigma Y_{j,m}\rangle_1=\frac{(p)^2}{j(j+1)(2j-1)(2j+3)}\Bigl(-3\sigma^2
  m^2+\nonumber \\
\!\!\!\!&{}\!\!\!\!&+j(j+1)\bigl(\sigma^2+m^2\bigr)+j(j+1)\bigl(j^2+j-1\bigr)
  \Bigr), \label{eq:A.4.4a}\\
\langle{}_\sigma Y_{j,m}\vert{\bf p}^3{\bf p}^3\vert{}_\sigma Y_{j,m}\rangle
  \!\!\!\!&{}\!\!\!\!&=\frac{(p)^2}{j(j+1)(2j-1)(2j+3)}\Bigl(6\sigma^2m^2-
  \nonumber \\
\!\!\!\!&{}\!\!\!\!&-2j(j+1)\bigl(\sigma^2+m^2\bigr)+j(j+1)\bigl(2j^2+2j-1
  \bigr)\Bigr); \label{eq:A.4.4b}
\end{eqnarray}
and all the other components of $\langle{}_\sigma Y_{j,m}\vert{\bf p}^i
{\bf p}^j\vert{}_\sigma Y_{j,m}\rangle$ are vanishing. Although at first 
sight these are singular for $\vert\sigma\vert=0,1/2$, but they are not; 
and they are $(p)^2/3$ is both cases. 

$\langle{}_{\pm s}Y_{s,m}\vert{\bf p}^i{\bf p}^j{\bf p}^k\vert{}_{\pm s}Y_{s,m}
\rangle_1$ can be calculated in a similar way, and for its non-zero matrix 
elements we obtain 
\begin{eqnarray}
\langle{}_{\pm s}Y_{s,m}\vert{\bf p}^1{\bf p}^1{\bf p}^3\vert{}_{\pm s}Y_{s,m}
  \rangle_1\!\!\!\!&=\!\!\!\!&\langle{}_{\pm s}Y_{s,m}\vert{\bf p}^2{\bf p}^2
  {\bf p}^3\vert{}_{\pm s}Y_{s,m}\rangle_1= \nonumber \\
\!\!\!\!&=\!\!\!\!&\pm\frac{(p)^3m}{(s+1)(s+2)(2s+3)}\bigl((s+1)^2-m^2\bigr), 
   \label{eq:A.4.5a}\\
\langle{}_{\pm s}Y_{s,m}\vert{\bf p}^3{\bf p}^3{\bf p}^3\vert{}_{\pm s}Y_{s,m}
   \rangle_1\!\!\!\!&=\!\!\!\!&\pm\frac{(p)^3m}{(s+1)(s+2)(2s+3)}\bigl(3s+4
   +2m^2\bigr). \label{eq:A.4.5b}
\end{eqnarray}
$\langle{}_{\pm s}Y_{s,m}\vert{\bf p}^i{\bf p}^j{\bf p}^k{\bf p}^l\vert{}_{\pm s}
Y_{s,m}\rangle_1$ could also be calculated in the same manner, but we need to 
know only that, for large $m=s$, 
\begin{equation}
\langle{}_{\pm s}Y_{s,s}\vert{\bf p}^i{\bf p}^j{\bf p}^k{\bf p}^l\vert{}_{\pm s}
Y_{s,s}\rangle_1=\Bigl(\frac{ps}{s+1}\Bigr)^4\delta^i_3\delta^j_3\delta^k_3
\delta^l_3+O(\frac{1}{\sqrt{s}}) \label{eq:A.4.6}
\end{equation}
holds, which can be derived directly from (\ref{eq:A.4.1a})-(\ref{eq:A.4.3c}), 
too. 

We also need to know $\langle{}_\sigma Y_{j,m}\vert m^i\bar m^j\vert{}_\sigma Y
_{j,m}\rangle_1$. Since $\eta^{ab}=p^ap^b/\mu^2-v^av^b-m^a\bar m^b-\bar m^am^b$, 
we have that $\delta^{ij}=-p^ip^j/\mu^2+v^iv^j+m^i\bar m^j+\bar m^im^j$, i.e. 
$m^i\bar m^j+\bar m^im^j=\delta^{ij}-p^ip^j/(p)^2$. Hence, 
\begin{equation}
\langle{}_\sigma Y_{j,m}\vert\frac{1}{2}(m^i\bar m^j+m^j\bar m^i)\vert
{}_\sigma Y_{j,m}\rangle_1=\frac{1}{2}\delta^{ij}-\frac{1}{2}\langle{}_\sigma Y
_{j,m}\vert\frac{{\bf p}^i{\bf p}^j}{(p)^2}\vert{}_\sigma Y_{j,m}\rangle_1.
\label{eq:A.4.7a}
\end{equation}
Using $m^i\bar m^j-m^j\bar m^i={\rm i}\delta^{ik}\delta^{jl}\varepsilon_{klr}
p^r/p$, a simple consequence of (\ref{eq:A.1.3}) and (\ref{eq:A.1.4a}), 
we obtain 
\begin{equation}
\langle{}_\sigma Y_{j,m}\vert\frac{1}{2}(m^i\bar m^j-m^j\bar m^i)\vert
{}_\sigma Y_{j,m}\rangle_1=\frac{\rm i}{2}\bigl(\delta^i_1\delta^j_2-\delta^i_2
\delta^j_1\bigr)\frac{\sigma m}{j(j+1)}, \label{eq:A.4.7b}
\end{equation}
where we used (\ref{eq:A.4.3a})-(\ref{eq:A.4.3c}), too. Finally, the sum 
of (\ref{eq:A.4.7a}) and (\ref{eq:A.4.7b}) together with (\ref{eq:A.4.4a}) 
and (\ref{eq:A.4.4b}) give the explicit form of $\langle{}_\sigma Y_{j,m}
\vert m^i\bar m^j\vert{}_\sigma Y_{j,m}\rangle_1$.


\end{document}